\documentclass[twoside]{article}

\usepackage[accepted]{template}

\usepackage{hyperref}
\usepackage{url}
\usepackage{booktabs}
\usepackage{graphicx}
\usepackage{subcaption}
\usepackage{pifont}
\usepackage{amsmath}
\usepackage{yfonts}
\usepackage{amssymb}
\usepackage{wrapfig}
\usepackage{mathtools}
\usepackage{amsthm}
\usepackage[dvipsnames]{xcolor}
\usepackage{amsmath}
\usepackage{dsfont}
\usepackage{derivative}  
\usepackage{algpseudocode}
\usepackage{algorithm}
\usepackage{mathtools}
\usepackage{enumitem}
\usepackage{multirow}
\usepackage{xr}  
\makeatletter
\newcommand*{\addFileDependency}[1]{
  \typeout{(#1)}
  \@addtofilelist{#1}
  \IfFileExists{#1}{}{\typeout{No file #1.}}
}
\makeatother

\theoremstyle{plain}
\newtheorem{theorem}{Theorem}[section]
\newtheorem{proposition}[theorem]{Proposition}
\newtheorem{lemma}[theorem]{Lemma}

\theoremstyle{definition}
\newtheorem{definition}[theorem]{Definition}
\newtheorem{assumption}[theorem]{Assumption}

\usepackage{bbold}

\newcommand\independent{\protect\mathpalette{\protect\independenT}{\perp}}
\def\independenT#1#2{\mathrel{\rlap{$#1#2$}\mkern2mu{#1#2}}}

\newcommand\ind{{ {{1}}\hspace{-0,8mm}{\mathrm I}}} 
\DeclareMathOperator*{\argmin}{argmin}

\DeclareMathOperator*{\sign}{sign}
\DeclareMathOperator*{\Var}{Var}
\DeclareMathOperator*{\E}{\mathbb{E}}

\usepackage[
    style=authoryear,
    uniquename=mininit,
  ]{biblatex}

\begin{document}

\twocolumn[

\conftitle{Sharp Bounds for Continuous-Valued Treatment Effects with Unobserved Confounders}

\confauthor{Jean-Baptiste Baitairian\textsuperscript{1,2,3,\Letter} \And Bernard Sebastien\textsuperscript{1} \And  Rana Jreich\textsuperscript{1}}

\vspace{0.1cm}

\confauthor{Sandrine Katsahian\textsuperscript{2,4} \And Agathe Guilloux\textsuperscript{2}}

\vspace{0.15cm}

\confaddress{\textsuperscript{1}Sanofi, Gentilly, France \\ \textsuperscript{2}Inria, Université Paris Cité, Inserm, HeKA, F-75015 Paris, France \\ \textsuperscript{3}Université Paris Cité, Paris, France \\ \textsuperscript{4}CIC-EC 1418 - Paris HEGP, Paris, France}
]

\begin{abstract}
In causal inference, treatment effects are typically estimated under the ignorability, or unconfoundedness, assumption, which is often unrealistic in observational data. By relaxing this assumption and conducting a sensitivity analysis, we introduce novel bounds and derive confidence intervals for the Average Potential Outcome (APO) -- a standard metric for evaluating continuous-valued treatment or exposure effects. We demonstrate that these bounds are sharp under a continuous sensitivity model, in the sense that they give the smallest possible interval under this model, and propose a doubly robust version of our estimators. In a comparative analysis with the method of \textcite{jesson2022scalable}, using both simulated and real datasets, we show that our approach not only yields sharper bounds but also achieves good coverage of the true APO, with significantly reduced computation times.
\end{abstract}

\section{INTRODUCTION}  \label{sec:introduction}

Estimating the causal effects of continuous interventions is crucial across many domains, including life sciences or economics. Understanding the impact of air pollutant concentration on cardiovascular mortality, drug concentration in plasma on tumor size, or income on demand for goods or services are examples of such applications of interest.
In particular, observational data, or more generally real-world data, offer a significant opportunity to enhance clinical drug development and support regulatory decisions by using, for instance, Electronic Health Records (EHR), medical claims data or measurements from wearable devices.
Since the 21\textsuperscript{st} Century Cures Act, the United States Food and Drug Administration (FDA) has even been requested to develop a program in order to evaluate how real-world evidence could be used for medical product approvals \parencite{ding2023sensitivity}.

To leverage observational data in the context of treatment effect estimation, several statistical methodologies have been developed for binary and continuous treatments. Most of these approaches rely on a strong assumption, known as \textit{ignorability}, \textit{unconfoundedness} or \textit{exogeneity} \parencite{rubin1974estimating} in causal inference. This assumption posits that, inside a subgroup of units that share similar observed characteristics, the treatment can be considered randomly assigned. In the case of continuous treatments, the \textit{Average Potential Outcome} (APO), sometimes named \textit{dose-response function}, stands out as a metric of reference to evaluate treatment effect. Among the various methods that assume ignorability, noteworthy estimation approaches of the APO include \textit{Inverse Probability Weighting}-like methods \parencite{imai2004causal, hirano2004propensity, kennedy2017non, kallus2018policy, colangelo2020double}, \textit{Bayesian Additive Regression Trees} (BART) \parencite{hill2011bayesian}, and \textit{Adversarial CounterFactual Regression} (ACFR) \parencite{kazemi2024adversarially}. The conditional counterpart of the APO, the \textit{Conditional Average Potential Outcome} (CAPO), can also be estimated via similar techniques.

Nevertheless, this hypothesis may be overly optimistic given the impossibility to observe all confounding variables, especially in observational studies. To overcome this issue, recent works have suggested using sensitivity models to bound the biased treatment effect estimates, providing intervals as solutions, after the seminal sensitivity analysis of \textcite{cornfield1959smoking} on smoking and lung cancer. In the binary treatment scenario, researchers have explored ways to relax the usual ignorability assumption using the \textit{Marginal Sensitivity Model} (MSM) \parencite{tan2006distributional, zhao2019sensitivity, dorn2022sharp, oprescu2023b, dorn2024doubly, tan2024model} or \textit{Rosenbaum's Sensitivity Model} (RSM) \parencite{rosenbaum2002observational, yadlowsky2022bounds}. More recently, variants of the MSM in the context of risk estimation and fairness assessment \parencite{rambachan2022counterfactual, byun2024auditing}, or in combination with a sensitivity model for the outcome \parencite{zhang2025enhanced} were also proposed.

Following these works, \textcite{jesson2022scalable} extended the MSM to the continuous case (\textit{Continuous Marginal Sensitivity Model}, or CMSM) and provided bounds for the APO.
In the meantime, \textcite{bonvini2022sensitivity} proposed a sensitivity analysis for binary, continuous and time-dependent treatments under the \textit{Structural Causal Model} (SCM) framework \parencite{pearl2009causality}, where a parametric model that depends on the treatment is assigned to the APO.
\textcite{frauen2024sharp} later proposed sharp bounds for a variety of causal effects under a generalized sensitivity model.
In this paper, we present a new methodology that combines the doubly robust kernel-based APO estimator from \textcite{kallus2018policy} with an additional constraint on a weight function, and derive sharp bounds under a sensitivity model introduced in Section~\ref{sec:problem_setting_notations}. Then, we deduce confidence intervals (CI) via the \textit{percentile bootstrap method}, as done in \textcite{zhao2019sensitivity}.
Specifically, this novel approach provides tighter and faster-to-compute bounds for the APO, as compared to \textcite{jesson2022scalable}.

The paper is organized as follows. Problem setting and notations are presented in Section~\ref{sec:problem_setting_notations}. Section~\ref{sec:bounds_capo_apo} provides a detailed description of the novel APO and CAPO bounds, along with convergence results.
Empirical results and comparisons with the method from \textcite{jesson2022scalable} are presented in Section~\ref{sec:experiments} on simulated and real datasets, supporting our theoretical findings while displaying the computational efficiency of our estimators.
Additional theoretical details, proofs and experimental results are provided in the appendices.

\section{PROBLEM SETTING AND NOTATIONS}  \label{sec:problem_setting_notations}

\subsection{Notations and Assumptions}

In the following, we consider the \textit{Neyman-Rubin potential outcome framework} \parencite{neyman1923application, rubin1974estimating} adapted to continuous treatments. We denote by $\mathbf{X} \in \mathcal{X} \subseteq \mathbb{R}^{p_\mathbf{X}}$ the vector of observed  confounders, with $p_\mathbf{X} \geq 1$, $T \in \mathcal{T} \subseteq \mathbb{R}$ the continuous treatment or exposition, and $Y(t) \in \mathcal{Y} \subseteq \mathbb{R}$ the potential outcome for a treatment value $t$. If $Y$ is the observed outcome and $T=t$, then we assume that $Y=Y(t)$ (\textit{consistency} and \textit{non-interference}). Estimated quantities are denoted with a hat $\hat \cdot$, unless stated otherwise. Conditional expectations are equally written with or without a subscript, e.g.\ $\mathbb{E}_{\mathbf{X}=\mathbf{x}}[Y] \coloneq \E[Y|\mathbf{X}=\mathbf{x}]$.

For a fixed treatment or exposition value $\tau \in \mathcal{T}$ and covariate vector $\mathbf{x} \in \mathcal{X}$, we are interested in estimating and bounding the APO $\theta(\tau)$ and CAPO $\theta(\tau, \mathbf{x})$, which are defined as
\begin{equation*}
    \theta(\tau) \coloneq \E [Y(\tau)] \quad \text{and} \quad \theta(\tau, \mathbf{x}) \coloneq \mathbb{E}_{\mathbf{X}=\mathbf{x}}[Y(\tau)].
\end{equation*}
The APO and CAPO are commonly estimated under the ignorability assumption for continuous treatments \parencite{hirano2004propensity, kennedy2017non, kallus2018policy}:
\begin{equation*}
    \forall t \in \mathcal{T}, \, Y(t) \independent T \mid \mathbf{X}.
\end{equation*}
For convenience, we refer to it as $\mathbf{X}$-\textit{ignorability}. Under this hypothesis and consistency of the outcomes, notice that
\begin{align}  \label{eqn:capo_eta_link_ign}
    \theta(\tau) & = \E[\theta(\tau, \mathbf{X})] = \E[\mathbb{E}[Y(\tau)|\mathbf{X}]] \nonumber \\
    & = \E[\mathbb{E}_{\mathbf{X}, T=\tau}[Y(T)]] = \E[\eta(\tau, \mathbf{X})]
\end{align}
where $\eta(t, \mathbf{x}) \coloneq \mathbb{E}_{\mathbf{X}=\mathbf{x}, T=t}[Y]$ is a conditional expectation that could be estimated by regression. \textcite{kallus2018policy} and \textcite{colangelo2020double} proposed simple, \textit{stabilized} and/or \textit{doubly robust} versions of a kernel-based estimator of the APO, all recalled in Appendix~\ref{sec:stab_augm_apo}. We focus here on the doubly robust estimator from \textcite{kallus2018policy}. For a fixed treatment value $\tau \in \mathcal{T}$, and considering an i.i.d. observed sample $\mathcal{D} = \{ (\mathbf{X}_i, T_i, Y_i) \}_{i=1}^{n}$ of size $n$, they express it as
\begin{align}  \label{eqn:estimator_under_ignorability}
\hat \theta_h(\tau) & = \frac{1}{n} \sum_{i=1}^n \frac{K_h(T_i-\tau)}{\hat f(T=T_i | \mathbf{X}=\mathbf{X}_i)} \bigl(Y_i - \hat \eta(T_i, \mathbf{X}_i) \bigr) \nonumber \\
& \quad + \hat \eta(\tau),
\end{align}
where $\hat \eta(\tau) = \sum_{i=1}^n \hat \eta(\tau, \mathbf{X}_i) / n$.
$K_h$ is defined as $K_h(s) = K(s/h) / h$, where $h>0$ is a \textit{bandwidth} and $K$ is a \textit{kernel}. Common choices of $K$ include the Epanechnikov or Gaussian kernels. The purpose of $K_h$ is to localize the estimation around the treatment of interest $\tau$. See Assumption~\ref{ass:kernel} in appendix for more details on the kernel.

The density of $T$ conditionally on $\mathbf{X}=\mathbf{x}$, $t \mapsto f(T=t | \mathbf{X}=\mathbf{x})$, is known as \textit{Generalized Propensity Score} (GPS) \parencite{hirano2004propensity}. Reweighting the observed outcomes $Y_i$s by the inverse of the GPS gives more importance to units that had less chance to be exposed to treatment $T_i$ and, thus, artificially rebalances the data. The GPS generalizes the common \textit{Propensity Score} \parencite{rosenbaum1983central} used in the binary treatment case. As in \textcite{kallus2018policy}, we assume that the GPS exists and that it is positive for all $\mathbf{x} \in \mathcal{X}$ and $t \in \mathcal{T}$ (see  Assumption~\ref{ass:positivity_cond_treatment} of \textit{positivity} in appendix). Nevertheless, we warn the reader that positivity can be a relatively strong assumption with continuous treatments, and we refer, for example, to \textcite{kennedy2017non} for a way to estimate the APO under near violations of this assumption.

However, as mentioned earlier, $\mathbf{X}$-ignorability is rarely satisfied in practice due to the presence of unobserved confounders $\mathbf{U} \in \mathcal{U} \subseteq \mathbb{R}^{p_\mathbf{U}}$, where $p_\mathbf{U} \geq 1$. A more reasonable assumption is to impose $(\mathbf{X}, \mathbf{U})$-\textit{ignorability}:
\begin{equation} \label{eqn:XU_ignorability}
    \forall t \in \mathcal{T}, \, Y(t) \independent T | \mathbf{X}, \mathbf{U}.
\end{equation}
It formalizes the idea that the unmeasured confounders $\mathbf{U}$ capture all common causes of $T$ and $Y(t)$ not included in $\mathbf{X}$ \parencite{kallus2021causal}. Under this assumption, the final equality in~\eqref{eqn:capo_eta_link_ign} no longer holds, meaning that $\theta(\tau) \neq \E[\eta(\tau, \mathbf{X})]$, and the estimator of the APO from Equation~\eqref{eqn:estimator_under_ignorability} becomes biased. In what follows, we show that, under $(\mathbf{X}, \mathbf{U})$-ignorability and a sensitivity model introduced in the next subsection, \textit{bounds} for the APO and CAPO can be derived and subsequently estimated. This is consistent with findings from~\textcite{jesson2022scalable}, except that we propose sharp bounds.

\subsection{Continuous Marginal Sensitivity Model}

As mentioned in Section~\ref{sec:introduction}, previous works suggested bounding binary causal treatment effects under the MSM from \textcite{tan2006distributional} in presence of unobserved confounders. This sensitivity model involves an odds ratio of propensity scores that is bounded by a user-defined sensitivity parameter.
Following this idea, a natural way to define a Continuous Marginal Sensitivity Model (CMSM) would be as in the following definition.
\begin{definition}{(CMSM, formulation with $\mathbf{U}$)}  \label{def:cmsm_U}
Under positivity Assumption~\ref{ass:positivity_cond_treatment}, there exists a \textit{sensitivity parameter} $\Gamma \geq 1$ such that, for a given treatment value $\tau \in \mathcal{T}$, and for all $(\mathbf{x}, \mathbf{u}) \in \mathcal{X} \times \mathcal{U}$,
\begin{equation}  \label{eqn:cmsm_U}
    \Gamma^{-1} \leq \frac{f(T=\tau|\mathbf{X}=\mathbf{x}, \mathbf{U}=\mathbf{u})}{f(T=\tau|\mathbf{X}=\mathbf{x})} \leq \Gamma.
\end{equation}
\end{definition}
The CMSM from Definition~\ref{def:cmsm_U} involves a conditional density with respect to $\mathbf{X}=\mathbf{x}$ and $\mathbf{U}=\mathbf{u}$, $t \mapsto f(T=t | \mathbf{X}=\mathbf{x}, \mathbf{U}=\mathbf{u})$, which is the true but inestimable Generalized Propensity Score, had we observed all possible confounders.
For the CAPO, the same definition applies but for a fixed treatment value $\tau$ and fixed vector of covariates $\mathbf{x}$.

On the other hand, \textcite{jesson2022scalable} chose to express their CMSM in terms of the potential outcome $Y(\tau)$ instead of $\mathbf{U}$, as in the definition below (see Appendix~\ref{sec:appendix_def} for the exact definition they used).
\begin{definition}{(CMSM, formulation with $Y(\tau)$)}  \label{def:cmsm_Y}
Under positivity Assumption~\ref{ass:positivity_cond_treatment}, there exists a \textit{sensitivity parameter} $\Gamma \geq 1$ such that, for a given treatment value $\tau \in \mathcal{T}$, and for all $(\mathbf{x}, y) \in \mathcal{X} \times \mathcal{Y}$,
\begin{equation}  \label{eqn:cmsm_Y}
    \Gamma^{-1} \leq \frac{f(T=\tau|\mathbf{X}=\mathbf{x}, Y(\tau)=y)}{f(T=\tau|\mathbf{X}=\mathbf{x})} \leq \Gamma.
\end{equation}
\end{definition}

We can show that the proposed CMSM~\eqref{eqn:cmsm_U} is equivalent to the one considered in \textcite{jesson2022scalable}, i.e.\ that it suffices to consider the case where $\mathbf{U}$ is equal to the potential outcome $Y(\tau)$ \parencite{robins2002covariance}. This idea is formalized in the following proposition.
\begin{proposition}  \label{prop:cmsm_equivalence}
    There exists a \textit{sensitivity parameter} $\Gamma \geq 1$ such that, for a given treatment value $\tau \in \mathcal{T}$, for all $(\mathbf{x}, \mathbf{u}, y) \in \mathcal{X} \times \mathcal{U} \times \mathcal{Y}$, under Assumptions~\ref{ass:positivity_cond_treatment} and \ref{ass:positivity_cond_outcome}, the CMSM from Definition~\ref{def:cmsm_U} is equivalent to the one from Definition~\ref{def:cmsm_Y}.
\end{proposition}
See Appendix \ref{sec:proof_cmsm_equivalence} for a proof. This result is similar to Lemma~1 from \textcite{tan2024model} with the MSM, in the binary treatment case. Note that $\Gamma$ can be seen as a user-defined sensitivity parameter which measures the deviation from the usual $\mathbf{X}$-ignorability assumption: if $\Gamma = 1$, the CMSM becomes equivalent to $\mathbf{X}$-ignorability, as if all confounders were observed; higher values of $\Gamma$ assume a greater effect of the unobserved confounders on the treatment $T$ and a deviation from $\mathbf{X}$-ignorability. $\Gamma$ is also sometimes referred to as a \textit{confounding strength} (e.g., \cite{jin2023sensitivity}). Practical choices of $\Gamma$ are detailed in Section~\ref{sec:real_dataset_exp}. Unless stated otherwise, all theoretical results from Section~\ref{sec:bounds_capo_apo} are now presented under the formulation from Definition~\ref{def:cmsm_Y}.

\section{BOUNDS FOR THE CAPO AND APO}  \label{sec:bounds_capo_apo}

In this section, we present novel bounds for the CAPO and APO in the presence of unobserved confounders. They rely on a constraint on a likelihood ratio that we leverage to reach sharp bounds under the CMSM~\eqref{eqn:cmsm_Y} in Theorem~\ref{theo:capo_bounds_and_sharpness}. Additionally, we show that our bounds are also sharper than the ones considered in \textcite{jesson2022scalable}.

\subsection{Weight Function}

To reach the desired solution, we start by working on the CAPO until Theorem~\ref{theo:capo_bounds_and_sharpness}, and extend the results to the APO afterwards. Notice first that, by positivity Assumption~\ref{ass:positivity_cond_outcome}, the CAPO can be rewritten
\begin{align*}
    \theta(\tau, \mathbf{x}) & \coloneq \mathbb{E}_{\mathbf{X}=\mathbf{x}}[Y(\tau)] = \int y f(Y(\tau)=y | \mathbf{X}=\mathbf{x}) \, \mathrm{d}y \\
    & = \int y \, w^\star(y, \mathbf{x}, \tau) f(Y=y | \mathbf{X}=\mathbf{x}, T=\tau) \, \mathrm{d}y \\
    & = \mathbb{E}_{\mathbf{X}=\mathbf{x}, T=\tau}[Y w^\star(Y, \mathbf{x}, \tau)],
\end{align*}
where, for all $(\mathbf{x}, t, y) \in \mathcal{X} \times \mathcal{T} \times \mathcal{Y}$,
\begin{equation*}
     w^\star(y, \mathbf{x}, t) \coloneq \frac{f(Y(t)=y | \mathbf{X}=\mathbf{x})}{f(Y=y | \mathbf{X}=\mathbf{x}, T=t)}
\end{equation*}
exists \parencite{jesson2022scalable}.
Note that the weight function, or likelihood ratio, $w^\star$ fulfills the constraint
\begin{equation}  \label{eqn:constraint_w_star}
    \mathbb{E}_{\mathbf{X}=\mathbf{x}, T=t}[w^\star(Y, \mathbf{x}, t)] = 1
\end{equation}
for all $(\mathbf{x}, t) \in \mathcal{X} \times \mathcal{T}$, and takes values within $[\Gamma^{-1}, \Gamma]$, according to Definition~\ref{def:cmsm_Y}. For computation details, please refer to Appendix~\ref{sec:details_weight_function}.

Using the conditional expectation $\eta$ and the previous constraint~\eqref{eqn:constraint_w_star}, the CAPO can also be written
\begin{align}  \label{eqn:augmented_capo}
    \theta(\tau, \mathbf{x}) & = \mathbb{E}_{\mathbf{X}=\mathbf{x}, T=\tau} \bigl[ \bigl( Y - \eta(\tau, \mathbf{x}) \bigr) w^\star(Y, \mathbf{x}, \tau) \bigr] \nonumber \\
    & \quad + \eta(\tau, \mathbf{x}).
\end{align}
In Equation~\eqref{eqn:augmented_capo}, the CAPO $\theta(\tau, \mathbf{x})$ can be decomposed into two terms: the first one is the estimation of $\theta(\tau, \mathbf{x})$ under the $\textbf{X}$-ignorability assumption, $\eta(\tau, \mathbf{x})$, which is biased in practice because unobserved confounders may exist; the second one (the conditional expectancy) leverages the residuals $Y - \eta(\tau, \mathbf{x})$ and can be seen as a correction of this biased estimate, had we taken unobserved confounders into account (i.e., through the likelihood ratio $w^\star$).

\subsection{Sharp Bounds for the CAPO and APO}  \label{sec:sharp_bounds}

Under the CMSM~\eqref{eqn:cmsm_Y}, we define the bounds for the CAPO as
\begin{align}
    \theta^-(\tau, \mathbf{x}) & \coloneq \inf_{w \in \mathcal W^\star_{\tau}} \mathbb E_{\mathbf{X}=\mathbf{x}, T=\tau} \bigl[ \bigl(Y - \eta(\tau, \mathbf{x})\bigr) w(Y, \mathbf{x}, \tau) \bigr] \nonumber \\
    & \quad + \eta(\tau, \mathbf{x}), \label{eqn:capo_lb} \\
    \theta^+(\tau, \mathbf{x}) & \coloneq \sup_{w \in \mathcal W^\star_{\tau}} \mathbb E_{\mathbf{X}=\mathbf{x}, T=\tau} \bigl[ \bigl(Y - \eta(\tau, \mathbf{x})\bigr) w(Y, \mathbf{x}, \tau) \bigr] \nonumber \\
    & \quad + \eta(\tau, \mathbf{x}), \label{eqn:capo_ub}
\end{align}
where $\mathcal{W}^\star_\tau$ is the set of functions $w : \mathcal{Y} \times \mathcal{X} \times \mathcal{T} \to [\Gamma^{-1},\Gamma]$ that satisfy Equation~\eqref{eqn:constraint_w_star}. In other words, these bounds correspond to the lowest and highest possible values for the CAPO when the weight function varies in $\mathcal{W}^\star_\tau$. Elements from convex analysis allow to solve the minimization and maximization problems and lead to the following Theorem~\ref{theo:capo_bounds_and_sharpness}. The resulting interval $[\theta^-(\tau, \mathbf{x}), \theta^+(\tau, \mathbf{x})]$ is called a \textit{partially identified set} for $\theta(\tau, \mathbf{x})$: the most informative bounds that could be obtained with an infinite amount of data under the sensitivity assumption (see e.g., \cite{manski2003partial} or \cite{dorn2022sharp}).
\begin{theorem}  \label{theo:capo_bounds_and_sharpness}
    Under positivity Assumption~\ref{ass:positivity_cond_treatment} and the CMSM~\eqref{eqn:cmsm_Y}, the solutions to the optimization problems~\eqref{eqn:capo_lb} and \eqref{eqn:capo_ub} are given by
    \begin{align}
        \theta^\pm(\tau, \mathbf{x}) & = \frac{2\gamma - 1}{\gamma} \mathbb E_{\mathbf{X}=\mathbf{x}, T=\tau} \Bigl[Y - \eta(\tau, \mathbf{x}) \Big| Q^\pm_{\mathbf{X}, \tau} \Bigr] \nonumber \\
        & \quad + \eta(\tau, \mathbf{x})  \label{eqn:capo_lb_ub_sol}
    \end{align}
    where $\pm$ stands for either $+$ or $-$, $\gamma = \Gamma / (1 + \Gamma)$, $Q^+_{\mathbf{X}, \tau} = \{ Y > q_\gamma^{\mathbf{X}, \tau} \}$, $Q^-_{\mathbf{X}, \tau} = \{ Y \leq q_{1-\gamma}^{\mathbf{X}, \tau} \}$, and $q_{\upsilon}^{\mathbf{x}, \tau} \coloneq Q(\upsilon; Y|\mathbf{X}=\mathbf{x}, T=\tau)$, the quantile of order $\upsilon$ of the distribution of $Y$ conditionally on $\mathbf{X}=\mathbf{x}$ and $T=\tau$ (see Definition~\ref{def:cond_quantile}).

    Moreover, the partially identified set $[\theta^-(\tau, \mathbf{x}), \theta^+(\tau, \mathbf{x})]$ is sharp under the CMSM~\eqref{eqn:cmsm_Y}, in the sense that it is the smallest possible interval under the CMSM for a given sensitivity parameter $\Gamma$.
\end{theorem}
We refer the reader to Appendix~\ref{proof:capo_bounds_and_sharpness} for a proof. An alternative demonstration of the optimal bounds is given in Appendix~\ref{proof:alt_capo_true_bounds}.
Observe that the conditional expectancy from Equation~\eqref{eqn:capo_lb_ub_sol} is linked to the \textit{Conditional Value at Risk} (CVaR), also known as \textit{Tail Value at Risk} (TVaR) or \textit{Expected Shortfall} (ES) in the financial literature (e.g., see \cite{rockafellar2000optimization}). The proof of Theorem~\ref{theo:capo_bounds_and_sharpness} relies on this connection and the ``Fenchel-Moreau-Rockafellar" dual representation of the CVaR (e.g., see \cite{herdegen2023elementary} for a formal statement of the dual problem). \textcite{dorn2024doubly} also found a connection with the CVaR in the binary treatment case.
Moreover, the bounds from Theorem~\ref{theo:capo_bounds_and_sharpness} are consistent with results stated in Theorem 1 from \textcite{frauen2024sharp} under Pearl's SCM.
Finally, Theorem~\ref{theo:capo_bounds_and_sharpness} implies that our bounds are sharper than the ones considered in \textcite{jesson2022scalable}. This result stems from the fact that we take an additional constraint on the weight function $w$ into consideration (Equation~\eqref{eqn:constraint_w_star}) when deriving our bounds. In the binary treatment case, \textcite{dorn2022sharp} already noticed that, without such an assumption on the binary propensity scores, the estimated bounds could possibly not be compatible with the observed data (Section~2.1.\ from \cite{dorn2022sharp}), and thus, could lead to conservative sensitivity bounds, such as the ones from \textcite{zhao2019sensitivity}. The result is the same in the continuous treatment case where the bounds for the APO can be conservative if this constraint is forgotten, such as in \textcite{jesson2022scalable}. Another way to understand the constraint is that it forces the complete generalized propensity score to, in a way, ``marginalize" to the nominal generalized propensity score.
In the following proposition, we suggest another comparison with the bounds from \textcite{jesson2022scalable} that does not rely on the CVaR. See Appendix~\ref{proof:comparison_us_vs_jesson} for a proof.
\begin{proposition}  \label{prop:comparison_us_vs_jesson}
    The bounds given in \textcite{jesson2022scalable} and defined as
    \begin{align}
        \bar \theta^-(\tau, \mathbf{x}) & = \underset{\kappa \in \mathcal{\bar K}}{\inf} \frac{\mathbb{E}_{\mathbf{X}=\mathbf{x}, T=\tau} [\kappa(Y, \mathbf{X}, \tau) (Y - \eta(\tau, \mathbf{X}))]}{\bigl( \Gamma^2 - 1 \bigr)^{-1} + \mathbb{E}_{\mathbf{X}=\mathbf{x}, T=\tau}[\kappa(Y, \mathbf{X}, \tau)]} \nonumber \\
        & + \eta(\tau, \mathbf{x})  \label{eqn:jesson_lb_capo} \\
        \bar \theta^+(\tau, \mathbf{x}) & = \underset{\kappa \in \mathcal{\bar K}}{\sup} \frac{\mathbb{E}_{\mathbf{X}=\mathbf{x}, T=\tau} [\kappa(Y, \mathbf{X}, \tau) (Y - \eta(\tau, \mathbf{X}))]}{\bigl( \Gamma^2 - 1 \bigr)^{-1} + \mathbb{E}_{\mathbf{X}=\mathbf{x}, T=\tau}[\kappa(Y, \mathbf{X}, \tau)]} \nonumber \\
        & + \eta(\tau, \mathbf{x})  \label{eqn:jesson_ub_capo}
    \end{align}
    with $\mathcal{\bar K} = \bigl\{ \kappa : \mathcal{Y} \times \mathcal{X} \times \mathcal{T} \to [0, 1] \bigr\}$, are sub-optimal in the sense that $\bar \theta^-(\tau, \mathbf{x}) \leq \theta^-(\tau, \mathbf{x})$ and $\theta^+(\tau, \mathbf{x}) \leq \bar \theta^+(\tau, \mathbf{x})$.
\end{proposition}

Similarly, for the APO, we can show that the interval $[\theta^-(\tau), \, \theta^+(\tau)]$ is sharp under the CMSM~\eqref{eqn:cmsm_Y}, using the relation $\theta^\pm(\tau) = \E[\theta^\pm(\tau, \mathbf{X})]$:
\begin{align}
    \theta^\pm(\tau) & = \frac{2\gamma - 1}{\gamma} \E \biggl[ \mathbb E_{\mathbf{X}, T=\tau} \Bigl[Y - \eta(\tau, \mathbf{X}) \Big| Q^\pm_{\mathbf{X}, \tau} \Bigr] \biggr] \nonumber \\
    & \quad + \eta(\tau),
 \label{eqn:apo_lb_ub_sol}
\end{align}
with $\eta(\tau) \coloneq \E[\eta(\tau, \mathbf{X})]$.

\subsection{Bound Estimation}  \label{sec:bound_estimation}

In binary treatment settings, the treatment $T$ is usually encoded as 0 for control units and 1 for treated units. IPW sensitivity analysis bounds for the average response for the treated and for the control involve indicator functions, respectively $\mathds{1}(T=1)$ and $\mathds{1}(T=0)$. As the treatment or exposition $T$ is continuous, estimating the conditional expectations in Equation~\eqref{eqn:apo_lb_ub_sol} requires a tool to localize the estimation around the treatment of interest $\tau$ because, if no unit had exactly received treatment $\tau$, then $\mathds{1}(T=\tau)$ would be equal to 0. Nonparametric kernel regression provides a solution. As in \textcite{kallus2018policy}, we use kernels and define below \textit{kernelized} versions of the bounds for the APO that are indexed by a bandwidth $h>0$:
\begin{align}
    & \theta_h^\pm(\tau) \coloneq \eta(\tau) \nonumber \\
    & + \frac{2 \gamma - 1}{\gamma} \E \biggl[ \mathbb{E} \biggl[ \frac{K_h(T - \tau)}{f(T|\mathbf{X})} (Y - \eta(\tau, \mathbf{X})) \bigg| \mathbf{X}, Q^\pm_{\mathbf{X}, T} \biggr] \biggr], \label{eqn:kernel_apo_lb_ub}
\end{align}
where $K_h$ is defined as in Equation~\eqref{eqn:estimator_under_ignorability} and can be seen as a smooth relaxation of the indicator function.

Therefore, by defining $\mathcal{I}_\mathcal{D_-} = \{ i \in [\![1, n]\!] \, ; \, Y_i \leq q_{1-\gamma}^{\mathbf{X}_i, T_i} \}$ of cardinality $n_-$, and $\mathcal{I}_\mathcal{D_+} = \{ i \in [\![1, n]\!] \, ; \, Y_i > q_\gamma^{\mathbf{X}_i, T_i} \}$ of cardinality $n_+$, we can obtain estimators of $\theta_h^-(\tau)$ and $\theta_h^+(\tau)$ as follows:
\begin{align}  \label{eqn:tilde_kernel_apo_ulb}
    \tilde \theta_h^\pm(\tau) & = \frac{2\gamma-1}{\gamma n_\pm} \sum_{i \in \mathcal{I}_\mathcal{D_\pm}} \frac{K_h(T_i - \tau)}{f(T_i|\mathbf{X}_i)} (Y_i - \eta(T_i, \mathbf{X}_i)) \nonumber \\
    & \quad + \tilde \eta(\tau),
\end{align}
where $\tilde \eta(\tau) = \frac{1}{n} \sum_{i=1}^n \eta(\tau, \mathbf{X}_i)$. See Appendix~\ref{sec:all_bounds} for another formulation of these estimators.
The following theorem gives rates of convergence of our estimators and implies their consistency.
\begin{theorem}  \label{theo:lim_tilde_theta_h_plus_moins}
    Under Assumptions~\ref{ass:positivity_cond_treatment} to \ref{ass:Q_c2_bounded}, the Mean Squared Error (MSE) of $\tilde \theta^+_h(\tau)$ can be upper bounded as follows:
    \begin{align*}
        \mathrm{MSE} \bigl( \tilde \theta^+_h(\tau) \bigr) \leq \frac{C_{\mathrm{Var}_1}}{nh} + \frac{C_{\mathrm{Var}_2}}{n} + h^4 C^2_{\text{bias}},
    \end{align*}
    where $C_{\mathrm{Var}_1}$, $C_{\mathrm{Var}_2}$ and $C_{\text{bias}}$ are constants defined in Appendix~\ref{proof:lim_tilde_theta_h_plus_moins}.
    The optimal bandwidth that minimizes the upper bound on the MSE of $\tilde \theta^-_{h_n}(\tau)$ and $\tilde \theta^+_{h_n}(\tau)$ is $h_n^\star = \mathcal{O}(n^{-1/5})$, as $n$ tends to $+\infty$. For this value, the optimal MSE is $\mathcal{O}(n^{-4/5})$, as $n$ tends to $+\infty$.
\end{theorem}
See Appendix~\ref{proof:lim_tilde_theta_h_plus_moins} for a proof. Notice that the previous rates of convergence are expected in nonparametric estimation (see, for instance, \cite{tsybakov2009nonparametric} or \cite{kallus2018policy}).

In practice, the densities, conditional quantiles and conditional expectations are estimated. Whatever the method, we denote by $\hat f$, $\hat Q$ and $\hat \eta$ their respective estimators, and by $\hat \theta_h^-(\tau)$ and $\hat \theta_h^+(\tau)$ the resulting bounds. See Section~\ref{sec:experiments} for different ways to compute the three nuisance parameters. We refer to $[\hat \theta_h^-(\tau), \hat \theta_h^+(\tau)]$ as a \textit{Point Estimate Interval} (PEI) (see e.g., \cite{zhao2019sensitivity}) for the APO.

\subsection{Partial and Double Robustness}

Double robustness is an interesting property that allows estimators to be unbiased even if $f(T=t|\mathbf{X}=\mathbf{x})$ or $\eta(t, \mathbf{x})$ is misspecified. We show in Proposition~\ref{prop:partial_robustness} that the suggested bounds can only achieve asymptotic \textit{partial} robustness, in a sense defined below, but that asymptotic doubly robust and \textit{doubly valid} \parencite{dorn2024doubly} ones can also be defined, as stated in Proposition~\ref{prop:double_robustness}.
\begin{proposition}  \label{prop:partial_robustness}
    If the conditional quantiles are correctly specified and under Assumptions~\ref{ass:kernel} to \ref{ass:Q_c2_bounded}, $\theta^\pm_h(\tau)$ are asymptotic partially robust bounds for $\theta^\pm(\tau)$, in the sense that, even if $\eta(t, \mathbf{x})$ is misspecified, $\theta^\pm_h(\tau) \to \theta^\pm(\tau)$, as $h$ tends to 0.
\end{proposition}
\begin{proposition}  \label{prop:double_robustness}
    If the conditional quantiles are correctly specified and under Assumptions~\ref{ass:positivity_cond_treatment}, \ref{ass:kernel} and \ref{ass:Theta_plus_c2_bounded},
    \begin{align}
        & \theta^{\pm, \mathrm{DR}}_h(\tau, \mathbf{x}) = \Theta^\pm(\tau, \mathbf{x}, q_\pm) \nonumber \\
        & + \mathbb{E}_{\mathbf{X}=\mathbf{x}} \biggl[ \frac{K_h(T-\tau)}{f(T|\mathbf{X}=\mathbf{x})} \Bigl[ q_\pm^{\mathbf{x}, T} + (Y - q_\pm^{\mathbf{x}, T}) \Gamma^{\pm \sign(Y - q_\pm^{\mathbf{x}, T})} \nonumber \\
        & \quad \quad \quad \quad \quad \quad \quad \quad \quad \quad - \Theta^\pm(T, \mathbf{x}, q_\pm) \Bigr] \biggr], \label{eqn:doubly_robust_est}
    \end{align}
    where
    \begin{align*}
        & \Theta^\pm : (t, \mathbf{x}, q_\pm) \mapsto q_\pm^{\mathbf{x}, t} 
        \\
        & + \mathbb{E}_{\mathbf{X}=\mathbf{x}, T=t} \Bigl[ (Y - q_\pm^{\mathbf{x}, t}) \Gamma^{\pm \sign(Y - q_\pm^{\mathbf{x}, t})} \Bigr],
    \end{align*}
    $q_+^{\mathbf{X}, t} = q_\gamma^{\mathbf{X}, t}$, and $q_-^{\mathbf{X}, t} = q_{1-\gamma}^{\mathbf{X}, t}$, are asymptotic doubly robust bounds for $\theta^\pm(\tau)$ as $h$ tends to 0, even if $f(T|\mathbf{X})$ or $\Theta^\pm(T, \mathbf{X}, q_\pm)$ is misspecified.

    If the conditional quantiles are not correctly specified and denoted $\bar q_\pm$, assuming $\Theta^\pm(T, \mathbf{X}, \bar q_\pm)$ satisfies Assumption~\ref{ass:Theta_plus_c2_bounded} even when misspecified, the bounds remain asymptotically doubly valid as $h$ tends to 0, in a sense defined in \textcite{dorn2024doubly}, i.e.\ $\theta^{-, \mathrm{DR}}_h(\tau, \mathbf{x}) \underset{h \to 0}{\to} \Theta^-(t, \mathbf{x}, \bar q_-) \leq \theta^-(\tau, \mathbf{x})$ and $\theta^{+, \mathrm{DR}}_h(\tau, \mathbf{x}) \underset{h \to 0}{\to} \Theta^+(t, \mathbf{x}, \bar q_+) \geq \theta^+(\tau, \mathbf{x})$.
\end{proposition}
See proof in Appendix~\ref{sec:proof_partial_robustness} and \ref{sec:proof_double_robustness}.

\subsection{Confidence Interval for the APO via the Percentile Bootstrap}

Finally, we follow \textcite{zhao2019sensitivity}, \textcite{dorn2022sharp} and \textcite{jesson2022scalable}, and build a $(1-\alpha)$-confidence interval for the upper and lower bounds of the APO under our CMSM via the \textit{percentile bootstrap method}. Consider $B$ bootstrap resamples, each of size $n$, obtained after sampling with replacement from the observed data, and denote by $\hat \theta_h^{-, b}(\tau)$ and $\hat \theta_h^{+, b}(\tau)$ the estimations of the lower and upper bounds of the APO on the $b$\textsuperscript{th} bootstrap resample. A two-sided $(1-\alpha)$-CI for the set $\bigl[\theta^-_{h}(\tau), \theta^+_{h}(\tau)\bigr]$ can be obtained after intersecting two one-sided $(1-\alpha/2)$-CIs for the lower and the upper bounds (see computation details in Appendix~\ref{sec:percentile_bootstrap}). The $(1-\alpha)$-CI is then given by $\bigl[ \hat \theta_{h}^{-, \lceil B(\alpha/2) \rceil}(\tau), \, \hat \theta_{h}^{+,\lceil B(1-\alpha/2) \rceil}(\tau) \bigr]$, with
\begin{align*}
    \hat \theta_{h}^{-, \lceil B(\alpha/2) \rceil}(\tau) = \hat Q\Bigl( \alpha/2 \, ; \, \bigl\{ \hat \theta_h^{-, b}(\tau) \bigr\}_{b=1}^B \Bigr) \quad \text{and} \\
    \hat \theta_{h}^{+, \lceil B(1-\alpha/2) \rceil}(\tau) = \hat Q\Bigl( 1-\alpha/2 \, ; \, \bigl\{ \hat \theta_h^{+, b}(\tau) \bigr\}_{b=1}^B \Bigr).
\end{align*}
Here, $\hat Q(\upsilon \, ; \, \{ \hat \theta_h^{-, b}(\tau) \}_{b=1}^B )$ denotes the empirical quantile of order $\upsilon$ of the sample $\{ \hat \theta_h^{-, b}(\tau) \}_{b=1}^B$. A visual representation of the method is given in Appendix~\ref{sec:percentile_bootstrap}. Notice that, for each bootstrap resample, we need to re-estimate $\hat f$ and $\hat \eta$. However, as in \textcite{dorn2022sharp}, we do not re-estimate $\hat Q$ to keep computations tractable.

\section{EXPERIMENTS}  \label{sec:experiments}

In the following, we detail our experiments on simulated and real datasets, where we show that our methods, especially the partially robust one, provide sharp bounds and outperform the existing methodology from \textcite{jesson2022scalable} in terms of computation time and tightness. Our implementation is publicly available at \href{https://github.com/Sanofi-Public/sharp-bounds-for-continuous-treatment-effects}{https://github.com/Sanofi-Public/sharp-bounds-for-continuous-treatment-effects}, under a non-commercial license.

\subsection{Implementation Details}

\subsubsection{Method from \texorpdfstring{\textcite{jesson2022scalable}}{}} \label{sec:jesson_method}

We recall that the lower and upper bounds for the CAPO, as defined by \textcite{jesson2022scalable}, are given by Equations~\eqref{eqn:jesson_lb_capo} and \eqref{eqn:jesson_ub_capo}. To solve the optimization problem, for example, for the upper bound, they prove that the formulation in Equation~\eqref{eqn:jesson_ub_capo} is equivalent to another one where the optimization is performed over the set $\mathcal{\bar K}^H = \{ \kappa(\cdot) : \: \forall y \in \mathcal{Y}, \, \kappa(y) = H(y - y_H) \}$, where $y_H \in \mathcal{Y}$ and $H$ is the Heaviside step function. The problem thus writes
\begin{align}
    \bar \theta^+(\tau, \mathbf{x}) & = \underset{\kappa \in \mathcal{\bar K}^H}{\sup} \frac{\mathrm{E}_{\mathbf{X}=\mathbf{x}, T=\tau} [\kappa(Y) (Y - \eta(\tau, \mathbf{X}))]}{\bigl( \Gamma^2 - 1 \bigr)^{-1} + \mathrm{E}_{\mathbf{X}=\mathbf{x}, T=\tau}[\kappa(Y)]} \nonumber \\
    & + \eta(\tau, \mathbf{x}). \label{eqn:jesson_ub_capo_2}
\end{align}
To solve the optimization problem in Equation~\eqref{eqn:jesson_ub_capo}, they perform a grid search in $\mathcal{\bar K}^H$ (see their Algorithm~1). More precisely, \textit{for each} treatment of interest $\tau$, they sample several points from the distribution $f(Y|\mathbf{X}=\mathbf{x}, T=\tau)$, which results in the set $\hat{\mathcal{Y}}_{\mathbf{x}, \tau}$. Then, they perform the following steps: they initialize the value of the upper bound to $-\infty$; for each unique value in $\hat{\mathcal{Y}}_{\mathbf{x}, \tau}$, they compute the new upper bound; if the new upper bound is greater than the previous one, the previous one is updated by this new value. Finally, they return the last value of the bound (which is also the biggest that was computed on $\hat{\mathcal{Y}}_{\mathbf{x}, \tau}$). During the procedure, each computation of a new upper bound involves two Monte-Carlo integrations over $\hat{\mathcal{Y}}_{\mathbf{x}, \tau}$ to estimate the two expectancies in Equation~\eqref{eqn:jesson_ub_capo}, and the estimation of $f(Y|\mathbf{X}=\mathbf{x}, T=\tau)$ to sample $\hat{\mathcal{Y}}_{\mathbf{x}, \tau}$ and estimate $\eta(\tau, \mathbf{x})$ (see Section~\ref{sec:density_quantile_est} for density estimation). More practical details are given in Appendix~\ref{sec:jesson_implementation}.

\subsubsection{Variance Reduction and Kernel Practical Issue}  \label{sec:var_red_and_kernel_issue}

In practice, as done in \textcite{kallus2018policy}, we work with stabilized versions of our estimators in order to avoid high variance due to extreme generalized propensity weights (see Equations~\eqref{eqn:stab_tilde_kernel_apo_lb_v1} to \eqref{eqn:stab_tilde_kernel_apo_ub_v2} in Appendix~\ref{sec:all_bounds}). In the real data case (Section~\ref{sec:real_dataset_exp}), we clip small propensity weights \parencite{kallus2018policy} by setting them to the 0.1 quantile of the estimated propensity scores if they fall below this value. This leads to a smaller variance as well, but increases bias.

As underlined in \textcite{kallus2018policy}, kernel estimations may become unstable near boundaries, beyond which no more data points can be observed. This is why we limit our estimations to values of $\tau$ between the 0.05 and 0.95 quantiles of the observed treatments. \textcite{kallus2018policy} suggest another solution by truncating and normalizing the kernel. Moreover, kernels require to specify a bandwidth $h$. To choose the best one, we use a nonparametric bootstrap approach, as detailed in Appendix~\ref{sec:bandwidth_estim}, instead of using the optimal $h$ from Theorem~\ref{theo:lim_tilde_theta_h_plus_moins}, as it involves intractable quantities.

\subsubsection{Density and Quantile Estimation} \label{sec:density_quantile_est}

As in \textcite{jesson2022scalable}, to estimate the GPS $f(T=t|\mathbf{X}=\mathbf{x})$ and conditional expectation $\eta(t, \mathbf{x})$, we model the conditional densities $\hat f(T=t|\mathbf{X}=\mathbf{x})$ and $\hat f(Y=y|\mathbf{X}=\mathbf{x}, T=t)$ by \textit{Mixture Density Networks (MDN)} \parencite{bishop1994mixture}, but other methods, such as the ones from \textcite{sugiyama2010conditional} or \textcite{rothfuss2019conditional}, could be considered as well. In particular, we take a weighted mixture of $K$ Gaussian components such that
\begin{align*}
    & \hat f(Y=y|\mathbf{X}=\mathbf{x}, T=t) \\
    & = \sum_{k=1}^K \hat \pi_k(\mathbf{x}, t) \mathcal{N}(y|\hat \mu_k(\mathbf{x}, t), \hat \sigma_k^2(\mathbf{x}, t)),
\end{align*}
where $\hat \pi_k(\mathbf{x}, t)$, $\hat \mu_k(\mathbf{x}, t)$ and $\hat \sigma_k^2(\mathbf{x}, t)$ are respectively, the estimated weight, mean and variance of the $k$\textsuperscript{th} component, and $\mathcal{N}(y|\mu, \sigma^2)$ is the density of a Gaussian distribution of mean $\mu$ and variance $\sigma^2$. \textcite{bishop1994mixture} shows that $\hat \eta(t, \mathbf{x}) = \sum_{k=1}^K \hat \pi_k(\mathbf{x}, t) \hat \mu_k(\mathbf{x}, t)$. $\hat \pi_k$, $\hat \mu_k$ and $\hat \sigma_k^2$ are estimated via neural networks and the GPS is modeled in the same way as $\hat f(Y=y|\mathbf{X}=\mathbf{x}, T=t)$ (see Appendix~\ref{sec:dens_estim_nn}).

To estimate the conditional quantile function $Q(\upsilon; Y=y|\mathbf{X}=\mathbf{X}_i, T=T_i)$, we suggest leveraging the estimation of $f(Y=y|\mathbf{X}=\mathbf{X}_i, T=T_i)$ by finding the root of $y \mapsto F(Y=y|\mathbf{X}=\mathbf{X}_i, T=T_i) - \upsilon$, where $F(Y=y|\mathbf{X}=\mathbf{X}_i, T=T_i)$ is the cumulative distribution function of $f(Y=y|\mathbf{X}=\mathbf{X}_i, T=T_i)$ (see Appendix~\ref{sec:cond_quant_est}). We could also use other methods like quantile regression via \textit{Generalized Random Forests} \parencite{athey2019generalized}, but they would not take advantage of the already estimated $\hat f(Y=y|\mathbf{X}=\mathbf{x}, T=t)$.

\subsection{Simulation Experiments}

\begin{figure}[t]
    \centering
    \vspace{.3in}
    \includegraphics[width=\linewidth]{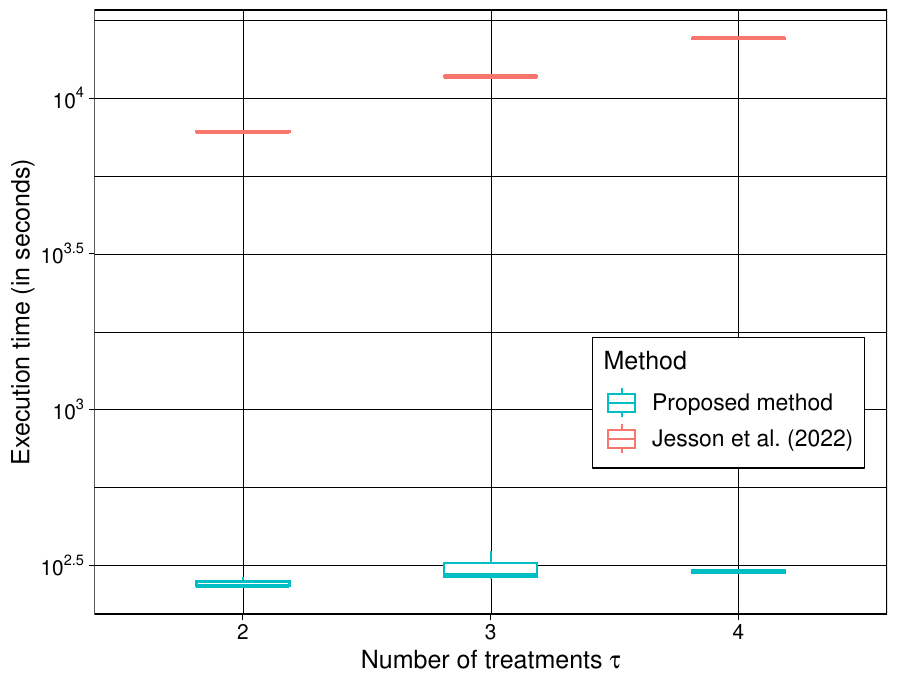}
    \vspace{.3in}
    \caption{Boxplots of execution times (in seconds) of the proposed (partially robust) method and \textcite{jesson2022scalable} on 3 Monte-Carlo samples. Sensitivity analyses were performed for $m=2$, 3 and then 4 values of treatment of interest $\tau$ (setup from Table~\ref{tab:simu_param_values}). The times do not include the fine-tuning step of the neural networks.}
    \label{fig:simu_exec_times_comparison}
\end{figure}

\begin{figure}[h]
    \centering
    \vspace{.3in}
    \includegraphics[width=\linewidth]{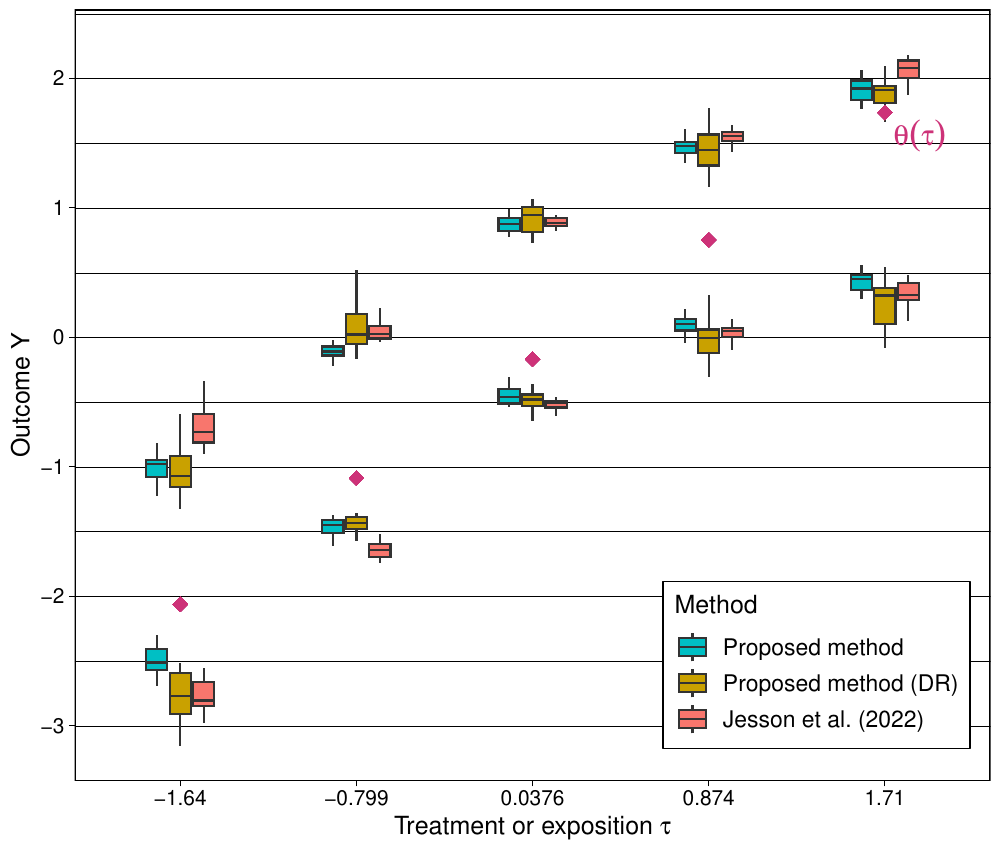}
    \vspace{.3in}
    \caption{Boxplots of 95\%-level confidence intervals for 20 Monte-Carlo samples. The partially robust bounds are in blue while the doubly robust (DR) ones are in golden, and the ones from the reference \parencite{jesson2022scalable} are in red. The true APO is represented by pink squares. The data are generated using the setup from Table~\ref{tab:simu_param_values}, with an estimated $\Gamma$ of 5.21.}
    \label{fig:simu_ci_comparison_with_dr}
\end{figure}

\begin{figure*}[t]
    \centering
    \vspace{.3in}
    \includegraphics[width=\linewidth]{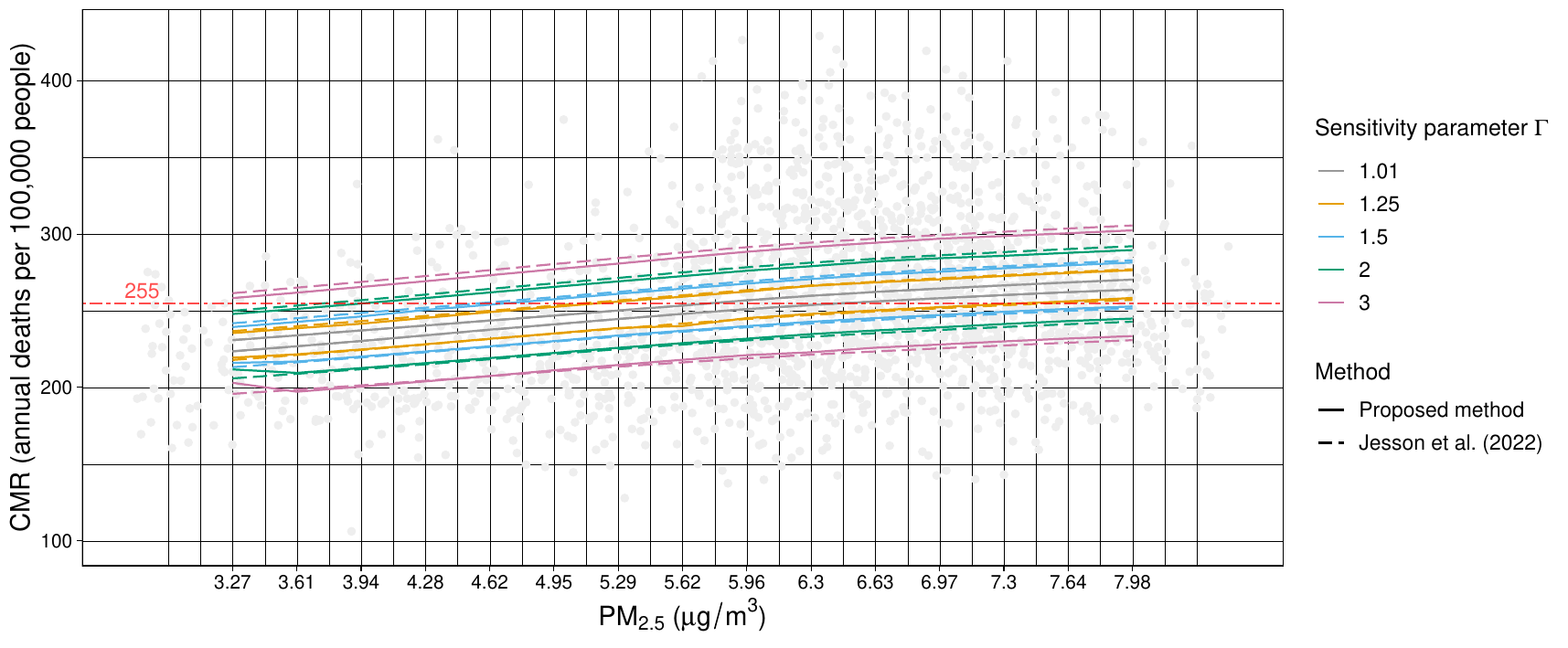}
    \vspace{.3in}
    \caption{Sensitivity analysis of the real dataset (95\%-level confidence intervals) with the proposed (partially robust) method and the one from \textcite{jesson2022scalable} for 5 values of $\Gamma$ and 15 values of exposition $\tau$ (PM2.5). The red dotted line corresponds to the average CMR (255 annual deaths per 100~000 people). The gray points are the real dataset with 84 observations removed to improve readability.}
    \label{fig:real_sensitivity_graph}
\end{figure*}

We first compare our method to the one from \textcite{jesson2022scalable} on  synthetic data. We recall that we consider $p_\mathbf{X}$ observed and $p_\mathbf{U}$ unobserved confounders. The joint distribution of $(\mathbf{X}, \mathbf{U})$ is a normal distribution $\mathcal{N}(\mathbf{0}, \mathbf{\Sigma})$, where
\begin{equation*}
    \mathbf{\Sigma} =
    \begin{pmatrix}
        \mathbf{\Sigma_X} & \mathbf{\Sigma_{XU}} \\
        \mathbf{\Sigma_{XU}}^\top & \mathbf{\Sigma_U}
    \end{pmatrix}.
\end{equation*}
$\mathbf{\Sigma_X}$ (resp., $\mathbf{\Sigma_U}$) is a tridiagonal matrix of size $p_\mathbf{X} \times p_\mathbf{X}$ (resp., $p_\mathbf{U} \times p_\mathbf{U}$), where the elements on the main diagonal are all equal to 1 and the elements on the subdiagonal and lower diagonal are all equal to $\rho_\mathbf{X} > 0$ (resp., $\rho_\mathbf{U} > 0$). $\mathbf{\Sigma_{XU}}$ is a $p_\mathbf{X} \times p_\mathbf{U}$ matrix with all coefficients equal to $\rho_\mathbf{XU} \geq 0$, where $\rho_\mathbf{XU} = \lambda \rho_\mathbf{XU}^\mathrm{max}$, with $0 \leq \lambda < 1$ and $\rho_\mathbf{XU}^\mathrm{max} = (1 - \rho_\mathbf{X})/p_\mathbf{U}$, to ensure that $\mathbf{\Sigma}$ is a diagonal dominant matrix and is, thus, invertible.
We define the treatment value $T$ conditionally on  $\mathbf{X}=\mathbf{x}$ and $\mathbf{U}=\mathbf{u}$ as $T = \langle \beta_\mathbf{X}, \mathbf{x} \rangle  + \langle \beta_\mathbf{U}, \mathbf{u} \rangle + \varepsilon_T$ where $\varepsilon_T \sim \mathcal N(0, \sigma_{\varepsilon_T}^2)$, with $\sigma_{\varepsilon_T} > 0$, $\beta_\mathbf{X} \in \mathbb R^{p_\mathbf{X}}$ and $\beta_\mathbf{U} \in \mathbb R^{p_\mathbf{U}}$.
Finally, for all $t \in \mathcal{T}$, we set the potential outcome to
\begin{equation*}
    Y(t) = t + \zeta  \langle \mathbf{X} , \gamma_\mathbf{X} \rangle \cdot e^{-t \langle \mathbf{X}, \gamma_\mathbf{X} \rangle} -  \langle \mathbf{U}, \gamma_\mathbf{U} \rangle \langle \mathbf{X}, \gamma_\mathbf{X} \rangle + \varepsilon_Y,
\end{equation*}
where $\varepsilon_Y \sim \mathcal{N}(0, \sigma_{\varepsilon_Y}^2)$, $\sigma_{\varepsilon_Y} > 0$, $\zeta \in \mathbb{R}$, $\gamma_\mathbf{X} \in \mathbb{R}^{p_\mathbf{X}}$ and $\gamma_\mathbf{U} \in \mathbb{R}^{p_\mathbf{U}}$. In this simulation scenario, the true APO has an explicit form
\begin{align*}
    \theta(t) & = \tau \Bigl(1 - \zeta \cdot \gamma_\mathbf{X}^\top \mathbf{\Sigma_X} \gamma_\mathbf{X} \cdot e^{\frac{\tau^2}{2} \gamma_\mathbf{X}^\top \mathbf{\Sigma_X} \gamma_\mathbf{X}} \Bigr) \\
    & \quad - \gamma_\mathbf{U}^\top \mathbf{\Sigma_{XU}}^\top \gamma_\mathbf{X}.
\end{align*}
During the simulation process, in order to avoid isolated data points and violations of the positivity assumption, the observations that correspond to the 10\% biggest hat values of the $(\mathbf{X}, T, Y)$ design matrix are removed. The complete simulation setup is given in Appendix~\ref{sec:details_sim_data}.

To assess the variability of the considered methods and avoid training and predicting on the same data, we perform 2-fold cross-fitting on several simulated Monte-Carlo (MC) samples, each of size $n=900$ after removing outliers. For each sample, we compute 95\%-level confidence intervals with $B=100$ bootstrap resamples for particular values of the treatment of interest $\tau$. To select the appropriate sensitivity parameter $\Gamma$, we consider a separate MC sample that acts as an external dataset where $\mathbf{U}$ is known and where we seek the lowest $\Gamma$ such that $[\Gamma^{-1}, \Gamma]$ would contain almost all, say a proportion $p_\Gamma = 99$\%, of the ratios $f(T=\tau|\mathbf{X}=\mathbf{X}_i, \mathbf{U}=\mathbf{U}_i)/f(T=\tau|\mathbf{X}=\mathbf{X}_i)$. This choice of $p_\Gamma$ ensures a good coverage of the confidence intervals because $\Gamma$ increases as the proportion $p_\Gamma$ gets bigger, and greater values of $\Gamma$ are associated with larger intervals. Additional results on the link between $\Gamma$ and the simulated dataset are provided in Appendix~\ref{sec:details_sim_data}.

First, we show that the order of magnitude of execution times for the method from \textcite{jesson2022scalable} is much larger than for the proposed method (partially robust bounds). Boxplots of execution times on each MC sample are given in Figure~\ref{fig:simu_exec_times_comparison}, where 3 sensitivity analyses are performed on 3 different MC samples (9 experiments in total): one analysis with only two treatments of interest $\tau$, one with three, and another one with four. To be clear, a sensitivity analysis with $m$ values of treatments of interest $\tau$ means that the algorithms compute confidence intervals only for these $m$ values (even if $T$ is continuous). Therefore, increasing $m$ means refining the analysis, or smoothing the sensitivity plots. Results in Figure~\ref{fig:simu_exec_times_comparison} show that the proposed method is on average 28 times faster than the concurrent method for a sensitivity analysis performed with $m=2$ values of $\tau$, and this gap increases as the number $m$ of treatments of interest grows. As explained in Section~\ref{sec:jesson_method}, the algorithm from \textcite{jesson2022scalable} involves a grid search step that depends directly on $\tau$. Therefore, for each value of $\tau$, a grid search must be reiterated, each grid search involving Monte-Carlo integrations. On the contrary, our method is almost insensitive to $m$ because it does not involve neither a grid search step, nor a Monte-Carlo integration.

In addition, the sharpness of the proposed bounds is displayed in Figure~\ref{fig:simu_ci_comparison_with_dr}. After performing a sensitivity analysis for 5 values of $\tau$ on 20 different Monte-Carlo samples, we report as boxplots the upper and lower bounds of 95\%-level CIs for the APO, computed with our methods and the one from \textcite{jesson2022scalable}. In average, our bounds (in blue) appear to be tighter than the ones from \textcite{jesson2022scalable} (in red). For each exposition $\tau$, the medians of the bounds obtained with the doubly robust estimator (in golden) are also located in the interval constituted by the medians of the bounds obtained with the method from \textcite{jesson2022scalable}. However, the doubly robust estimator has a slightly greater variance, most likely due to the estimation of the modified outcome regression nuisance parameter $\Theta^\pm$. Nevertheless, our bounds keep a good coverage of the true APO, as the pink squares, representing the true unknown APO for each value of $\tau$, remain in the confidence intervals. Finally, practical considerations about the validity of the percentile bootstrap method to compute the confidence intervals on the simulated data are detailed in Appendix~\ref{sec:validity_perc_boot}.

\subsection{Real Dataset Experiments} \label{sec:real_dataset_exp}

We further illustrate the behavior of our method on a \href{https://catalog.data.gov/dataset/annual-pm2-5-and-cardiovascular-mortality-rate-data-trends-modified-by-county-socioeconomi}{publicly available dataset from the U.S.\ Environmental Protection Agency} that studies 
the impact of annual PM2.5 level on cardiovascular mortality rate (CMR) in 2,132 U.S.\ counties between 1990 and 2010 \parencite{wyatt2020annual}. PM2.5 fine particulate matter level corresponds to the continuous exposition $T$ and is measured in $\mu g / m^3$. CMR corresponds to the observed outcome $Y$ and is measured in annual deaths per 100,000 people.  As in \textcite{bahadori2022end}, we restrict the data to year 2010 to simplify the study and we consider 10 continuous variables as observed confounders $\mathbf{X}$. Finally, as in the simulated dataset, we remove 10\% of extreme observations which leads to an effective sample size of $n = 1918$. See Appendix~\ref{sec:details_real_data} for additional details on the preprocessing step.

In the following, the results are displayed for a user-defined range of sensitivity parameters $\Gamma$. This is common in sensitivity analyses: in the continuous case, see \textcite{jesson2022scalable}; in the binary case, see \textcite{zhao2019sensitivity} or \textcite{dorn2022sharp}. $\Gamma$ can also be compared to a cutoff or a \textit{critical value} $\Gamma_c$, the lowest value of $\Gamma$ consistent with no treatment effect or any value of interest (see e.g., \cite{tan2006distributional}, \cite{yadlowsky2022bounds} or \cite{dorn2022sharp}). Assuming $\Gamma$ is finite, which we will never be sure of, we can give an estimate $\hat \Gamma$ by computing the density ratio from the CMSM if \textit{measured} confounders were omitted from the data and deduce acceptable values for $\Gamma$ by assuming $\mathbf{U}$ has not a bigger strength than already observed covariates \parencite{kallus2018confounding, yadlowsky2022bounds, dorn2022sharp}. This method is known as \textit{informal benchmarking} \parencite{cinelli2020making} and is presented in details in Appendix~\ref{sec:informal_benchmarking}. If $\hat{\Gamma} < \Gamma_c$, then the sensitivity analysis is robust to unobserved confounders and we can conclude to an effect of $T$ on $Y$ with great certainty. If $\hat{\Gamma} \geq \Gamma_c$, then the sensitivity analysis is not robust to unobserved confounders and we cannot conclude about an effect of $T$ on $Y$ because the null effect is included in the sensitivity bounds.

After performing 2-fold cross-fitting on the data, 95\%-level CIs are obtained with the proposed (partially robust) and concurrent methods for 15 different values of $\tau$ and 5 values of $\Gamma$ in Figure~\ref{fig:real_sensitivity_graph}. In total, 5 simulations were ran, one for each value of $\Gamma$.
First, as expected, we can observe that the confidence intervals grow as the sensitivity parameter $\Gamma$ increases. The gray curves, where $\Gamma = 1.01$, correspond to results close to the $\mathbf{X}$-ignorability assumption, when $\Gamma = 1$, i.e.\ results that could be obtained with estimator~\eqref{eqn:estimator_under_ignorability} from \textcite{kallus2018policy}. As $\Gamma$ increases, we deviate from $\mathbf{X}$-ignorability and assume that there are unobserved confounders. The more they have an effect on $T$, the more the uncertainty about the estimation of the treatment effect grows and the CIs become larger. Moreover, notice that, as the level of fine particles PM2.5 increases, there is a shift of the CIs towards greater values of the CMR. This observation supports the conclusion that cardiovascular mortality rate rises when the level of fine particles increases. However, if this result is significant when $\Gamma = 1.01$, it is not true for larger values of $\Gamma$. Indeed, with $\Gamma = 3$, the horizontal line representing the average CMR of 255 annual deaths per 100,000 people goes through all the CIs between 3.27 and 7.98 $\mu g/m^3$, and it could potentially be the true APO function. Thus, $\Gamma_c = 3$ is the critical value, as it would indicate no significant effect of PM2.5 level on the CMR between 3.27 and 7.98 $\mu g/m^3$ of PM2.5. The results from the informal benchmarking (see Appendix~\ref{sec:informal_benchmarking}) indicate that all observed covariates lead to an estimation of $\Gamma$ greater than $\Gamma_c$. As a consequence, the sensitivity analysis shows that the no-effect assumption could not significantly be rejected from these data.
Finally, it is also immediately clear that the confidence intervals obtained with our method are smaller than the ones from \textcite{jesson2022scalable}. Thus, with our method, a greater $\Gamma$ would be needed to move away from the hypothesis of effect of PM2.5 on CMR than with the concurrent method. A figure similar to Figure~\ref{fig:real_sensitivity_graph}, where the method from \textcite{jesson2022scalable} is compared to the doubly robust estimator, is given in Appendix~\ref{sec:details_real_data} (Figure~\ref{fig:real_sensitivity_graph_dr}).

\section{CONCLUSION AND PERSPECTIVES}

We presented novel bounds for the APO after introducing a new continuous sensitivity model and leveraging a constraint on a likelihood ratio, derived confidence intervals from them, and showed that our algorithm outperforms existing methodologies in terms of computation times and sharpness. These results were demonstrated on a simulated dataset and real data from the U.S.\ Environmental Protection Agency. Additionally, we proposed doubly robust estimators of the bounds and performed an exploratory analysis on the variation of $\Gamma$ with respect to certain generation parameters of the dataset in appendix. To go further, it could be interesting to extend our method to the case where $T$ is multivariate, for example, to study drug antagonism or synergism. Moreover, derivation of the bounds under Pearl's SCM could be another avenue, as this framework is also popular in causal inference but involves different assumptions. Additionally, in this article, we did not explain how to estimate the bounds for the CAPO from the data because it requires localizing the estimation around the covariates of interest, which is a case of high-dimensional regression. The estimation and interpretation of $\Gamma$ also requires further investigation and is still an active domain of research. For example, \textcite{kallus2018confounding} suggested estimating a lower bound for $\Gamma$ through negative controls \parencite{lipsitch2010negative}. Finally, the proposed doubly robust estimators can be promising but more thorough examination could be needed to improve them.

\section{DISCLOSURES}

Jean-Baptiste Baitairian, Bernard Sebastien and Rana Jreich are Sanofi employees and may hold shares and/or stock options in the company. Agathe Guilloux is employed by INRIA. Sandrine Katsahian is employed by Université Paris-Cité and Assistance Publique -- Hôpitaux de Paris (AP-HP). This work was supported by Sanofi and INRIA (Institut National de Recherche en Informatique et en Automatique).

\setlength\bibitemsep{2\itemsep}
\subsubsection*{References}  
\printbibliography[heading=none]

@article{herdegen2023elementary,
  title={An elementary proof of the dual representation of Expected Shortfall},
  author={Herdegen, Martin and Munari, Cosimo},
  journal={Mathematics and Financial Economics},
  volume={17},
  number={4},
  pages={655--662},
  year={2023},
  publisher={Springer}
}

@article{rockafellar2000optimization,
  title={Optimization of conditional value-at-risk},
  author={Rockafellar, R Tyrrell and Uryasev, Stanislav and others},
  journal={Journal of risk},
  volume={2},
  pages={21--42},
  year={2000},
  publisher={Citeseer}
}

@article{rothfuss2019conditional,
  title={Conditional density estimation with neural networks: Best practices and benchmarks},
  author={Rothfuss, Jonas and Ferreira, Fabio and Walther, Simon and Ulrich, Maxim},
  journal={arXiv preprint arXiv:1903.00954},
  year={2019}
}

@inproceedings{sugiyama2010conditional,
  title={Conditional density estimation via least-squares density ratio estimation},
  author={Sugiyama, Masashi and Takeuchi, Ichiro and Suzuki, Taiji and Kanamori, Takafumi and Hachiya, Hirotaka and Okanohara, Daisuke},
  booktitle={Proceedings of the Thirteenth International Conference on Artificial Intelligence and Statistics},
  pages={781--788},
  year={2010},
  organization={JMLR Workshop and Conference Proceedings}
}

@article{dorn2022sharp,
  title={Sharp sensitivity analysis for inverse propensity weighting via quantile balancing},
  author={Dorn, Jacob and Guo, Kevin},
  journal={Journal of the American Statistical Association},
  pages={1--13},
  year={2022},
  publisher={Taylor \& Francis}
}

@article{zhao2019sensitivity,
  title={Sensitivity analysis for inverse probability weighting estimators via the percentile bootstrap},
  author={Zhao, Qingyuan and Small, Dylan S and Bhattacharya, Bhaswar B},
  journal={Journal of the Royal Statistical Society Series B: Statistical Methodology},
  volume={81},
  number={4},
  pages={735--761},
  year={2019},
  publisher={Oxford University Press}
}

@article{tan2006distributional,
  title={A distributional approach for causal inference using propensity scores},
  author={Tan, Zhiqiang},
  journal={Journal of the American Statistical Association},
  volume={101},
  number={476},
  pages={1619--1637},
  year={2006},
  publisher={Taylor \& Francis}
}

@article{rosenbaum1983central,
  title={The central role of the propensity score in observational studies for causal effects},
  author={Rosenbaum, Paul R and Rubin, Donald B},
  journal={Biometrika},
  volume={70},
  number={1},
  pages={41--55},
  year={1983},
  publisher={Oxford University Press}
}

@article{jin2023sensitivity,
  title={Sensitivity analysis of individual treatment effects: A robust conformal inference approach},
  author={Jin, Ying and Ren, Zhimei and Cand{\`e}s, Emmanuel J},
  journal={Proceedings of the National Academy of Sciences},
  volume={120},
  number={6},
  pages={e2214889120},
  year={2023},
  publisher={National Acad Sciences}
}

@article{jesson2022scalable,
  title={Scalable sensitivity and uncertainty analyses for causal-effect estimates of continuous-valued interventions},
  author={Jesson, Andrew and Douglas, Alyson and Manshausen, Peter and Solal, Ma{\"e}lys and Meinshausen, Nicolai and Stier, Philip and Gal, Yarin and Shalit, Uri},
  journal={Advances in Neural Information Processing Systems},
  volume={35},
  pages={13892--13907},
  year={2022}
}

@inproceedings{kallus2018policy,
  title={Policy evaluation and optimization with continuous treatments},
  author={Kallus, Nathan and Zhou, Angela},
  booktitle={International conference on artificial intelligence and statistics},
  pages={1243--1251},
  year={2018},
  organization={PMLR}
}

@article{kallus2018confounding,
  title={Confounding-robust policy improvement},
  author={Kallus, Nathan and Zhou, Angela},
  journal={Advances in neural information processing systems},
  volume={31},
  year={2018}
}

@article{yadlowsky2022bounds,
  title={Bounds on the conditional and average treatment effect with unobserved confounding factors},
  author={Yadlowsky, Steve and Namkoong, Hongseok and Basu, Sanjay and Duchi, John and Tian, Lu},
  journal={The Annals of Statistics},
  volume={50},
  number={5},
  pages={2587--2615},
  year={2022},
  publisher={Institute of Mathematical Statistics}
}

@book{rosenbaum2002observational,
  title={Observational Studies},
  author={Rosenbaum, P.R.},
  isbn={9780387989679},
  lccn={2001049264},
  series={Springer Series in Statistics},
  url={https://books.google.fr/books?id=K0OglGXtpGMC},
  year={2002},
  publisher={Springer}
}

@article{hill2011bayesian,
  title={Bayesian nonparametric modeling for causal inference},
  author={Hill, Jennifer L},
  journal={Journal of Computational and Graphical Statistics},
  volume={20},
  number={1},
  pages={217--240},
  year={2011},
  publisher={Taylor \& Francis}
}

@article{imai2004causal,
  title={Causal inference with general treatment regimes: Generalizing the propensity score},
  author={Imai, Kosuke and Van Dyk, David A},
  journal={Journal of the American Statistical Association},
  volume={99},
  number={467},
  pages={854--866},
  year={2004},
  publisher={Taylor \& Francis}
}

@article{hirano2004propensity,
  title={The propensity score with continuous treatments},
  author={Hirano, Keisuke and Imbens, Guido W},
  journal={Applied Bayesian modeling and causal inference from incomplete-data perspectives},
  volume={226164},
  pages={73--84},
  year={2004}
}

@inproceedings{kazemi2024adversarially,
  title={Adversarially Balanced Representation for Continuous Treatment Effect Estimation},
  author={Kazemi, Amirreza and Ester, Martin},
  booktitle={Proceedings of the AAAI Conference on Artificial Intelligence},
  volume={38},
  number={12},
  pages={13085--13093},
  year={2024}
}

@article{rubin1974estimating,
  title={Estimating causal effects of treatments in randomized and nonrandomized studies.},
  author={Rubin, Donald B},
  journal={Journal of educational Psychology},
  volume={66},
  number={5},
  pages={688},
  year={1974},
  publisher={American Psychological Association}
}

@article{wyatt2020annual,
  title={Annual PM2. 5 and cardiovascular mortality rate data: Trends modified by county socioeconomic status in 2,132 US counties},
  author={Wyatt, Lauren H and Peterson, Geoffrey Colin L and Wade, Tim J and Neas, Lucas M and Rappold, Ana G},
  journal={Data in brief},
  volume={30},
  pages={105318},
  year={2020},
  publisher={Elsevier}
}

@misc{colangelo2020double,
  title={Double Debiased Machine Learning Nonparametric Inference with Continuous Treatments}, 
  author={Kyle Colangelo and Ying-Ying Lee},
  year={2023},
  eprint={2004.03036},
  archivePrefix={arXiv},
  primaryClass={econ.EM},
  url={https://arxiv.org/abs/2004.03036}, 
}

@article{neyman1923application,
  title={On the application of probability theory to agricultural experiments. Essay on principles},
  author={Neyman, Jerzy},
  journal={Ann. Agricultural Sciences},
  pages={1--51},
  year={1923}
}

@article{dorn2024doubly,
  title={Doubly-valid/doubly-sharp sensitivity analysis for causal inference with unmeasured confounding},
  author={Dorn, Jacob and Guo, Kevin and Kallus, Nathan},
  journal={Journal of the American Statistical Association},
  pages={1--12},
  year={2024},
  publisher={Taylor \& Francis}
}

@techreport{bishop1994mixture,
title = "Mixture density networks",
author = "Bishop, {Christopher M.}",
year = "1994",
language = "English",
publisher = "Aston University",
type = "WorkingPaper",
institution = "Aston University",
}

@article{athey2019generalized,
author = {Susan Athey and Julie Tibshirani and Stefan Wager},
title = {{Generalized random forests}},
volume = {47},
journal = {The Annals of Statistics},
number = {2},
publisher = {Institute of Mathematical Statistics},
pages = {1148 -- 1178},
year = {2019},
doi = {10.1214/18-AOS1709},
URL = {https://doi.org/10.1214/18-AOS1709}
}

@article{ding2023sensitivity,
  title={Sensitivity Analysis for Unmeasured Confounding in Medical Product Development and Evaluation Using Real World Evidence},
  author={Ding, Peng and Fang, Yixin and Faries, Doug and Gruber, Susan and Lee, Hana and Lee, Joo-Yeon and Mishra-Kalyani, Pallavi and Shan, Mingyang and van der Laan, Mark and Yang, Shu and others},
  journal={arXiv preprint arXiv:2307.07442},
  year={2023}
}

@article{tsybakov2009nonparametric,
  title={Nonparametric estimators},
  author={Tsybakov, Alexandre B},
  journal={Introduction to Nonparametric Estimation},
  pages={1--76},
  year={2009},
  publisher={Springer}
}

@article{kennedy2017non,
  title={Non-parametric methods for doubly robust estimation of continuous treatment effects},
  author={Kennedy, Edward H and Ma, Zongming and McHugh, Matthew D and Small, Dylan S},
  journal={Journal of the Royal Statistical Society Series B: Statistical Methodology},
  volume={79},
  number={4},
  pages={1229--1245},
  year={2017},
  publisher={Oxford University Press}
}

@article{ascoli2018limitations,
  title={Limitations Of Richardson Extrapolation For Kernel Density Estimation},
  author={Ascoli, Ruben G},
  journal={arXiv preprint arXiv:1812.08619},
  year={2018}
}

@inproceedings{bahadori2022end,
  title={End-to-end balancing for causal continuous treatment-effect estimation},
  author={Bahadori, Taha and Tchetgen, Eric Tchetgen and Heckerman, David},
  booktitle={International Conference on Machine Learning},
  pages={1313--1326},
  year={2022},
  organization={PMLR}
}

@Manual{rstatisticalsoftware,
    title = {R: A Language and Environment for Statistical Computing},
    author = {{R Core Team}},
    organization = {R Foundation for Statistical Computing},
    address = {Vienna, Austria},
    year = {2023},
    url = {https://www.R-project.org/},
}

@Manual{torch,
    title={torch: Tensors and Neural Networks with 'GPU' Acceleration},
    author = {Daniel Falbel and Javier Luraschi},
    note = {R package version 0.12.0, https://github.com/mlverse/torch},
    year = {2023},
    url = {https://torch.mlverse.org/docs},
}

@Book{ggplot2,
    author = {Hadley Wickham},
    title = {ggplot2: Elegant Graphics for Data Analysis},
    publisher = {Springer-Verlag New York},
    year = {2016},
    isbn = {978-3-319-24277-4},
    url = {https://ggplot2.tidyverse.org},
}

@Manual{foreach,
    title = {foreach: Provides Foreach Looping Construct},
    author = {{Microsoft} and Steve Weston},
    year = {2022},
    note = {R package version 1.5.2},
    url = {https://CRAN.R-project.org/package=foreach},
}

@Manual{latex2exp,
    title = {latex2exp: Use LaTeX Expressions in Plots},
    author = {Stefano Meschiari},
    year = {2022},
    note = {R package version 0.9.6},
    url = {https://CRAN.R-project.org/package=latex2exp},
}

@Manual{tictoc,
    title = {tictoc: Functions for Timing R Scripts, as Well as Implementations of
"Stack" and "StackList" Structures},
    author = {Sergei Izrailev},
    year = {2024},
    note = {R package version 1.2.1},
    url = {https://CRAN.R-project.org/package=tictoc},
}

@Manual{scales,
    title = {scales: Scale Functions for Visualization},
    author = {Hadley Wickham and Dana Seidel},
    year = {2022},
    note = {R package version 1.2.1},
    url = {https://CRAN.R-project.org/package=scales},
}

@article{faraway1990bootstrap,
 ISSN = {01621459, 1537274X},
 URL = {http://www.jstor.org/stable/2289609},
 author = {Julian J. Faraway and Myoungshic Jhun},
 journal = {Journal of the American Statistical Association},
 number = {412},
 pages = {1119--1122},
 publisher = {[American Statistical Association, Taylor & Francis, Ltd.]},
 title = {Bootstrap Choice of Bandwidth for Density Estimation},
 urldate = {2024-09-03},
 volume = {85},
 year = {1990}
}

@article{silverman1987bootstrap,
  title={The bootstrap: to smooth or not to smooth?},
  author={Silverman, BW and Young, GA},
  journal={Biometrika},
  volume={74},
  number={3},
  pages={469--479},
  year={1987},
  publisher={Oxford University Press}
}

@article{goldenshluger2011bandwidth,
author = {Alexander Goldenshluger and Oleg Lepski},
title = {{Bandwidth selection in kernel density estimation: Oracle inequalities and adaptive minimax optimality}},
volume = {39},
journal = {The Annals of Statistics},
number = {3},
publisher = {Institute of Mathematical Statistics},
pages = {1608 -- 1632},
year = {2011},
doi = {10.1214/11-AOS883},
URL = {https://doi.org/10.1214/11-AOS883}
}

@article{tan2024model,
  title={Model-assisted sensitivity analysis for treatment effects under unmeasured confounding via regularized calibrated estimation},
  author={Tan, Zhiqiang},
  journal={Journal of the Royal Statistical Society Series B: Statistical Methodology},
  pages={qkae034},
  year={2024},
  publisher={Oxford University Press UK}
}

@article{frauen2024sharp,
  title={Sharp bounds for generalized causal sensitivity analysis},
  author={Frauen, Dennis and Melnychuk, Valentyn and Feuerriegel, Stefan},
  journal={Advances in Neural Information Processing Systems},
  volume={36},
  year={2024}
}

@article{cornfield1959smoking,
  title={Smoking and lung cancer: recent evidence and a discussion of some questions},
  author={Cornfield, Jerome and Haenszel, William and Hammond, E Cuyler and Lilienfeld, Abraham M and Shimkin, Michael B and Wynder, Ernst L},
  journal={Journal of the National Cancer institute},
  volume={22},
  number={1},
  pages={173--203},
  year={1959},
  publisher={Oxford University Press}
}

@book{pearl2009causality,
  title={Causality},
  author={Pearl, Judea},
  year={2009},
  publisher={Cambridge university press}
}

@article{cinelli2020making,
  title={Making sense of sensitivity: Extending omitted variable bias},
  author={Cinelli, Carlos and Hazlett, Chad},
  journal={Journal of the Royal Statistical Society Series B: Statistical Methodology},
  volume={82},
  number={1},
  pages={39--67},
  year={2020},
  publisher={Oxford University Press}
}

@article{lipsitch2010negative,
  title={Negative controls: a tool for detecting confounding and bias in observational studies},
  author={Lipsitch, Marc and Tchetgen, Eric Tchetgen and Cohen, Ted},
  journal={Epidemiology},
  volume={21},
  number={3},
  pages={383--388},
  year={2010},
  publisher={LWW}
}

@book{manski2003partial,
  title={Partial identification of probability distributions},
  author={Manski, Charles F},
  year={2003},
  publisher={Springer Science \& Business Media}
}

@article{bonvini2022sensitivity,
  title={Sensitivity analysis for marginal structural models},
  author={Bonvini, Matteo and Kennedy, Edward and Ventura, Valerie and Wasserman, Larry},
  journal={arXiv preprint arXiv:2210.04681},
  year={2022}
}

@article{kallus2021causal,
  title={Causal inference under unmeasured confounding with negative controls: A minimax learning approach},
  author={Kallus, Nathan and Mao, Xiaojie and Uehara, Masatoshi},
  journal={arXiv preprint arXiv:2103.14029},
  year={2021}
}

@article{robins2002covariance,
  title={[Covariance Adjustment in Randomized Experiments and Observational Studies]: Comment},
  author={Robins, James M},
  journal={Statistical Science},
  volume={17},
  number={3},
  pages={309--321},
  year={2002},
  publisher={JSTOR}
}

@article{rambachan2022counterfactual,
  title={Counterfactual risk assessments under unmeasured confounding},
  author={Rambachan, Ashesh and Coston, Amanda and Kennedy, Edward},
  journal={arXiv preprint arXiv:2212.09844},
  year={2022}
}

@inproceedings{byun2024auditing,
  title={Auditing Fairness under Unobserved Confounding},
  author={Byun, Yewon and Sam, Dylan and Oberst, Michael and Lipton, Zachary and Wilder, Bryan},
  booktitle={International Conference on Artificial Intelligence and Statistics},
  pages={4339--4347},
  year={2024},
  organization={PMLR}
}

@inproceedings{oprescu2023b,
  title={B-learner: Quasi-oracle bounds on heterogeneous causal effects under hidden confounding},
  author={Oprescu, Miruna and Dorn, Jacob and Ghoummaid, Marah and Jesson, Andrew and Kallus, Nathan and Shalit, Uri},
  booktitle={International Conference on Machine Learning},
  pages={26599--26618},
  year={2023},
  organization={PMLR}
}

@article{zhang2025enhanced,
  title={Enhanced Marginal Sensitivity Model and Bounds},
  author={Zhang, Yi and Xu, Wenfu and Tan, Zhiqiang},
  journal={arXiv preprint arXiv:2504.08301},
  year={2025}
}

@misc{helwig2017bootstrap,
  title={Bootstrap confidence intervals},
  author={Helwig, Nathaniel E},
  journal={University of Minnesota},
  year={2017}
}

\newpage

\onecolumn

\appendix

\section{ADDITIONAL DEFINITIONS, ASSUMPTIONS, DETAILS AND PROOFS}

In the following, we provide additional definitions used in the paper, along with proofs and their related assumptions.

\subsection{Definitions}  \label{sec:appendix_def}

Definition~\ref{def:cmsm_jesson} corresponds to the sensitivity model that was considered in \textcite{jesson2022scalable}.
\begin{definition}[CMSM from \cite{jesson2022scalable}]  \label{def:cmsm_jesson}
    For a given treatment value $\tau \in \mathcal{T}$, for all $(\mathbf{x}, y) \in \mathcal{X} \times \mathcal{Y}$, assuming $f(T=\tau|\mathbf{X}=\mathbf{x})$ is absolutely continuous with respect to $f(T=\tau|\mathbf{X}=\mathbf{x}, Y(\tau)=y)$, there exists a sensitivity parameter $\Gamma \geq 1$ such that
    \begin{equation}
        \Gamma^{-1} \leq \frac{f(T=\tau|\mathbf{X}=\mathbf{x})}{f(T=\tau|\mathbf{X}=\mathbf{x}, Y(\tau)=y)} \leq \Gamma.
    \end{equation}
\end{definition}

Theorem~\ref{theo:capo_bounds_and_sharpness} naturally leads to a conditional quantile. The following definition is the one we use in this paper.
\begin{definition}[Conditional quantile]  \label{def:cond_quantile}
    The generalized conditional quantile of order $\upsilon \in [0, 1]$ of the distribution of $Y$ conditionally on $\mathbf{X}=\mathbf{x}$ and $T=t$ can be defined as:
    \begin{equation*}
        Q(\upsilon; Y|\mathbf{X}=\mathbf{x}, T=t) \overset{\mathrm{not.}}{=} q_{\upsilon}^{\mathbf{x}, \mathbf{t}} \coloneq \inf\{ q \in \mathbb{R}, \, \mathbb{P}(Y \leq q|\mathbf{X}=\mathbf{x}, T=t) > \upsilon \}.
    \end{equation*}
\end{definition}

\subsection{Assumptions}

Assumption~\ref{ass:positivity_cond_treatment} ensures that the quotients in sensitivity models~\eqref{eqn:cmsm_U} and \eqref{eqn:cmsm_Y} exist. Assumptions~\ref{ass:positivity_cond_treatment} and \ref{ass:positivity_cond_outcome} are equivalent to assuming two by two absolute continuity of the densities. See Proposition 1 in \textcite{jesson2022scalable} for another use of the absolute continuity assumption.
\begin{assumption}[Positivity and existence of the conditional densities for the treatment]  \label{ass:positivity_cond_treatment}
    $\forall (\mathbf{x}, \mathbf{u}, t, y) \in \mathcal{X} \times \mathcal{U} \times \mathcal{T} \times \mathcal{Y}, \, f(T=t|\mathbf{X}=\mathbf{x}), \, f(T=t|\mathbf{X}=\mathbf{x}, Y(t)=y) \text{ and } f(T=t|\mathbf{X}=\mathbf{x}, \mathbf{U}=\mathbf{u})$ exist and are positive. We will also assume that $\exists m_f > 0, \, \forall (\mathbf{x}, t) \in \mathcal{X} \times \mathcal{T}, \, m_f \leq f(T=t|\mathbf{X}=\mathbf{x})$.
\end{assumption}

\begin{assumption}[Positivity and existence of the conditional densities for the outcome]  \label{ass:positivity_cond_outcome}
    $\forall (\mathbf{x}, t, y) \in \mathcal{X} \times \mathcal{T} \times \mathcal{Y}, \, f(Y=y|\mathbf{X}=\mathbf{x}) \text{ and } f(Y=y|\mathbf{X}=\mathbf{x}, T=t)$ exist and are positive.
\end{assumption}

The following hypotheses concern the user-defined kernel. They are usual in nonparametric estimation.

\begin{assumption}[Kernel]  \label{ass:kernel}
    We assume that $K : \mathbb{R} \to \mathbb{R}_+$ is a symmetric and integrable \textit{kernel}, with $\int_{-1}^1 K(u) \, \mathrm{d}u = 1$. We assume in addition that
    \begin{enumerate}[label=\textnormal{(\roman*)}, topsep=0pt]
        \item $K$ is squared-integrable. In particular, $\exists M_{K^2} \in \mathbb{R}, \, \int K^2(u) \, \mathrm{d}u \leq M_{K^2} < + \infty$. \label{ass:kernel_add_1}
        \item $K$ is a kernel of order 1, i.e.\ $\exists  M_{u^2K} \in \mathbb{R}, \, \int u^2 K(u) \, \mathrm{d}u \leq M_{u^2K} < +\infty$. \label{ass:kernel_add_2}
    \end{enumerate}
\end{assumption}

Assumption~\ref{ass:bounded_outcomes} is classical in treatment effect estimation (see e.g., \cite{kallus2018policy}, \cite{jesson2022scalable}, and \cite{dorn2024doubly}).
\begin{assumption}  \label{ass:bounded_outcomes}
    The potential outcomes are bounded: $\exists M_Y \in \mathbb{R}, \, |Y| \leq M_Y$.
\end{assumption}

The following assumptions are used in Theorem~\ref{theo:lim_tilde_theta_h_plus_moins} to get bounds on the MSE of our estimators. For example, \textcite{ascoli2018limitations} uses similar hypotheses with kernels.

\begin{assumption}  \label{ass:f_outcome_c2_bounded}
    For all $y \in \mathcal Y$ and $x \in \mathcal X$, the function $t \mapsto f(Y=y|\mathbf{X}= \mathbf{x}, T=t)$ is $\mathcal{C}^2$ on $\mathcal{T}$. We also assume that, for a fixed treatment value $\tau \in \mathcal{T}$, for all $\mathbf{x} \in \mathcal{X}$, $\underset{y \in \mathcal{Y}}{\sup} |f(Y=y|\mathbf{X}=\mathbf{x}, T=\tau)| = \lVert f \rVert_\infty < \infty $. In addition, there exists $M_{\partial^2 f} \geq 0$ such that,
    \begin{equation*}
         \forall (\mathbf{x}, y, t) \in \mathcal{X} \times \mathcal{Y} \times \mathcal{T}, \, \left| \pdv[order=2]{f}{T}(Y=y | \mathbf{X}=\mathbf{x}, T=t) \right| \leq M_{\partial^2 f}.
    \end{equation*}
\end{assumption}

\begin{assumption}  \label{ass:Q_c2_bounded}
   For all $(\mathbf{x}, y, \upsilon)$ in $\mathcal{X} \times \mathcal{Y} \times [0, 1]$, the function $t \mapsto Q(\upsilon; Y|\mathbf{X}= \mathbf{x}, T=t)$ is $\mathcal{C}^2$ on $\mathcal{T}$ and there exists
    $M_{\partial^2 Q} \geq 0$ such that
    \begin{equation*}
      \forall (\mathbf{x}, t, \upsilon) \in \mathcal{X} \times \mathcal{T} \times [0, 1], \, \left| \pdv[order=2]{Q}{T}(\upsilon;Y|\mathbf{X}=\mathbf{x},T=t) \right| \leq M_{\partial^2 Q}.
    \end{equation*}
\end{assumption}

Assumption~\ref{ass:Theta_plus_c2_bounded} is used to demonstrate the double robustness property in Proposition~\ref{prop:double_robustness}.

\begin{assumption}  \label{ass:Theta_plus_c2_bounded}
   For all $\mathbf{x}$ in $\mathcal{X}$ and $\bar q_\gamma \in \mathcal{Y}$, the function $t \mapsto \Theta^+(t, \mathbf{x}, \bar q_\gamma)$, as defined in Equation~\eqref{eqn:Theta_plus_def}, is $\mathcal{C}^2$ on $\mathcal{T}$ and there exists $M_{\partial^2 \Theta^+} \geq 0$ such that
    \begin{equation*}
      \forall (\mathbf{x}, t, \bar q_\gamma) \in \mathcal{X} \times \mathcal{T} \times \mathcal{Y}, \, \left| \pdv[order=2]{\Theta^+}{T}(t, \mathbf{x}, \bar q_\gamma) \right| \leq M_{\partial^2 \Theta^+}.
    \end{equation*}
\end{assumption}

\subsection{Stabilized and Doubly Robust Estimations of the APO from \texorpdfstring{\textcite{kallus2018policy}}{Kallus and Zhou (2018)}}  \label{sec:stab_augm_apo}

The stabilized (or normalized) estimator from \textcite{kallus2018policy} can be written as:
\begin{equation}  \label{eqn:stab_apo}
\hat \theta^{\text{stab}}_h(\tau) = \sum_{i=1}^n \frac{K_h(T_i-\tau) Y_i}{\hat f(T=T_i | \mathbf{X}=\mathbf{X}_i)} \Big/ \sum_{j=1}^n \frac{K_h(T_j-\tau)}{\hat f(T=T_j | \mathbf{X}=\mathbf{X}_j)}.
\end{equation}

The doubly robust (or augmented) estimator from \textcite{kallus2018policy} can be written as:
\begin{equation}  \label{eqn:augm_apo}
\hat \theta^{\text{augm}}_h(\tau) = \hat \eta(\tau) + \frac{1}{n} \sum_{i=1}^n \frac{K_h(T_i-\tau)}{\hat f(T=T_i | \mathbf{X}=\mathbf{X}_i)} (Y_i - \hat \eta(T_i, \mathbf{X}_i)),
\end{equation}
where $\hat \eta(T_i, \mathbf{X}_i)$ is an estimator of $\E[Y|T=T_i, \mathbf{X}=\mathbf{X}_i]$ and $\hat \eta(\tau) = \sum_{i=1}^n \hat \eta(\tau, \mathbf{X}_i) / n$.

It is also possible to combine stabilization with double-robustness in one estimator. The estimator becomes:
\begin{equation}  \label{eqn:stab_augm_apo}
\hat \theta^{\text{stab}, \text{augm}}_h(\tau) = \hat \eta(\tau) + \frac{\sum_{i=1}^n \frac{K_h(T_i-\tau)}{\hat f(T=T_i | \mathbf{X}=\mathbf{X}_i)} (Y_i - \hat \eta(T_i, \mathbf{X}_i))}{\sum_{j=1}^n \frac{K_h(T_j-\tau)}{\hat f(T=T_j | \mathbf{X}=\mathbf{X}_j)}},
\end{equation}

Stabilization aims at reducing variance while double robustness ensures robustness to model misspecification: only $\hat f(T=T_i | \mathbf{X}=\mathbf{X}_i)$ or $\hat \eta(\tau, \mathbf{X}_i)$ needs to be correctly specified to ensure that the estimator is asymptotically unbiased.

\subsection{Details for the Weight Function \texorpdfstring{$w^\star$}{}}  \label{sec:details_weight_function}

We can rewrite the CAPO as
\begin{align*}
    \theta(\tau, \mathbf{x}) & \coloneq \mathbb{E}_{\mathbf{X}=\mathbf{x}}[Y(\tau)] \\
    & = \int y f(Y(\tau)=y | \mathbf{X}=\mathbf{x}) \, \mathrm{d}y \\
    & = \int y \frac{f(Y(\tau)=y | \mathbf{X}=\mathbf{x})}{f(Y=y | \mathbf{X}=\mathbf{x}, T=\tau)} f(Y=y | \mathbf{X}=\mathbf{x}, T=\tau) \, \mathrm{d}y \quad \text{by Assumption~\ref{ass:positivity_cond_outcome}} \\
    & = \mathbb{E}_{\mathbf{X}=\mathbf{x}, T=\tau}[Y w^\star(Y, \mathbf{x}, \tau)],
\end{align*}
where, for all $(\mathbf{x}, t, y) \in \mathcal{X} \times \mathcal{T} \times \mathcal{Y}$, the likelihood ratio $w^\star$ is defined as
\begin{equation*}
     w^\star(y, \mathbf{x}, t) = \frac{f\bigl(Y(t)=y \big| \mathbf{X}=\mathbf{x} \bigr)}{f\bigl(Y=y \big| \mathbf{X}=\mathbf{x}, T=t \bigr)}.
\end{equation*}

Moreover, for all $(\mathbf{x}, t) \in \mathcal{X} \times \mathcal{T}$,
\begin{align*}
    \mathbb{E}_{\mathbf{X}=\mathbf{x}, T=t}[w^\star(Y, \mathbf{x}, t)] & = \int \frac{f\bigl(Y(t)=y \big| \mathbf{X}=\mathbf{x} \bigr)}{f\bigl(Y=y |\mathbf{X}=\mathbf{x}, T=t \bigr)} f\bigl(Y=y \big| \mathbf{X}=\mathbf{x}, T=t \bigr) \, \mathrm{d}y \\
    & = \int f\bigl(Y(t)=y \big| \mathbf{X}=\mathbf{x} \bigr) \, \mathrm{d}y = 1,
\end{align*}
which leads to Equation~\eqref{eqn:constraint_w_star}.

\subsection{All Bounds for the CAPO and APO}  \label{sec:all_bounds}

\subsubsection{Bounds for the CAPO}

From the proof of Theorem~\ref{theo:capo_bounds_and_sharpness}, we can get the true bounds for the CAPO:
\begin{align}
    \theta^-(\tau, \mathbf{x}) & = \eta(\tau, \mathbf{x}) + \frac{2\gamma - 1}{\gamma} \mathbb E_{\mathbf{X}=\mathbf{x}, T=\tau} \Bigl[Y - \eta(\tau, \mathbf{x}) \Big| Y \leq q_{1-\gamma}^{\mathbf{x}, \tau} \Bigr]  \label{eqn:capo_lb_v1} \\
    & = \eta(\tau, \mathbf{x}) + \mathbb{E}_{\mathbf{X}=\mathbf{x}, T=\tau} \Bigl[ (Y - \eta(\tau, \mathbf{x})) \Gamma^{- \sign(Y - q_{1-\gamma}^{\mathbf{x}, \tau})} \Bigr]  \label{eqn:capo_lb_v2}
\end{align}
and
\begin{align}
    \theta^+(\tau, \mathbf{x}) & = \eta(\tau, \mathbf{x}) + \frac{2\gamma - 1}{\gamma} \mathbb E_{\mathbf{X}=\mathbf{x}, T=\tau} \Bigl[Y - \eta(\tau, \mathbf{x}) \Big| Y > q_\gamma^{\mathbf{x}, \tau} \Bigr] \label{eqn:capo_ub_v1} \\
    & = \eta(\tau, \mathbf{x}) + \mathbb{E}_{\mathbf{X}=\mathbf{x}, T=\tau} \Bigl[ (Y - \eta(\tau, \mathbf{x})) \Gamma^{\sign(Y - q_\gamma^{\mathbf{x}, \tau})} \Bigr]  \label{eqn:capo_ub_v2}
\end{align}
where $\gamma = \Gamma/(1+\Gamma)$ and $q_\upsilon^{\mathbf{x}, \tau} = Q(\upsilon ; Y|\mathbf{X}=\mathbf{x}, T=\tau)$.

In subsection~\ref{sec:bound_estimation}, the kernelized versions of the bounds are given by
\begin{align}
    \theta_h^-(\tau, \mathbf{x}) & = \eta(\tau, \mathbf{x}) + \frac{2\gamma - 1}{\gamma} \mathbb{E}_{\mathbf{X}=\mathbf{x}} \biggl[ \frac{K_h(T - \tau)}{f(T|\mathbf{X}=\mathbf{x})} (Y - \eta(\tau, \mathbf{x})) \bigg| Y \leq q_{1-\gamma}^{\mathbf{x}, T} \biggr] \label{eqn:kernel_capo_lb_v1} \\
    & = \eta(\tau, \mathbf{x}) + \mathbb{E}_{\mathbf{X}=\mathbf{x}} \biggl[ \frac{K_h(T - \tau)}{f(T|\mathbf{X}=\mathbf{x})} (Y - \eta(\tau, \mathbf{x})) \Gamma^{- \sign(Y - q_{1-\gamma}^{\mathbf{x}, T})} \biggr] \label{eqn:kernel_capo_lb_v2}
\end{align}
and
\begin{align}
    \theta_h^+(\tau, \mathbf{x}) & = \eta(\tau, \mathbf{x}) + \frac{2\gamma - 1}{\gamma} \mathbb{E}_{\mathbf{X}=\mathbf{x}} \biggl[ \frac{K_h(T - \tau)}{f(T|\mathbf{X}=\mathbf{x})} (Y - \eta(\tau, \mathbf{x})) \bigg| Y > q_\gamma^{\mathbf{x}, T} \biggr] \label{eqn:kernel_capo_ub_v1} \\
    & = \eta(\tau, \mathbf{x}) + \mathbb{E}_{\mathbf{X}=\mathbf{x}} \biggl[ \frac{K_h(T - \tau)}{f(T|\mathbf{X}=\mathbf{x})} (Y - \eta(\tau, \mathbf{x})) \Gamma^{\sign(Y - q_\gamma^{\mathbf{x}, T})} \biggr] \label{eqn:kernel_capo_ub_v2}
\end{align}
where $K_h$ is defined as in Equation~\eqref{eqn:estimator_under_ignorability}.

\subsubsection{Bounds for the APO}

From the proof of Theorem~\ref{theo:capo_bounds_and_sharpness} and the relation $\theta(\tau) = \E[\theta(\tau, \mathbf{X})]$, we can get the true bounds for the APO:
\begin{align}
    \theta^-(\tau) & = \eta(\tau) + \frac{2\gamma - 1}{\gamma} \E \biggl[ \mathbb E_{\mathbf{X}, T=\tau} \Bigl[Y - \eta(\tau, \mathbf{X}) \Big| Y \leq q_{1-\gamma}^{\mathbf{X}, \tau} \Bigr] \biggr]  \label{eqn:apo_lb_v1} \\
    & = \eta(\tau) + \E \biggl[ \mathbb{E}_{\mathbf{X}, T=\tau} \Bigl[ (Y - \eta(\tau, \mathbf{X})) \Gamma^{- \sign(Y - q_{1-\gamma}^{\mathbf{X}, \tau})} \Bigr] \biggr]  \label{eqn:apo_lb_v2}
\end{align}
and
\begin{align}
    \theta^+(\tau) & = \eta(\tau) + \frac{2\gamma - 1}{\gamma} \E \biggl[ \mathbb E_{\mathbf{X}, T=\tau} \Bigl[Y - \eta(\tau,\mathbf{X}) \Big| Y > q_\gamma^{\mathbf{X}, \tau} \Bigr] \biggr]  \label{eqn:apo_ub_v1} \\
    & = \eta(\tau) + \E \biggl[ \mathbb{E}_{\mathbf{X}, T=\tau} \Bigl[ (Y - \eta(\tau, \mathbf{X})) \Gamma^{\sign(Y - q_\gamma^{\mathbf{X}, \tau})} \Bigr] \biggr],  \label{eqn:apo_ub_v2}
\end{align}
where $\eta(\tau) = \E[\eta(\tau, \mathbf{X})]$, $\gamma = \Gamma/(1+\Gamma)$ and $q_\upsilon^{\mathbf{X}, \tau} = Q(\upsilon ; Y|\mathbf{X}, T=\tau)$.

In subsection~\ref{sec:bound_estimation}, the kernelized versions of the bounds are given by
\begin{align}
    \theta_h^-(\tau) & = \eta(\tau) + \frac{2\gamma - 1}{\gamma} \E \biggl[ \mathbb{E} \biggl[ \frac{K_h(T - \tau)}{f(T|\mathbf{X})} (Y - \eta(\tau, \mathbf{X})) \bigg| \mathbf{X}, Y \leq q_{1-\gamma}^{\mathbf{X}, T} \biggr] \biggr] \label{eqn:kernel_apo_lb_v1} \\
    & = \eta(\tau) + \E \biggl[ \frac{K_h(T - \tau)}{f(T|\mathbf{X})} (Y - \eta(\tau, \mathbf{X})) \Gamma^{- \sign(Y - q_{1-\gamma}^{\mathbf{X}, T})} \biggr] \label{eqn:kernel_apo_lb_v2}
\end{align}
and
\begin{align}
    \theta_h^+(\tau) & = \eta(\tau) + \frac{2\gamma - 1}{\gamma} \E \biggl[ \mathbb{E} \biggl[ \frac{K_h(T - \tau)}{f(T|\mathbf{X})} (Y - \eta(\tau, \mathbf{X})) \bigg| \mathbf{X}, Y > q_\gamma^{\mathbf{X}, T} \biggr] \biggr] \label{eqn:kernel_apo_ub_v1} \\
    & = \eta(\tau) + \E \biggl[ \frac{K_h(T - \tau)}{f(T|\mathbf{X})} (Y - \eta(\tau, \mathbf{X})) \Gamma^{\sign(Y - q_\gamma^{\mathbf{X}, T})} \biggr], \label{eqn:kernel_apo_ub_v2}
\end{align}
where $K_h$ is defined as in Equation~\eqref{eqn:estimator_under_ignorability}.

By defining $\mathcal{I}_\mathcal{D_-} = \{ i \in [\![1, n]\!] \, ; \, Y_i \leq q_{1-\gamma}^{\mathbf{X}_i, T_i} \}$ of cardinality $n_-$, and $\mathcal{I}_\mathcal{D_+} = \{ i \in [\![1, n]\!] \, ; \, Y_i > q_\gamma^{\mathbf{X}_i, T_i} \}$ of cardinality $n_+$, we can obtain estimators of $\theta_h^-(\tau)$ and $\theta_h^+(\tau)$ as follows, using Equations~\eqref{eqn:kernel_apo_lb_v1} and \eqref{eqn:kernel_apo_ub_v1}:
\begin{equation}  \label{eqn:tilde_kernel_apo_lb_v1}
    \tilde \theta_h^-(\tau) = \tilde \eta(\tau) + \frac{2\gamma-1}{\gamma n_-} \sum_{i \in \mathcal{I}_\mathcal{D_-}} \frac{K_h(T_i - \tau)}{f(T_i|\mathbf{X}_i)} (Y_i - \eta(T_i, \mathbf{X}_i))
\end{equation}
and
\begin{equation}  \label{eqn:tilde_kernel_apo_ub_v1}
    \tilde \theta_h^+(\tau) = \tilde \eta(\tau) + \frac{2\gamma-1}{\gamma n_+} \sum_{i \in \mathcal{I}_\mathcal{D_+}} \frac{K_h(T_i - \tau)}{f(T_i|\mathbf{X}_i)} (Y_i - \eta(T_i, \mathbf{X}_i)),
\end{equation}
where $\tilde \eta(\tau) = \frac{1}{n} \sum_{i=1}^n \eta(\tau, \mathbf{X}_i)$.

Using Equations~\eqref{eqn:kernel_apo_lb_v2} and \eqref{eqn:kernel_apo_ub_v2}, we can also get estimators via
\begin{equation}  \label{eqn:tilde_kernel_apo_lb_v2}
    \tilde \theta_h^-(\tau) = \tilde \eta(\tau) + \frac{1}{n} \sum_{i=1}^n \frac{K_h(T_i-\tau) (Y_i - \eta(T_i, \mathbf{X}_i))}{f(T_i|\mathbf{X}_i)} \Gamma^{- \sign \left(Y_i - q_{1-\gamma}^{\mathbf{X}_i, T_i} \right)}
\end{equation}
and
\begin{equation}  \label{eqn:tilde_kernel_apo_ub_v2}
    \tilde \theta_h^+(\tau) = \tilde \eta(\tau) + \frac{1}{n} \sum_{i=1}^n \frac{K_h(T_i-\tau) (Y_i - \eta(T_i, \mathbf{X}_i))}{f(T_i|\mathbf{X}_i)} \Gamma^{\sign \left(Y_i - q_\gamma^{\mathbf{X}_i, T_i} \right)},
\end{equation}
where $\tilde \eta(\tau) = \frac{1}{n} \sum_{i=1}^n \eta(\tau, \mathbf{X}_i)$.

In practice, we use stabilized versions of our estimators to reduce their variance. Thus, Equations~\eqref{eqn:tilde_kernel_apo_lb_v1} and \eqref{eqn:tilde_kernel_apo_ub_v1} become
\begin{equation}  \label{eqn:stab_tilde_kernel_apo_lb_v1}
    \tilde \theta_h^{-, \mathrm{stab}}(\tau) = \tilde \eta(\tau) + \frac{2\gamma-1}{\gamma} \cdot \frac{\sum_{i \in \mathcal{I}_\mathcal{D_-}} \frac{K_h(T_i - \tau)}{f(T_i|\mathbf{X}_i)} (Y_i - \eta(T_i, \mathbf{X}_i))}{\sum_{j \in \mathcal{I}_\mathcal{D_-}} \frac{K_h(T_j - \tau)}{f(T_j|\mathbf{X}_j)}}
\end{equation}
and
\begin{equation}  \label{eqn:stab_tilde_kernel_apo_ub_v1}
    \tilde \theta_h^{+, \mathrm{stab}}(\tau) = \tilde \eta(\tau) + \frac{2\gamma-1}{\gamma} \cdot \frac{\sum_{i \in \mathcal{I}_\mathcal{D_+}} \frac{K_h(T_i - \tau)}{f(T_i|\mathbf{X}_i)} (Y_i - \eta(T_i, \mathbf{X}_i))}{\sum_{j \in \mathcal{I}_\mathcal{D_+}} \frac{K_h(T_j - \tau)}{f(T_j|\mathbf{X}_j)}},
\end{equation}
and Equations~\eqref{eqn:tilde_kernel_apo_lb_v2} and \eqref{eqn:tilde_kernel_apo_ub_v2} become
\begin{equation}  \label{eqn:stab_tilde_kernel_apo_lb_v2}
    \tilde \theta_h^{-, \mathrm{stab}}(\tau) = \tilde \eta(\tau) + \frac{\sum_{i=1}^n \frac{K_h(T_i-\tau) (Y_i - \eta(T_i, \mathbf{X}_i))}{f(T_i|\mathbf{X}_i)} \Gamma^{- \sign \left(Y_i - q_{1-\gamma}^{\mathbf{X}_i, T_i} \right)}}{\sum_{j=1}^n \frac{K_h(T_j-\tau)}{f(T_j|\mathbf{X}_j)} \Gamma^{- \sign \left(Y_j - q_{1-\gamma}^{\mathbf{X}_j, T_j} \right)}}
\end{equation}
and
\begin{equation}  \label{eqn:stab_tilde_kernel_apo_ub_v2}
    \tilde \theta_h^{+, \mathrm{stab}}(\tau) = \tilde \eta(\tau) + \frac{\sum_{i=1}^n \frac{K_h(T_i-\tau) (Y_i - \eta(T_i, \mathbf{X}_i))}{f(T_i|\mathbf{X}_i)} \Gamma^{\sign \left(Y_i - q_\gamma^{\mathbf{X}_i, T_i} \right)}}{\sum_{j=1}^n \frac{K_h(T_j-\tau)}{f(T_j|\mathbf{X}_j)} \Gamma^{\sign \left(Y_j - q_\gamma^{\mathbf{X}_j, T_j} \right)}},
\end{equation}
where $\tilde \eta(\tau) = \frac{1}{n} \sum_{i=1}^n \eta(\tau, \mathbf{X}_i)$.

\subsection{Percentile Bootstrap Method}  \label{sec:percentile_bootstrap}

\begin{figure}[H]
    \centering
    \vspace{.3in}
    \includegraphics[width=0.8\linewidth]{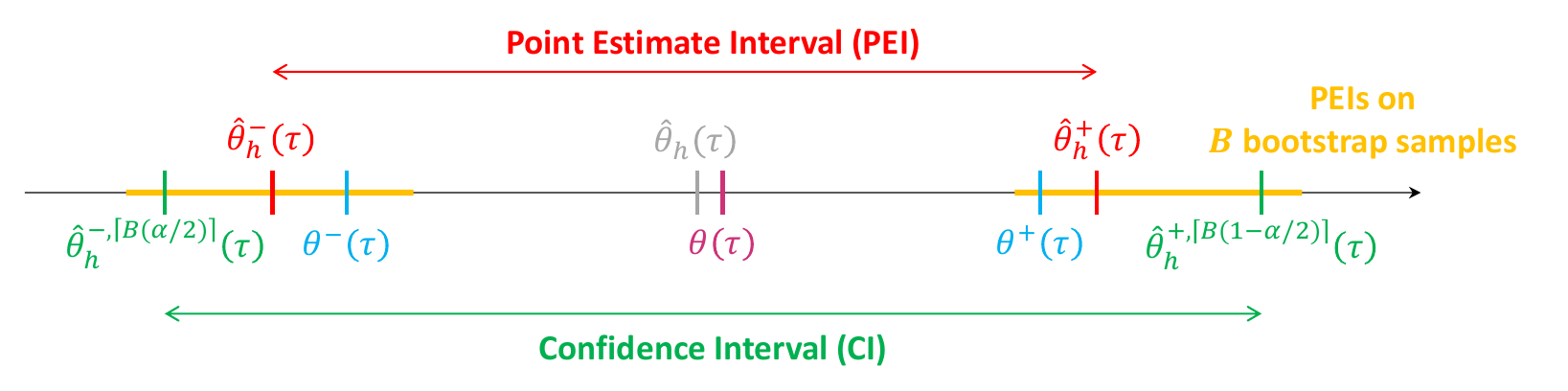}
    \vspace{.3in}
    \caption{Percentile bootstrap method.}
    \label{fig:percentile_bootstrap}
\end{figure}

From the bounds obtained in the previous subsection, we can derive confidence intervals via the percentile bootstrap, as done in \textcite{zhao2019sensitivity}. Figure~\ref{fig:percentile_bootstrap} gives an intuitive visualization of this method.

In Figure~\ref{fig:percentile_bootstrap}, the true unknown APO is represented in violet. The APO estimated via Equation~\eqref{eqn:estimator_under_ignorability} is in gray. The two true bounds on the APO from Equation~\eqref{eqn:apo_lb_ub_sol} are represented in blue (partially identified set). If we estimate these bounds on the whole dataset, we can get the Point Estimate Interval in red. Instead of using the whole dataset, if we compute the PEI on $B$ bootstrap samples and then take the quantiles of order $\alpha/2$ and $1-\alpha/2$ of, respectively, the lower and upper bounds, we can get a $(1-\alpha)$-level confidence interval in green. The ranges of lower bounds and upper bounds on the $B$ bootstrap samples are represented in yellow.

As mentioned in the main text, a two-sided $(1-\alpha)$-CI for the set $\bigl[\theta^-_{h}(\tau), \theta^+_{h}(\tau)\bigr]$ can be obtained after intersecting two one-sided $(1-\alpha/2)$-CIs for the lower and the upper bounds. Indeed, by the union bound,
\begin{align*}
    & \mathbb{P} \Bigl( \bigl[ \theta_h^-(\tau), \, \theta_h^+(\tau) \bigr] \not\subset \bigl[ \hat \theta_h^{-, \lceil B(\alpha/2) \rceil}(\tau), \, \hat \theta_h^{+, \lceil B(1-\alpha/2) \rceil}(\tau) \bigr] \Bigr) \\
    & = \mathbb{P} \Bigl( \bigl\{ \theta_h^-(\tau) < \hat \theta_h^{-, \lceil B(\alpha/2) \rceil}(\tau) \bigr\} \cup \bigl\{ \hat \theta_h^{+, \lceil B(1-\alpha/2) \rceil}(\tau) < \theta_h^+(\tau) \bigr\} \Bigr) \\
    & \leq \mathbb{P} \Bigl( \theta_h^-(\tau) < \hat \theta_h^{-, \lceil B(\alpha/2) \rceil}(\tau) \Bigr) + \mathbb{P} \Bigl( \hat \theta_h^{+, \lceil B(1-\alpha/2) \rceil}(\tau) < \theta_h^+(\tau) \Bigr) \\
    & \leq \frac{\alpha}{2} + \frac{\alpha}{2} = \alpha.
\end{align*}
Therefore,
\begin{equation*}
    \mathbb{P} \Bigl( \bigl[ \theta_h^-(\tau), \, \theta_h^+(\tau) \bigr] \subset \bigl[ \hat \theta_h^{-, \lceil B(\alpha/2) \rceil}(\tau), \, \hat \theta_h^{+, \lceil B(1-\alpha/2) \rceil}(\tau) \bigr] \Bigr) \geq 1 - \alpha.
\end{equation*}

\subsection{Proof of Proposition~\ref{prop:cmsm_equivalence}}  \label{sec:proof_cmsm_equivalence}

($\Longrightarrow$) We start to show that the CMSM from Definition~\ref{def:cmsm_U} implies the one from Definition~\ref{def:cmsm_Y}. Under Assumptions~\ref{ass:positivity_cond_treatment} and \ref{ass:positivity_cond_outcome}, simple computations show that the ratio considered in the sensitivity model from \textcite{jesson2022scalable} (see Definition~\ref{def:cmsm_jesson} or their Definition 1) satisfies, for a fixed treatment value $\tau \in \mathcal{T}$,
\begin{equation}  \label{eqn:ratio_eq_jesson}
    \forall (\mathbf{x}, y) \in \mathcal{X} \times \mathcal{Y}, \, \frac{f\bigl(T=\tau \big| \mathbf{X}=\mathbf{x}, Y(\tau)=y \bigr)}{f\bigl(T=\tau \big| \mathbf{X}=\mathbf{x} \bigr)} = \frac{f\bigl(Y(\tau)=y \big| \mathbf{X}=\mathbf{x}, T=\tau \bigr)}{f\bigl(Y(\tau)=y \big| \mathbf{X}=\mathbf{x} \bigr)}
\end{equation}
by Bayes' theorem.
Now, notice that
\begin{equation*}
    \forall (\mathbf{x}, \mathbf{u}) \in \mathcal{X} \times \mathcal{U}, \, \upsilon(\mathbf{u}, \mathbf{x}, \tau) \coloneq \frac{f\bigl(\mathbf{U}=\mathbf{u} \big| \mathbf{X}=\mathbf{x} \bigr)}{f\bigl(\mathbf{U}=\mathbf{u} \big| \mathbf{X}=\mathbf{x}, T=\tau \bigr)} = \frac{f\bigl(T=\tau \big| \mathbf{X}=\mathbf{x} \bigr)}{f\bigl(T=\tau \big| \mathbf{X}=\mathbf{x}, \mathbf{U}=\mathbf{u} \bigr)}
\end{equation*}
is in $[\Gamma^{-1}, \Gamma]$ under the CMSM from Definition~\ref{def:cmsm_U}. Moreover, $(\mathbf{X}, \mathbf{U})$-ignorability implies that
\begin{align*}
    f\bigl(Y(\tau)=y \big| \mathbf{X}=\mathbf{x} \bigr) & = \int_{\mathcal U} f\bigl(Y(\tau)=y, \mathbf{U}=\mathbf{u} \big| \mathbf{X}=\mathbf{x} \bigr) \, \mathrm{d}\mathbf{u} \\
    & = \int_{\mathcal U} f\bigl(Y(\tau)=y \big| \mathbf{X}=\mathbf{x}, \mathbf{U}=\mathbf{u} \bigr) f\bigl(\mathbf{U}=\mathbf{u} \big| \mathbf{X}=\mathbf{x} \bigr) \, \mathrm{d}\mathbf{u} \\
    & = \int_{\mathcal U} f\bigl(Y(\tau)=y \big| \mathbf{X}=\mathbf{x}, \mathbf{U}=\mathbf{u}, T=\tau \bigr) f\bigl(\mathbf{U}=\mathbf{u} \big| \mathbf{X}=\mathbf{x}, T=\tau \bigr) \upsilon(\mathbf{u}, \mathbf{x}, \tau) \, \mathrm{d}\mathbf{u}.
\end{align*}
Therefore, by Bayes' theorem and as $\upsilon(\mathbf{u}, \mathbf{x}, \tau)$ is in $[\Gamma^{-1}, \Gamma]$ for all $\mathbf{x}$ and $\mathbf{u}$,
\begin{align*}
    & \Gamma^{-1} \int_{\mathcal U} f\bigl(Y(\tau)=y, \mathbf{U}=\mathbf{u} \big| \mathbf{X}=\mathbf{x}, T=\tau \bigr) \, \mathrm{d}\mathbf{u} \leq f\bigl(Y(\tau)=y \big| \mathbf{X}=\mathbf{x} \bigr) \leq \Gamma \int_{\mathcal U} f\bigl(Y(\tau)=y, \mathbf{U}=\mathbf{u} \big| \mathbf{X}=\mathbf{x}, T=\tau \bigr) \, \mathrm{d}\mathbf{u} \\
    & \Rightarrow \Gamma^{-1} f\bigl(Y(\tau)=y \big| \mathbf{X}=\mathbf{x}, T=\tau \bigr) \leq f\bigl(Y(\tau)=y \big| \mathbf{X}=\mathbf{x} \bigr) \leq \Gamma f\bigl(Y(\tau)=y \big| \mathbf{X}=\mathbf{x}, T=\tau \bigr) \\
    & \Rightarrow \Gamma^{-1} \leq \frac{f\bigl(Y(\tau)=y \big| \mathbf{X}=\mathbf{x} \bigr)}{f\bigl(Y(\tau)=y \big| \mathbf{X}=\mathbf{x}, T=\tau \bigr)} \leq \Gamma
\end{align*}
Finally, using Equation~\eqref{eqn:ratio_eq_jesson}, we just proved that our sensitivity model~\eqref{eqn:cmsm_U} implies the one from \textcite{jesson2022scalable}.

($\Longleftarrow$) To show that the CMSM from Definition~\ref{def:cmsm_Y} implies the one from Definition~\ref{def:cmsm_U}, suppose that Equation~\eqref{eqn:cmsm_Y} is true and consider the case where $\mathbf{U}$ is defined as $Y(\tau)$. This is a valid confounder as it satisfies Equation~\eqref{eqn:XU_ignorability} (($\mathbf{X}, \mathbf{U}$)-ignorability). Then, the ratio considered in Equation~\eqref{eqn:cmsm_U} is in $[\Gamma^{-1}, \Gamma]$.

\subsection{Proof of Theorem~\ref{theo:capo_bounds_and_sharpness}} \label{proof:capo_bounds_and_sharpness}

The proof focuses on the lower bound for the CAPO $\theta^-(\tau, \mathbf{x})$ but a similar reasoning can be used for the upper bound $\theta^+(\tau, \mathbf{x})$.

Notice that the minimization problem~\eqref{eqn:capo_lb} can be written:
\begin{align}
    \theta^-(\tau, \mathbf{x}) & = \eta(\tau, \mathbf{x}) + \inf_{w \in \mathcal W^\star_{\tau}} \mathbb  E_{\mathbf{X}=\mathbf{x}, T=\tau} \bigl[\bigl(Y - \eta(\tau, \mathbf{x})\bigr) w(Y, \mathbf{x}, \tau) \bigr] \nonumber \\
    & = \Gamma^{-1} \eta(\tau, \mathbf{x}) + (1 - \Gamma^{-1}) \underset{w \in \mathcal{W}^\star_\tau}{\inf} \mathbb{E}_{\mathbf{X}=\mathbf{x}, T=\tau} \Biggl[ Y\frac{w(Y, \mathbf{x}, \tau) - \Gamma^{-1}}{1 - \Gamma^{-1}} \Biggr]  \label{eqn:capo_lb_infimum}
\end{align}

Now, rewrite the expectancy
\begin{align*}
    \mathbb{E}_{\mathbf{X}=\mathbf{x}, T=\tau} \Biggl[ Y \frac{w(Y, \mathbf{x}, \tau) - \Gamma^{-1}}{1 - \Gamma^{-1}} \Biggr] & = \int y \frac{w(y, \mathbf{x}, \tau) - \Gamma^{-1}}{1 - \Gamma^{-1}} f(Y=y | \mathbf{X}=\mathbf{x}, T=\tau) \, \mathrm{d}y \\
    & = \int y g(y, \mathbf{x}, \tau) f(Y=y | \mathbf{X}=\mathbf{x}, T=\tau) \, \mathrm{d}y
\end{align*}
where $g(y, \mathbf{x}, \tau) \coloneq (w(y, \mathbf{x}, \tau) - \Gamma^{-1}) / (1 - \Gamma^{-1})$ is a density ratio because $f(Y=y|\mathbf{X}=\mathbf{x}, T=\tau) g(y, \mathbf{x}, \tau)$ is a density. Indeed,
\begin{align*}
    \int f(Y=y|\mathbf{X}=\mathbf{x}, T=\tau) g(y, \mathbf{x}, \tau) \, \mathrm{d}y & = \frac{1}{1 - \Gamma^{-1}} \int f(Y=y|\mathbf{X}=\mathbf{x}, T=\tau) \left( \frac{f(Y(\tau)=y|\mathbf{X}=\mathbf{x})}{f(Y=y|\mathbf{X}=\mathbf{x}, T=\tau)} - \Gamma^{-1} \right) \, \mathrm{d}y \\
    & = \frac{1}{1 - \Gamma^{-1}} \int f(Y=y|\mathbf{X}=\mathbf{x}, T=\tau) - \Gamma^{-1} f(Y(\tau)=y|\mathbf{X}=\mathbf{x}) \, \mathrm{d}y \\
    & = \frac{1}{1-\Gamma^{-1}} - \frac{\Gamma^{-1}}{1-\Gamma^{-1}} = 1
\end{align*}
and, as $w \in \mathcal W^\star(\tau)$, it takes its values in $[\Gamma^{-1}, \Gamma]$, so
\begin{align*}
    g(y, \mathbf{x}, \tau) = \frac{1}{1-\Gamma^{-1}} \left( w(y, \mathbf{x}, \tau) - \Gamma^{-1} \right) \in [0, \Gamma+1] = [0, (1-\gamma)^{-1}].
\end{align*}
As $f(Y=y|\mathbf{X}=\mathbf{x}, T=\tau)$ is nonnegative, $f(Y=y|\mathbf{X}=\mathbf{x}, T=\tau) g(y, \mathbf{x}, \tau)$ is also nonnegative.

Therefore, we can rewrite the infimum from Equation~\eqref{eqn:capo_lb_infimum} as an infimum over the set of density ratios $g$ that are lower than $1/(1-\gamma)$:
\begin{equation*}
    \underset{w \in \mathcal{W}^\star_\tau}{\inf} \mathbb{E}_{\mathbf{X}=\mathbf{x}, T=\tau} \Biggl[ Y\frac{w(Y, \mathbf{x}, \tau) - \Gamma^{-1}}{1 - \Gamma^{-1}} \Biggr] = \underset{g(Y, \mathbf{x}, \tau) \leq \frac{1}{1-\gamma}}{\inf} \mathbb{E}_{\mathbf{X}=\mathbf{x}, T=\tau} [Y g(Y, \mathbf{x}, \tau)].
\end{equation*}

The above equation can be rewritten in terms of the Expected Shortfall $\mathrm{ES}_{1-\gamma}(Y) \coloneq \mathbb E_{\mathbf{X}=\mathbf{x}, T=\tau} [-Y | Y \leq q_{1-\gamma}^{\mathbf{x}, \tau}]$. Indeed, the ``Fenchel-Moreau-Rockafellar" dual representation of the ES gives (e.g., see \cite{herdegen2023elementary}):
\begin{equation*}
    \mathrm{ES}_{1-\gamma}(Y) = \underset{g(Y, \mathbf{x}, \tau) \leq \frac{1}{1-\gamma}}{\sup} \mathbb{E}_{\mathbf{X}=\mathbf{x}, T=\tau} [- Y g(Y, \mathbf{x}, \tau)] = - \underset{g(Y, \mathbf{x}, \tau) \leq \frac{1}{1-\gamma}}{\inf} \mathbb{E}_{\mathbf{X}=\mathbf{x}, T=\tau} [Y g(Y, \mathbf{x}, \tau)]
\end{equation*}
    
By plugging the previous results in Equation~\eqref{eqn:capo_lb_infimum}, it leads to the following equality:
\begin{equation}  \label{eqn:capo_lb_expected_shortfall}
    \theta^-(\tau, \mathbf{x}) = \Gamma^{-1} \eta(\tau, \mathbf{x}) + (\Gamma^{-1} - 1) \mathbb E_{\mathbf{X}=\mathbf{x}, T=\tau} [-Y | Y \leq q_{1-\gamma}^{\mathbf{x}, \tau}].
\end{equation}

Finally, to get a dependence only in $\gamma$, we can rewrite the lower bound for the CAPO:
\begin{align*}
    \theta^-(\tau, \mathbf{x}) & = \Gamma^{-1} \eta(\tau, \mathbf{x}) + (\Gamma^{-1} - 1) \mathbb E_{\mathbf{X}=\mathbf{x}, T=\tau} [-Y | Y \leq q_{1-\gamma}^{\mathbf{x}, \tau}] \\
    & = \frac{1 - \gamma}{\gamma} \eta(\tau, \mathbf{x}) + \frac{1 - 2\gamma}{\gamma} \mathbb E_{\mathbf{X}=\mathbf{x}, T=\tau} [-Y | Y \leq q_{1-\gamma}^{\mathbf{x}, \tau}] \\
    & = \eta(\tau, \mathbf{x}) + \frac{1 - 2\gamma}{\gamma} \mathbb E_{\mathbf{X}=\mathbf{x}, T=\tau} [-Y | Y \leq q_{1-\gamma}^{\mathbf{x}, \tau}] + \frac{1 - 2\gamma}{\gamma} \mathbb E_{\mathbf{X}=\mathbf{x}, T=\tau} [\eta(\tau, \mathbf{x}) | Y \leq q_{1-\gamma}^{\mathbf{x}, \tau}] \\
    & = \eta(\tau, \mathbf{x}) + \frac{2\gamma - 1}{\gamma} \mathbb E_{\mathbf{X}=\mathbf{x}, T=\tau} [Y - \eta(\tau, \mathbf{x}) | Y \leq q_{1-\gamma}^{\mathbf{x}, \tau}].
\end{align*}

Similarly, the upper bound for the CAPO is given by
\begin{equation*}
    \theta^+(\tau, \mathbf{x}) = \eta(\tau, \mathbf{x}) + \frac{2\gamma - 1}{\gamma} \mathbb E_{\mathbf{X}=\mathbf{x}, T=\tau} [Y - \eta(\tau, \mathbf{x}) | Y > q_\gamma^{\mathbf{x}, \tau}].
\end{equation*}

As the proof relies on the ``Fenchel-Moreau-Rockafellar" dual representation, which is an optimum, the interval $[\theta^-(\tau, \mathbf{x}), \theta^+(\tau, \mathbf{x})]$ is also sharp under the CMSM.

\subsection{Alternative Proof of the Bounds \texorpdfstring{$\theta^-(\tau, \mathbf{x})$}{} and \texorpdfstring{$\theta^+(\tau, \mathbf{x})$}{} from Theorem~\ref{theo:capo_bounds_and_sharpness}}  \label{proof:alt_capo_true_bounds}

In this alternative proof of the bounds for the CAPO, we leverage the constraint on the weight function from Equation~\eqref{eqn:constraint_w_star}, which reminds the \textit{Quantile Balancing} condition from \textcite{dorn2022sharp} in the binary treatment case.
    
We first show that the minimizer and maximizer of the optimization problems~\eqref{eqn:capo_lb} and \eqref{eqn:capo_ub} are
\begin{equation*}
    w^-(Y, \mathbf{x}, \tau) = \Gamma^{- \sign \left( Y - Q \left(1-\gamma ; \, Y|\mathbf{X}=\mathbf{x}, T=\tau \right) \right)} \text{ and } w^+(Y, \mathbf{x}, \tau) = \Gamma^{\sign \left( Y - Q \left(\gamma ; \, Y|\mathbf{X}=\mathbf{x}, T=\tau \right) \right)},
\end{equation*}
where $\sign(\cdot)$ is the sign function.
    
We focus on the maximization problem~\eqref{eqn:capo_ub}. First, $w^+(Y, \mathbf{x}, \tau)$ fulfills Equation~\eqref{eqn:constraint_w_star} because
\begin{align*}
    \mathbb{E}_{\mathbf{X}=\mathbf{x}, T=\tau} [w^+(Y, \mathbf{x}, \tau)] & = \int_{\mathcal Y} \Gamma^{\sign(y - Q(\gamma;Y|\mathbf{X}=\mathbf{x},T=\tau))} f(Y=y | \mathbf{X}=\mathbf{x}, T=\tau) \, \mathrm{d}y \\
    & = \Gamma \int_{\mathcal Y} \mathds{1}(y > Q(\gamma;Y|\mathbf{X}=\mathbf{x},T=\tau)) f(Y=y | \mathbf{X}=\mathbf{x}, T=\tau) \, \mathrm{d}y \\
    & \quad + \Gamma^{-1} \int_{\mathcal Y} \mathds{1}(y \leq Q(\gamma;Y|\mathbf{X}=\mathbf{x},T=\tau)) f(Y=y | \mathbf{X}=\mathbf{x}, T=\tau) \, \mathrm{d}y \\
    & = \Gamma (1 - \gamma) + \Gamma^{-1} \gamma = 1,
\end{align*}
by definition of the conditional quantile. The same applies to $w^-(Y, \mathbf{x}, \tau)$. Moreover, $w^+(Y, \mathbf{x}, \tau)$ and $w^-(Y, \mathbf{x}, \tau)$ are in $[\Gamma^{-1}, \Gamma]$, so they are in $\mathcal{W}^\star_\tau$.
    
Then, notice that, for all $w \in \mathcal{W}^\star_\tau$,
\begin{align*}
    \mathbb{E}_{\mathbf{X}=\mathbf{x}, T=\tau} \left[ Y w(Y, \mathbf{x}, \tau) \right] & = \mathbb{E}_{\mathbf{X}=\mathbf{x}, T=\tau} [Y w(Y, \mathbf{x}, \tau)] \\
    & = \mathbb{E}_{\mathbf{X}=\mathbf{x}, T=\tau} \left[ \bigl(Y - Q(\gamma;Y|\mathbf{X}=\mathbf{x},T=\tau)\bigr) w(Y, \mathbf{x}, \tau) \right] + Q(\gamma;Y|\mathbf{X}=\mathbf{x},T=\tau)
\end{align*}
thanks to Equation~\eqref{eqn:constraint_w_star}.
Solving the maximization problem is then equivalent to maximizing the following expression with respect to $w$:
\begin{equation*}
    \mathbb{E}_{\mathbf{X}=\mathbf{x}, T=\tau} \left[ \bigl( Y - Q(\gamma;Y|\mathbf{X}=\mathbf{x},T=\tau) \bigr) w(Y,\mathbf{x},\tau) \right].
\end{equation*}
To maximize the expectancy, we want $w$ to be the biggest possible when $Y - Q(\gamma;Y|\mathbf{X}=\mathbf{x},T=\tau)$ is positive, i.e.\ be equal to $\Gamma$, and the lowest possible when $Y - Q(\gamma;Y|\mathbf{X}=\mathbf{x},T=\tau)$ is negative, i.e.\ be equal to $\Gamma^{-1}$. Thus, the only possible maximizer is $w^+(Y, \mathbf{x}, \tau) = \Gamma^{\sign \left( Y - Q \left(\gamma ; \, Y|\mathbf{X}=\mathbf{x}, T=\tau \right) \right)}$.

The same reasoning applies for the minimization problem~\eqref{eqn:capo_lb}. Finally, as the problem is linear in $w$, $[\theta^-(\tau, \mathbf{x}), \theta^+(\tau, \mathbf{x})]$ is an interval. Therefore, the bounds are:
\begin{align}
    \theta^-(\tau, \mathbf{x}) & = \eta(\tau, \mathbf{x}) + \mathbb{E}_{\mathbf{X}=\mathbf{x}, T=\tau} \left[ (Y - \eta(\tau, \mathbf{x})) \Gamma^{-\sign(Y - q_{1-\gamma}^{\mathbf{x}, \tau})} \right] \label{eqn:capo_minus_second_form} \\
    \theta^+(\tau, \mathbf{x}) & = \eta(\tau, \mathbf{x}) + \mathbb{E}_{\mathbf{X}=\mathbf{x}, T=\tau} \left[ (Y - \eta(\tau, \mathbf{x})) \Gamma^{\sign(Y - q_\gamma^{\mathbf{x}, \tau})} \right] \label{eqn:capo_plus_second_form}
\end{align}
To get their final form, simply notice that
\begin{align*} 
    \mathbb E_{ \mathbf{X}=\mathbf{x}, \, T=\tau} \Bigl[\bigl(Y - \eta(\tau, \mathbf{x}) \bigr) w^+(Y, \mathbf{x}, \tau)  \Bigr] & = \Gamma  \mathbb E_{ \mathbf{X}=\mathbf{x}, \, T=\tau} \Bigl[\bigl(Y - \eta(\tau, \mathbf{x}) \bigr) \ind_{Y > Q \left(\gamma ; \, Y|\mathbf{X}=\mathbf{x}, T=\tau\right)}\Bigr] \\
    & \quad + \Gamma^{-1}  \mathbb E_{ \mathbf{X}=\mathbf{x}, \, T=\tau} \Bigl[\bigl(Y - \eta(\tau, \mathbf{x}) \bigr) \ind_{Y \leq Q \left(\gamma ; \, Y|\mathbf{X}=\mathbf{x}, T=\tau\right)}\Bigr] \\
    & = \left( \frac{\gamma}{1-\gamma} - \frac{1-\gamma}{\gamma} \right) \mathbb E_{ \mathbf{X}=\mathbf{x}, \, T=\tau} \Bigl[ \bigl(Y - \eta(\tau, \mathbf{x}) \bigr)  \ind_{Y > Q \left(\gamma ; \, Y|\mathbf{X}=\mathbf{x}, T=\tau\right)} \Bigr] \\
    & = \frac{2\gamma - 1}{\gamma (1-\gamma)} \mathbb E_{ \mathbf{X}=\mathbf{x}, \, T=\tau} \Bigl[\bigl(Y - \eta(\tau, \mathbf{x}) \bigr) \ind_{Y > Q \left(\gamma ; \, Y|\mathbf{X}=\mathbf{x}, T=\tau\right)}\Bigr] \\
    & = \frac{2\gamma - 1}{\gamma} \mathbb E_{ \mathbf{X}=\mathbf{x}, \, T=\tau} \Bigl[ Y - \eta(\tau, \mathbf{x}) \Big| Y > q_\gamma^{\mathbf{x}, \tau} \Bigr]
\end{align*}
because $\mathbb{P}(Y > Q \left(\gamma ; \, Y|\mathbf{X}=\mathbf{x}, T=\tau\right)) = 1-\gamma$, and
\begin{align*} 
    \mathbb E_{ \mathbf{X}=\mathbf{x}, \, T=\tau} \Bigl[ \bigl(Y - \eta(\tau, \mathbf{x}) \bigr) w^-(Y, \mathbf{x}, \tau) \Bigr] & = \Gamma^{-1} \mathbb E_{ \mathbf{X}=\mathbf{x}, \, T=\tau} \Bigl[ \bigl(Y - \eta(\tau, \mathbf{x}) \bigr) \ind_{Y > Q \left(1-\gamma ; \, Y|\mathbf{X}=\mathbf{x}, T=\tau\right)} \Bigr] \\
    & \quad + \Gamma  \mathbb E_{ \mathbf{X}=\mathbf{x}, \, T=\tau} \Bigl[ \bigl(Y - \eta(\tau, \mathbf{x}) \bigr) \ind_{Y \leq Q \left(1-\gamma ; \, Y|\mathbf{X}=\mathbf{x}, T=\tau\right)} \Bigr] \\
    & = \frac{2\gamma - 1}{\gamma}\mathbb E_{ \mathbf{X}=\mathbf{x}, \, T=\tau} \Bigl[Y - \eta(\tau, \mathbf{x}) \Big| Y \leq q_{1-\gamma}^{\mathbf{x}, \tau} \Bigr].
\end{align*}
because $\mathbb{P}(Y \leq Q \left(1-\gamma ; \, Y|\mathbf{X}=\mathbf{x}, T=\tau\right)) = 1-\gamma$.

\subsection{Alternative Comparison Between our Bounds and the Ones from \texorpdfstring{\textcite{jesson2022scalable}}{}}  \label{proof:comparison_us_vs_jesson}

Using our notations, we can reformulate the main steps leading to the bounds proposed in \textcite{jesson2022scalable} and demonstrate that they are larger than ours.

Recall that, from Equation~\eqref{eqn:augmented_capo}
\begin{align*}
    \theta(\tau, \mathbf{x}) & = \eta(\tau, \mathbf{x}) + \mathbb{E}_{\mathbf{X}=\mathbf{x}} [Y(\tau) - \eta(\tau, \mathbf{x})] \\
    & = \eta(\tau, \mathbf{x}) + \mathbb{E}_{\mathbf{X}=\mathbf{x}, T=\tau} [(Y - \eta(\tau, \mathbf{x})) w^\star(Y, \mathbf{x}, \tau)].
    &&
\end{align*}
Now, define the weight function considered in \textcite{jesson2022scalable} (denoted there $w(y, \mathbf{x})$) as
\begin{equation*}
    \kappa^\star(y, \mathbf{x}, \tau) = \frac{w^\star(y, \mathbf{x}, \tau) - \Gamma^{-1}}{\Gamma - \Gamma^{-1}},
\end{equation*}
or, equivalently,
\begin{equation*}
    w^\star(y, \mathbf{x}, \tau) = \Gamma^{-1} + \kappa^\star(y, \mathbf{x}, \tau) \bigl( \Gamma - \Gamma^{-1} \bigr).
\end{equation*} 
$\kappa^\star$ takes its values in $[0,1]$ (because $w^\star$ takes its values in $[\Gamma^{-1}, \Gamma]$). Equation~\eqref{eqn:constraint_w_star} ensures that
\begin{equation*}
    \mathbb{E}_{\mathbf{X}=\mathbf{x}, T=\tau} [\kappa^\star(Y, \mathbf{x}, \tau)] = \frac{1 - \Gamma^{-1}}{\Gamma - \Gamma^{-1}} = \frac{1}{\Gamma + 1}.
\end{equation*}
Then, we can write
\begin{align*}
    \theta(\tau, \mathbf{x}) & = \eta(\tau, \mathbf{x}) + \Gamma^{-1} \overbrace{\mathbb{E}_{\mathbf{X}=\mathbf{x}, T=\tau} [Y - \eta(\tau, \mathbf{x})]}^{= 0} + \bigl( \Gamma - \Gamma^{-1} \bigr) \mathbb{E}_{\mathbf{X}=\mathbf{x}, T=\tau} [(Y - \eta(\tau, \mathbf{x})) \kappa^\star(Y, \mathbf{x}, \tau)] \\
    & = \eta(\tau, \mathbf{x}) + \bigl( \Gamma - \Gamma^{-1} \bigr) \mathbb{E}_{\mathbf{X}=\mathbf{x}, T=\tau} [(Y - \eta(\tau, \mathbf{x})) \kappa^\star(Y, \mathbf{x}, \tau)]. 
\end{align*}
Moreover, the definition of $\kappa^{\star}$ ensures that
\begin{align*}
    \bigl( \Gamma^2 - 1 \bigr)^{-1} + \int \kappa^{\star}(y, \mathbf{x}, \tau) f(y | \mathbf{X}=\mathbf{x}, T=\tau) \, \mathrm{d}y & = \bigl( \Gamma^2 - 1 \bigr)^{-1} + \mathbb{E}_{\mathbf{X}=\mathbf{x}, T=\tau}[ \kappa^{\star}(Y, \mathbf{x}, \tau)] \\
    & = \bigl( \Gamma^2 - 1 \bigr)^{-1} + \frac{1}{\Gamma + 1} \\
    & = \frac{1}{\Gamma - \Gamma^{-1}}.
\end{align*}
This proves that 
\begin{align*}
    \theta(\tau, \mathbf{x}) & = \eta(\tau, \mathbf{x}) + \bigl( \Gamma - \Gamma^{-1} \bigr) \mathbb{E}_{\mathbf{X}=\mathbf{x}, T=\tau} [(Y - \eta(\tau, \mathbf{x})) \kappa^\star(Y, \mathbf{x}, \tau)] \\
    & = \eta(\tau, \mathbf{x}) + \frac{\mathbb{E}_{\mathbf{X}=\mathbf{x}, T=\tau} [(Y - \eta(\tau, \mathbf{x})) \kappa^\star(Y, \mathbf{x}, \tau)]}{\bigl( \Gamma^2 - 1 \bigr)^{-1} + \mathbb{E}_{\mathbf{X}=\mathbf{x}, T=\tau}[ \kappa^{\star}(Y, \mathbf{x}, \tau)]}.
\end{align*}
We can now give an alternative definition of our lower and upper bounds on the CAPO using the rescaled weight function $\kappa$,
\begin{equation*}
    \theta^-(\tau, \mathbf{x}) = \eta(\tau, \mathbf{x}) + \underset{\kappa \in \mathcal{K}^\star_\tau}{\inf} \frac{\mathbb{E}_{\mathbf{X}=\mathbf{x}, T=\tau} [(Y - \eta(\tau, \mathbf{x})) \kappa(Y, \mathbf{x}, \tau)]}{\bigl( \Gamma^2 - 1 \bigr)^{-1} + \mathbb{E}_{\mathbf{X}=\mathbf{x}, T=\tau}[ \kappa(Y, \mathbf{x}, \tau)]}
\end{equation*}
and
\begin{equation*}
    \theta^+(\tau, \mathbf{x}) = \eta(\tau, \mathbf{x}) + \underset{\kappa \in \mathcal{K}^\star_\tau}{\sup} \frac{\mathbb{E}_{\mathbf{X}=\mathbf{x}, T=\tau} [(Y - \eta(\tau, \mathbf{x})) \kappa(Y, \mathbf{x}, \tau)]}{\bigl( \Gamma^2 - 1 \bigr)^{-1} + \mathbb{E}_{\mathbf{X}=\mathbf{x}, T=\tau}[ \kappa(Y, \mathbf{x}, \tau)]}
\end{equation*}
where $\mathcal{K}^\star_\tau = \bigl\{ \kappa : \mathcal{Y} \times \mathcal{X} \times \mathcal{T} \to [0, 1] \, ; \, \forall \mathbf{x} \in \mathcal{X}, \, \mathbb{E}_{\mathbf{X}=\mathbf{x}, T=\tau}[\kappa(Y, \mathbf{x}, \tau)] = (\Gamma + 1)^{-1} \bigr\}$.

As $\mathcal{K}^\star_\tau \subset \mathcal{\bar K}$, the following inclusion holds: $[\theta^-(\tau, \mathbf{x}), \theta^+(\tau, \mathbf{x})] \subset \bigl[ \bar \theta^-(\tau, \mathbf{x}), \bar \theta^+(\tau, \mathbf{x}) \bigr]$. This concludes the proof.

\subsection{Proof of Theorem~\ref{theo:lim_tilde_theta_h_plus_moins}}  \label{proof:lim_tilde_theta_h_plus_moins}

The proof of Theorem~\ref{theo:lim_tilde_theta_h_plus_moins} is divided into two parts: we first study the variance, then the bias of the estimators. From that, we deduce the order of the optimal bandwidth $h^\star$ and Mean Squared Error (MSE) of the estimators of the bounds. In the following, we only focus on the upper bound $\tilde \theta^+_h(\tau)$. A similar reasoning can be made with the lower bound of the APO $\tilde \theta^-_h(\tau)$. In the proof, we use the alternative form of $\tilde \theta^+_h(\tau)$ given by Equation~\eqref{eqn:tilde_kernel_apo_lb_v2}.

\subsubsection{Variance of \texorpdfstring{$\tilde \theta^+_h(\tau)$}{}}

We first study the variance of our estimator of the upper bound for the APO.
\begin{align*}
    \Var \bigl( \tilde \theta^+_h(\tau) \bigr) & = \Var \left( \frac{1}{n} \sum_{i=1}^n \frac{K_h(T_i-\tau) (Y_i - \eta(T_i, \mathbf{X}_i))}{f(T_i|\mathbf{X}_i)} \Gamma^{\sign(Y_i - Q(\gamma ; Y|\mathbf{X}_i,T_i))} + \eta(\tau, \mathbf{X}_i) \right) \\
    & = \frac{1}{n} \Var \left( \frac{K_h(T-\tau) (Y - \eta(T, \mathbf{X}))}{f(T|\mathbf{X})} \Gamma^{\sign(Y - Q(\gamma ; Y|\mathbf{X},T))} + \eta(\tau, \mathbf{X}) \right) \quad \text{by i.i.d. assumption} \\
    & \leq \frac{1}{n} \E \left[ \left( \frac{K_h(T-\tau) (Y - \eta(T, \mathbf{X}))}{f(T|\mathbf{X})} \Gamma^{\sign(Y - Q(\gamma ; Y|\mathbf{X},T))} + \eta(\tau, \mathbf{X}) \right)^2 \right]
\end{align*}
Using the fact that $\forall (x, y) \in \mathbb{R}^2, \, (x+y)^2 \leq 2x^2 + 2y^2$, we get
\begin{equation*}
    \Var \bigl( \tilde \theta^+_h(\tau) \bigr) \leq \frac{2}{n} \underbrace{\E 
    \Biggl[ \E \left[ \frac{K_h(T-\tau)^2 (Y - \eta(T, \mathbf{X}))^2}{f(T|\mathbf{X})^2} \Gamma^{2 \sign(Y - Q(\gamma ; Y|\mathbf{X},T))} \middle| \mathbf{X} \right] \Biggr]}_\mathbf{A} + \frac{2}{n} \underbrace{\E \left[ \eta(\tau, \mathbf{X})^2 \right]}_\mathbf{B}
\end{equation*}

By Assumption~\ref{ass:bounded_outcomes}, we know that $\eta(T, \mathbf{X})$ is upper bounded by $M_Y$, so $\mathbf{B}$ is upper bounded by $M_Y^2$.

We now study Term $\mathbf{A}$:
\begin{align*}
    \mathbf{A} & = \E \Biggl[ \iint \frac{K_h(t-\tau)^2 (y - \eta(t, \mathbf{X}))^2}{f(T=t|\mathbf{X})^2} \Gamma^{2 \sign(y - Q(\gamma ; Y|\mathbf{X},t))} \underbrace{f(Y=y,T=t|\mathbf{X})}_{\mathclap{=f(Y=y|\mathbf{X},T=t)f(T=t|\mathbf{X})}} \, \mathrm{d}y \, \mathrm{d}t \Biggr] \\
    & = \Gamma^2 \E \Biggl[ \int \frac{K_h(t-\tau)^2}{f(T=t|\mathbf{X})} \Var[Y|\mathbf{X}, t] \, \mathrm{d}t \Biggr] \\
    & \quad + \underbrace{(\Gamma^{-2} - \Gamma^2)}_{\leq 0} \underbrace{\mathbb{E} \left[ \int \frac{K_h(t-\tau)^2}{f(T=t|\mathbf{X})} \int (y - \eta(t, \mathbf{X}))^2 f(Y=y|\mathbf{X},T=t) \mathds{1}(y < Q(\gamma;Y|\mathbf{X},T=t)) \, \mathrm{d}y \, \mathrm{d}t \right]}_{\geq 0} \\
    & \leq \Gamma^2 \E \left[ \int \frac{K_h(t-\tau)^2}{f(T=t|\mathbf{X})} \E[Y^2|\mathbf{X}, t] \, \mathrm{d}t \right] \quad \text{by $\Var[Y|\mathbf{X}, t] \leq \E[Y^2|\mathbf{X}, t]$} \\
    & \leq \frac{\Gamma^2 M_Y^2}{h m_f} \int K(s)^2 \, \mathrm{d}s \quad \text{by Assumptions~\ref{ass:positivity_cond_treatment} and \ref{ass:bounded_outcomes}} \\
    & \leq \frac{\Gamma^2 M_{K^2} M_Y^2}{m_f} \cdot \frac{1}{h} \quad \text{by Assumption~\ref{ass:kernel}.\ref{ass:kernel_add_1}}
\end{align*}

Finally,
\begin{align*}
    \Var \bigl( \tilde \theta^+_h(\tau) \bigr) & \leq \frac{C_{\mathrm{Var}_1}}{nh} + \frac{C_{\mathrm{Var}_2}}{n},
\end{align*}
where $C_{\mathrm{Var}_1} = \frac{2 \Gamma^2 M_{K^2} M_Y^2}{m_f}$ and $C_{\mathrm{Var}_2} = 2 M_Y^2$.

\subsubsection{Bias of \texorpdfstring{$\tilde \theta^+_h(\tau)$}{}}

Define the bias for the estimator of the upper bound of the APO:
\begin{equation*}
    b^+_h(\tau) \coloneq \mathbb{E} \left[ \tilde \theta^+_h(\tau) \right] - \theta^+(\tau)
\end{equation*}
The expectancy can be written
\begin{align*}
    \mathbb{E} \left[ \tilde \theta^+_h(\tau) \right] & = \mathbb{E} \left[ \frac{1}{n} \sum_{i=1}^n \frac{K_h(T_i-\tau) (Y_i - \eta(T_i, \mathbf{X}_i))}{f(T_i|\mathbf{X}_i)} \Gamma^{\sign(Y_i - Q(\gamma ; Y|\mathbf{X}_i,T_i))} + \eta(\tau, \mathbf{X}_i) \right] \\
    & = \mathbb{E} \Biggl[ \underbrace{\mathbb{E} \left[ \frac{K_h(T-\tau) (Y - \eta(T, \mathbf{X}))}{f(T|\mathbf{X})} \Gamma^{\sign(Y - Q(\gamma ; Y|\mathbf{X},T))} \middle| \mathbf{X} \right]}_{\mathbf{A}} \Biggr]+ \eta(\tau)\quad \text{by i.i.d. Assumption.}
\end{align*}

We focus on the first term:
\begin{align*}
    \mathbf{A} & = \mathbb{E} \left[ \frac{K_h(T-\tau) (Y - \eta(T, \mathbf{X}))}{f(T|\mathbf{X})} \Gamma^{\sign(Y - Q(\gamma ; Y|\mathbf{X},T))} \middle| \mathbf{X} \right] \\
    & =  \iint \frac{K_h(t-\tau) (y - \eta(t, \mathbf{X}))}{f(T=t|\mathbf{X})} \Gamma^{\sign(y - Q(\gamma ; Y|\mathbf{X},t))} \underbrace{f(Y=y,T=t|\mathbf{X})}_{\mathclap{= f(Y=y|\mathbf{X},T=t)f(T=t|\mathbf{X})}} \, \mathrm{d}y \, \mathrm{d}t \\
    & = \iint K(s) (y - \eta(\tau + sh, \mathbf{X})) \Gamma^{\sign(y - Q(\gamma ; Y|\mathbf{X}, \tau + sh))} f(Y=y|\mathbf{X},T=\tau + sh) \, \mathrm{d}y \, \mathrm{d}s  \\
    & = \Gamma  \iint K(s) (y - \eta(\tau + sh, \mathbf{X})) f(Y=y|\mathbf{X},T=\tau + sh) \, \mathrm{d}y \, \mathrm{d}s  \\
    & \quad + \left( \Gamma^{-1} - \Gamma \right)  \iint K(s) (y - \eta(\tau + sh, \mathbf{X})) \mathds{1}(y \leq Q(\gamma;Y|\mathbf{X},\tau + sh)) f(Y=y|\mathbf{X},T=\tau + sh) \, \mathrm{d}y \, \mathrm{d}s \\
    & = \Gamma  \int K(s) \int y f(Y=y|\mathbf{X},T=\tau + sh) \, \mathrm{d}y \, \mathrm{d}s  \\
    & \quad - \Gamma  \int K(s) \eta(\tau + sh, \mathbf{X}) \underbrace{\int f(Y=y|\mathbf{X},T=\tau + sh) \, \mathrm{d}y}_{= 1} \, \mathrm{d}s \\
    & \quad + \left( \Gamma^{-1} - \Gamma \right)  \iint K(s) y \mathds{1}(y \leq Q(\gamma;Y|\mathbf{X},\tau + sh)) f(Y=y|\mathbf{X},T=\tau + sh) \, \mathrm{d}y \, \mathrm{d}s  \\
    & \quad - \left( \Gamma^{-1} - \Gamma \right)  \int K(s) \eta(\tau + sh, \mathbf{X}) \underbrace{\int \mathds{1}(y \leq Q(\gamma;Y|\mathbf{X},\tau + sh)) f(Y=y|\mathbf{X},T=\tau + sh) \, \mathrm{d}y}_{= \gamma} \, \mathrm{d}s \\
    & = (\Gamma - 1) \int K(s) \int y f(Y=y|\mathbf{X},T=\tau + sh) \, \mathrm{d}y \, \mathrm{d}s \\
    & \quad + \left( \Gamma^{-1} - \Gamma \right) \int K(s) \int_{- \infty}^{Q(\gamma;Y|\mathbf{X},\tau + sh)} y f(Y=y|\mathbf{X},T=\tau + sh) \, \mathrm{d}y \, \mathrm{d}s
\end{align*}

The same lines of decomposition apply to $\theta^+(\tau)$, for which we get
\begin{align*}
    \theta^+(\tau) & = \eta(\tau) + (\Gamma-1) \eta(\tau) + \left( \Gamma^{-1} - \Gamma \right) \E \left[ \int_{- \infty}^{Q(\gamma;Y|\mathbf{X},\tau )} y f(Y=y|\mathbf{X},T=\tau) \, \mathrm{d}y \right].
\end{align*}
Thus, the bias is
\begin{align*}
   b^+_h(\tau) & = (\Gamma - 1) \E \Biggl[ \underbrace{ \int K(s) \int y f(Y=y|\mathbf{X},T=\tau + sh) \, \mathrm{d}y \, \mathrm{d}s - \eta(\tau)}_{\mathbf{B}} \Biggr] \\
   & \quad + \left( \Gamma^{-1} - \Gamma \right) \E \Biggl[ \underbrace{ \int K(s) \int_{- \infty}^{Q(\gamma;Y|\mathbf{X},\tau + sh)} y f(Y=y|\mathbf{X},T=\tau + sh) \, \mathrm{d}y \, \mathrm{d}s - \int_{- \infty}^{Q(\gamma;Y|\mathbf{X},\tau )} y f(Y=y|\mathbf{X}, T=\tau) \, \mathrm{d}y}_{\mathbf{C}} \Biggr]
\end{align*}
We first study Term $\mathbf{B}$. Under Assumption~\ref{ass:f_outcome_c2_bounded}, we use a Taylor expansion of order 2 around $T=\tau$:
\begin{equation*}
    \exists a \in [0, 1], \, f(Y=y|\mathbf{X}, T=\tau+sh) = f(Y=y|\mathbf{X}, T=\tau) + sh \pdv[order=1]{f}{T}(Y=y|\mathbf{X}, T=\tau) + \frac{s^2 h^2}{2} \pdv[order=2]{f}{T}(Y=y|\mathbf{X}, T=\tau+sha),
\end{equation*}
so that, using the symmetry of the kernel (Assumption~\ref{ass:kernel}),
\begin{align*}
    \mathbf{B} = \frac{h^2}{2}\int s^2 K(s)\int y \pdv[order=2]{f}{T}(Y=y|\mathbf{X}, T=\tau+sha) \, \mathrm{d}y \, \mathrm{d}s.
\end{align*}
We now use the fact that $\pdv[order=2]{f}{T}(Y=y|\mathbf{X}=\mathbf{x}, T=t)$ is $M_{\partial^2 f}$-bounded (Assumption~\ref{ass:f_outcome_c2_bounded}) to write
\begin{equation*}
    \left| \int y \pdv[order=2]{f}{T}(Y=y|\mathbf{X}, T=\tau+sha) \, \mathrm{d}y \right| \leq M_Y^2 M_{\partial^2 f}.
\end{equation*}

Finally, Assumption~\ref{ass:kernel}.\ref{ass:kernel_add_2} ensures that 
\begin{equation*}
    |\mathbf{B}| \leq \frac{h^2}{2} M_Y^2 M_{\partial^2 f} M_{u^2 K}.
\end{equation*}
Turning now to term $\mathbf{C}$, we first use the same Taylor expansion to write
\begin{align*}
    \mathbf{C} & = \int K(s) \int_{- \infty}^{Q(\gamma;Y|\mathbf{X},\tau + sh)} y f(Y=y|\mathbf{X},T=\tau + sh) \, \mathrm{d}y \, \mathrm{d}s -  \int_{- \infty}^{Q(\gamma;Y|\mathbf{X},\tau )} y f(Y=y|\mathbf{X},T=\tau) \, \mathrm{d}y \\
    & = \frac{h^2}{2} \int s^2 K(s) \int_{- \infty}^{Q(\gamma;Y|\mathbf{X},\tau)} y \pdv[order=2]{f}{T}(Y=y|\mathbf{X}, T=\tau+sha) \, \mathrm{d}y \, \mathrm{d}s \\
    & \quad + \int K(s) \int_{Q(\gamma;Y|\mathbf{X},\tau )}^{Q(\gamma;Y|\mathbf{X},\tau + sh)} y f(Y=y|\mathbf{X}, T=\tau+sh) \, \mathrm{d}y \, \mathrm{d}s.
\end{align*}
The absolute value of the first term is bounded by $h^2 M_Y^2 M_{\partial^2 f} M_{u^2 K} / 2$.
By means of a Taylor expansion of order 2 of the conditional quantile around $T=\tau$, the absolute value of the second term can be upper bounded as follows:
\begin{align*}
    & \int K(s) \int_{Q(\gamma;Y|\mathbf{X},\tau)}^{Q(\gamma;Y|\mathbf{X},\tau + sh)} |y| |f(Y=y|\mathbf{X},T=\tau+sh)| \, \mathrm{d}y \, \mathrm{d}s  \\
    & \leq M_Y \lVert f \rVert_{\infty} \int K(s) \bigl( Q(\gamma;Y|\mathbf{X},\tau + sh) - Q(\gamma;Y|\mathbf{X},\tau) \bigr) \, \mathrm{d}s \quad \text{by Assumptions~\ref{ass:bounded_outcomes} and \ref{ass:f_outcome_c2_bounded}} \\
    & \leq \frac{h^2 M_Y \lVert f \rVert_{\infty}}{2} \int s^2 K(s) \pdv[order=2]{Q}{T}(\gamma;Y|\mathbf{X},\tau + sha^{\prime}) \, \mathrm{d}s \quad \text{by Assumptions~\ref{ass:kernel} and \ref{ass:Q_c2_bounded}, with $a^{\prime} \in [0, 1]$} \\
    & \leq \frac{h^2}{2}  M_Y \lVert f \rVert_{\infty} M_{\partial^2 Q} M_{u^2 K} \quad \text{by Assumptions~\ref{ass:kernel}.\ref{ass:kernel_add_2} and \ref{ass:Q_c2_bounded}}.
\end{align*}
By bringing together our previous results, we get the following bound for the bias:
\begin{equation*}
    |b^+_h(\tau)| \leq \frac{h^2}{2} M_Y M_{u^2 K} \left[ (\Gamma - 1) M_Y M_{\partial^2 f} + \left( \Gamma - \Gamma^{-1} \right) (M_Y M_{\partial^2 f} + \lVert f \rVert_{\infty} M_{\partial^2 Q}) \right] = h^2 C_{\text{bias}},
\end{equation*}
where $C_{\text{bias}} \coloneq M_Y M_{u^2 K} [ (\Gamma - 1) M_Y M_{\partial^2 f} + \left( \Gamma - \Gamma^{-1} \right) (M_Y M_{\partial^2 f} + \lVert f \rVert_{\infty} M_{\partial^2 Q}) ] / 2$.

\subsubsection{Mean Squared Error of \texorpdfstring{$\tilde \theta^+_h(\tau)$}{} and Optimal Bandwidth \texorpdfstring{$h^\star$}{}}

The last step of the proof is to bound the Mean Squared Error,
\begin{equation*}
    \Var \bigl( \tilde \theta^+_h(\tau) \bigr) + |b^+_h(\tau)|^2 \leq \frac{C_{\mathrm{Var}_1}}{nh} + \frac{C_{\mathrm{Var}_2}}{n} + h^4 C^2_{\text{bias}},
\end{equation*}
which, then, gives us the order of magnitude of the optimal bandwidth $h^\star = \mathcal{O}(n^{-1/5})$, as $n$ tends to infinity. For this choice, the MSE is $\mathcal{O}(n^{-4/5})$, which is usual is nonparametric estimation (see, for instance, \cite{tsybakov2009nonparametric}) and is of the same order as the MSE obtained in \textcite{kallus2018policy}.

\textit{Note:} more general results can be obtained by considering Hölder- or Lipschitz-continuous functions regarding the second order partial derivatives of the conditional densities for the outcome and the conditional quantiles (see \cite{tsybakov2009nonparametric}). We could even develop a bandwidth selection method in the spirit of \textcite{goldenshluger2011bandwidth} for automatically adapting to the unknown regularity.

\subsection{Proof of Proposition~\ref{prop:partial_robustness}}  \label{sec:proof_partial_robustness}

In the following, we assume that the conditional quantiles are correctly specified and we focus on the upper bounds, but the reasoning is the same for the lower bounds. Moreover, for simplicity, we demonstrate the properties for the CAPO instead of the APO. Results for the APO can be obtained by means of the tower property. For the rest of the proof, notice that $\theta^+(\tau, \mathbf{x})$ from Equation~\eqref{eqn:capo_ub_v2} is also equal to
\begin{equation}  \label{eqn:capo_ub_v3}
    \theta^+(\tau, \mathbf{x}) = \mathbb{E}_{\mathbf{X}=\mathbf{x}, T=\tau} \Bigl[ Y \Gamma^{\sign(Y - q_\gamma^{\mathbf{x}, \tau})} \Bigr]
\end{equation}
by Equation~\eqref{eqn:constraint_w_star}, as $\Gamma^{\sign(Y - q_\gamma^{\mathbf{x}, \tau})} \in \mathcal{W}^\star_\tau$.

Recall that
\begin{align*}
    \theta_h^+(\tau, \mathbf{x}) = \eta(\tau, \mathbf{x}) + \mathbb{E}_{\mathbf{X}=\mathbf{x}} \biggl[ \frac{K_h(T - \tau)}{f(T|\mathbf{X}=\mathbf{x})} (Y - \eta(\tau, \mathbf{x})) w^+(Y, \mathbf{x}, T) \biggr]
\end{align*}
where $w^+(Y, \mathbf{x}, T) = \Gamma^{\sign(Y - q_\gamma^{\mathbf{x}, T})} \in \mathcal{W}^\star_\tau$.
\begin{itemize}
    \item If $f(T|\mathbf{X}=\mathbf{x})$ is correctly specified but not $\eta(\tau, \mathbf{x})$, which is misspecified by $\bar \eta(\tau, \mathbf{x})$, then:
    \begin{align*}
    \theta_h^+(\tau, \mathbf{x}) & = \bar \eta(\tau, \mathbf{x}) + \mathbb{E}_{\mathbf{X}=\mathbf{x}} \biggl[ \frac{K_h(T - \tau)}{f(T|\mathbf{X}=\mathbf{x})} (Y - \bar \eta(\tau, \mathbf{x})) w^+(Y, \mathbf{x}, T) \biggr] \\
    & = \bar \eta(\tau, \mathbf{x}) + \mathbb{E}_{\mathbf{X}=\mathbf{x}} \biggl[ \frac{K_h(T - \tau)}{f(T|\mathbf{X}=\mathbf{x})} Y w^+(Y, \mathbf{x}, T) \biggr] - \mathbb{E}_{\mathbf{X}=\mathbf{x}} \biggl[ \frac{K_h(T - \tau)}{f(T|\mathbf{X}=\mathbf{x})} \bar \eta(\tau, \mathbf{x}) \underbrace{\mathbb{E}_{\mathbf{X}=\mathbf{x}, T=\tau} [w^+(Y, \mathbf{x}, T)]}_{=1 \text{ by Eq.~\eqref{eqn:constraint_w_star}}} \biggr].
\end{align*}
As
\begin{align*}
    \mathbb{E}_{\mathbf{X}=\mathbf{x}} \biggl[ \frac{K_h(T - \tau)}{f(T|\mathbf{X}=\mathbf{x})} \bar \eta(\tau, \mathbf{x}) \biggr] & = \bar \eta(\tau, \mathbf{x}) \mathbb{E}_{\mathbf{X}=\mathbf{x}} \biggl[ \frac{K_h(T - \tau)}{f(T|\mathbf{X}=\mathbf{x})} \biggr] = \bar \eta(\tau, \mathbf{x}) \int K_h(t-\tau) \frac{f(T=t|\mathbf{X}=\mathbf{x})}{f(T=t|\mathbf{X}=\mathbf{x})} \, \mathrm{d}t = \bar \eta(\tau, \mathbf{x}),
\end{align*}
where the last equality comes from Assumption~\ref{ass:kernel}, we can write
\begin{align*}
    \theta_h^+(\tau, \mathbf{x}) = \mathbb{E}_{\mathbf{X}=\mathbf{x}} \biggl[ \frac{K_h(T - \tau)}{f(T|\mathbf{X}=\mathbf{x})} Y w^+(Y, \mathbf{x}, T) \biggr] \underset{h \to 0}{\to} \theta^+(\tau, \mathbf{x}).
\end{align*}
after using similar arguments as in the proof of the bias of $\tilde \theta^+_h(\tau)$ (proof of Theorem~\ref{theo:lim_tilde_theta_h_plus_moins} with Assumptions~\ref{ass:kernel} to \ref{ass:Q_c2_bounded}).
    \item If $\eta(\tau, \mathbf{x})$ is correctly specified but not $f(T|\mathbf{X}=\mathbf{x})$, which is misspecified by $\bar f(T|\mathbf{X}=\mathbf{x})$, then:
    \begin{align*}
        \theta_h^+(\tau, \mathbf{x}) & = \eta(\tau, \mathbf{x}) + \mathbb{E}_{\mathbf{X}=\mathbf{x}} \biggl[ \frac{K_h(T - \tau)}{\bar f(T|\mathbf{X}=\mathbf{x})} (Y - \eta(\tau, \mathbf{x})) w^+(Y, \mathbf{x}, T) \biggr] \\
        & = \eta(\tau, \mathbf{x}) + \mathbb{E}_{\mathbf{X}=\mathbf{x}} \biggl[ \frac{K_h(T - \tau)}{\bar f(T|\mathbf{X}=\mathbf{x})} \mathbb{E}_{\mathbf{X}=\mathbf{x}, T}[(Y - \eta(\tau, \mathbf{x})) w^+(Y, \mathbf{x}, T)] \biggr] \quad \text{by tower property} \\
        & = \eta(\tau, \mathbf{x}) + \int_\mathcal{T} \frac{K_h(t - \tau)}{\bar f(T=t|\mathbf{X}=\mathbf{x})} f(T=t|\mathbf{X}=\mathbf{x}) \underbrace{\int_\mathcal{Y} (y - \eta(\tau, \mathbf{x})) w^+(y, \mathbf{x}, t) f(Y=y|T=t, \mathbf{X}=\mathbf{x}) \, \mathrm{d}y}_{= \int_\mathcal{Y} y w^+(y, \mathbf{x}, t) f(Y=y|T=t, \mathbf{X}=\mathbf{x}) \, \mathrm{d}y - \eta(\tau, \mathbf{x}) \text{ by Eq.~\eqref{eqn:constraint_w_star}}} \, \mathrm{d}t \\
        & = \eta(\tau, \mathbf{x}) + \int_\mathcal{T} \frac{K_h(t - \tau)}{\bar f(T=t|\mathbf{X}=\mathbf{x})} f(T=t|\mathbf{X}=\mathbf{x}) \bigr[ \theta^+(t, \mathbf{x}) - \eta(\tau, \mathbf{x}) \bigl] \, \mathrm{d}t.
    \end{align*}
    Double robustness would be reached if, except for $T=\tau$, $\bar f(T=t|\mathbf{X}=\mathbf{x})$ was equal to $f(T=t|\mathbf{X}=\mathbf{x})$. Therefore, as it is not the case, we only have a partial robustness with respect to $\eta(\tau, \mathbf{x})$.
\end{itemize}

\subsection{Proof of Proposition~\ref{prop:double_robustness}}  \label{sec:proof_double_robustness}

In the following, we focus on the upper bound, but the reasoning is the same for the lower bound. Moreover, for simplicity, we demonstrate the properties for the CAPO instead of the APO. Results for the APO can be obtained by means of the tower property.

We want to show that
\begin{equation*}
    \theta^{+, \mathrm{DR}}_h(\tau, \mathbf{x}) =  \Theta^+(\tau, \mathbf{x}, q_\gamma) + \mathbb{E}_{\mathbf{X}=\mathbf{x}} \biggl[ \frac{K_h(T-\tau)}{f(T|\mathbf{X}=\mathbf{x})} \Bigl[ q_\gamma^{\mathbf{x}, T} + (Y - q_\gamma^{\mathbf{x}, T}) \Gamma^{\sign(Y - q_\gamma^{\mathbf{x}, T})} - \Theta^+(T, \mathbf{x}, q_\gamma) \Bigr] \biggr]
\end{equation*}
where
\begin{equation}  \label{eqn:Theta_plus_def}
    \Theta^+ : (t, \mathbf{x}, \bar q_\gamma) \mapsto \bar q_\gamma^{\mathbf{x}, t} 
    + \mathbb{E}_{\mathbf{X}=\mathbf{x}, T=t} \Bigl[ (Y - \bar q_\gamma^{\mathbf{x}, t}) \Gamma^{\sign(Y - \bar q_\gamma^{\mathbf{x}, t})} \Bigr]
\end{equation}
is a doubly robust estimator of $\theta^+(\tau, \mathbf{x})$. 

Recall that Equation~\eqref{eqn:capo_ub_v2} gives
\begin{align*}
    \theta^+(\tau, \mathbf{x}) = \eta(\tau, \mathbf{x}) + \mathbb{E}_{\mathbf{X}=\mathbf{x}, T=\tau} \Bigl[ (Y - \eta(\tau, \mathbf{x})) \underbrace{\Gamma^{\sign(Y-q_\gamma^{\mathbf{x}, \tau})}}_{= w^+(Y, \mathbf{x}, \tau)} \Bigr].
\end{align*}
By constraint~\eqref{eqn:constraint_w_star} on $w^+$, we also have
\begin{align*}
    \theta^+(\tau, \mathbf{x}) & = \mathbb{E}_{\mathbf{X}=\mathbf{x}, T=\tau} \Bigl[ Y \Gamma^{\sign(Y-q_\gamma^{\mathbf{x}, \tau})} \Bigr] \\
    & = q_\gamma^{\mathbf{x}, \tau} + \mathbb{E}_{\mathbf{X}=\mathbf{x}, T=\tau} \Bigl[ (Y - q_\gamma^{\mathbf{x}, \tau}) \Gamma^{\sign(Y-q_\gamma^{\mathbf{x}, \tau})} \Bigr].
\end{align*}
Therefore, we notice that, when the conditional quantile is correctly specified, $\Theta^+(\tau, \mathbf{x}, q_\gamma) = \theta^+(\tau, \mathbf{x})$.
Using the following Lemma~\ref{lem:minimum_cq}, we can show that, for all $\bar q_\gamma \in \mathcal{Y}$,
\begin{equation}  \label{eqn:minimizer_Theta_plus}
    \Theta^+(\tau, \mathbf{x}, \bar q_\gamma) \geq \Theta^+(\tau, \mathbf{x}, q_\gamma) = \theta^+(\tau, \mathbf{x}).
\end{equation}
\begin{lemma}[\cite{rockafellar2000optimization}]  \label{lem:minimum_cq}
    For all $\mathbf{x} \in \mathcal{X}$ and $t \in \mathcal{T}$, the application
    \begin{equation*}
        g : b \mapsto b + \mathbb{E}_{\mathbf{X}=\mathbf{x}, T=t} \Bigl[ (Y - b) \Gamma^{\sign(Y-b)} \Bigr]
    \end{equation*}
    is minimal in $b^\star = q_\gamma^{\mathbf{x}, t}$.
\end{lemma}
\begin{proof}
    To lighten the notations, we drop the conditioning on $\mathbf{X}=\mathbf{x}$ and $T=t$ of the expectancy. For all $b \in \mathcal{Y}$, rewrite $g$ as
    \begin{align*}
        g(b) & = b + \mathbb{E} \Bigl[ (Y - b) \Gamma^{\sign(Y-b)} \Bigr] \\
        & = b + \int_{-\infty}^b (y-b) \Gamma^{-1} f(y) \, \mathrm{d}y + \int_b^{+\infty} (y-b) \Gamma f(y) \, \mathrm{d}y \\
        & = b + \int_{-\infty}^b y \Gamma^{-1} f(y) \, \mathrm{d}y - b \int_{-\infty}^b \Gamma^{-1} f(y) \, \mathrm{d}y + \int_b^{+\infty} y \Gamma f(y) \, \mathrm{d}y - b \int_b^{+\infty} \Gamma f(y) \, \mathrm{d}y.
    \end{align*}
    Therefore, the minimizer $b^\star$ of $g$ is solution to
    \begin{align*}
        & g^\prime(b^\star) = 0 \\
        & \Leftrightarrow 1 + b^\star \Gamma^{-1} f(b^\star) - \int_{-\infty}^{b^\star} \Gamma^{-1} f(y) \, \mathrm{d}y - b^\star \Gamma^{-1} f(b^\star) - b^\star \Gamma f(b^\star) - \int_{b^\star}^{+\infty} \Gamma f(y) \, \mathrm{d}y + b^\star \Gamma f(b^\star) = 0 \\
        & \Leftrightarrow 1 - \int_{-\infty}^{b^\star} \Gamma^{-1} f(y) \, \mathrm{d}y - \Gamma \biggl( 1 - \int_{-\infty}^{b^\star} f(y) \, \mathrm{d}y \biggr) = 0 \\
        & \Leftrightarrow 1 - \Gamma + \bigl( \Gamma - \Gamma^{-1} \bigr) \int_{-\infty}^{b^\star} f(y) \, \mathrm{d}y = 0 \\
        & \Leftrightarrow \int_{-\infty}^{b^\star} f(y) \, \mathrm{d}y = F(b^\star) = \frac{\Gamma - 1}{\Gamma - \Gamma^{-1}} = \frac{\Gamma}{\Gamma + 1} = \gamma \\
        & \Leftrightarrow b^\star = F^{-1}(\gamma) = q_\gamma.
    \end{align*}
\end{proof}

In the case where the conditional quantiles are possibly misspecified (denoted then $\bar q_\gamma$), we can show that $\theta^{+, \mathrm{DR}}_h(\tau, \mathbf{x}) \geq \theta^+(\tau, \mathbf{x})$, whether $f(T|\mathbf{X})$ or $\Theta^+$ is misspecified, what \textcite{dorn2024doubly} call \textit{double validity}:
\begin{itemize}
    \item if $f(T|\mathbf{X})$ is correctly specified but not $\Theta^+$ (denoted $\bar \Theta^+$, to emphasize the fact that the regression model is misspecified):
    \begin{align*}
        \theta^{+, \mathrm{DR}}_h(\tau, \mathbf{x}) & = \bar \Theta^+(\tau, \mathbf{x}, \bar q_\gamma) + \mathbb{E}_{\mathbf{X}=\mathbf{x}} \biggl[ \frac{K_h(T-\tau)}{f(T|\mathbf{X}=\mathbf{x})} \Bigl[ \bar q_\gamma^{\mathbf{x}, T} + (Y - \bar q_\gamma^{\mathbf{x}, T}) \Gamma^{\sign(Y - \bar q_\gamma^{\mathbf{x}, T})} - \bar \Theta^+(T, \mathbf{x}, \bar q_\gamma) \Bigr] \biggr] \\
        & = \bar \Theta^+(\tau, \mathbf{x}, \bar q_\gamma) - \mathbb{E}_{\mathbf{X}=\mathbf{x}} \biggl[ \frac{K_h(T-\tau)}{f(T|\mathbf{X}=\mathbf{x})} \bar \Theta^+(T, \mathbf{x}, \bar q_\gamma) \biggr] \\
        & \quad + \underbrace{\mathbb{E}_{\mathbf{X}=\mathbf{x}} \biggl[ \frac{K_h(T-\tau)}{f(T|\mathbf{X}=\mathbf{x})} \Bigl[ \bar q_\gamma^{\mathbf{x}, T} + (Y - \bar q_\gamma^{\mathbf{x}, T}) \Gamma^{\sign(Y - \bar q_\gamma^{\mathbf{x}, T})} \Bigr] \biggr]}_{= \Theta_h^+(\tau, \mathbf{x}, \bar q_\gamma)} \\
        & = \bar \Theta^+(\tau, \mathbf{x}, \bar q_\gamma) - \int K(s) \bar \Theta^+(\tau+sh, \mathbf{x}, \bar q_\gamma) \, \mathrm{d}s + \Theta_h^+(\tau, \mathbf{x}, \bar q_\gamma)
    \end{align*}

\textit{Note:} $\bar \Theta^+(\tau, \mathbf{x}, \bar q_\gamma)$ means that the regression model of $\Theta^+$ \textit{and} $\bar q_\gamma$ are misspecified, whereas $\Theta^+(\tau, \mathbf{x}, \bar q_\gamma)$ means that the regression model of $\Theta^+$ is correctly specified, but $\bar q_\gamma$ is misspecified.

First, under Assumptions~\ref{ass:Theta_plus_c2_bounded} (that we suppose true even when $\Theta^+$ is misspecified) and \ref{ass:kernel}, we can show that
    \begin{align*}
        \bar \Theta^+(\tau, \mathbf{x}, \bar q_\gamma) - \int K(s) \bar \Theta^+(\tau+sh, \mathbf{x}, \bar q_\gamma) \, \mathrm{d}s \underset{h \to 0}{\to} 0.
    \end{align*}
    Indeed, with a Taylor expansion of $\bar \Theta^+$ at order 2 around $T = \tau$, $\exists a \in [0, 1]$,
    \begin{align*}
        \Bigl| \bar \Theta^+(\tau, \mathbf{x}, \bar q_\gamma) - \int K(s) \bar \Theta^+(\tau+sh, \mathbf{x}, \bar q_\gamma) \Bigr| \, \mathrm{d}s & \leq \frac{h^2}{2} \int s^2 K(s) \Bigl| \pdv[order=2]{\bar \Theta^+}{T}(\tau+sha, \mathbf{x}, \bar q_\gamma) \Bigr| \, \mathrm{d}s \\
        & \leq \frac{h^2}{2} M_{\partial^2 \bar \Theta^+} M_{u^2 K} \underset{h \to 0}{\to} 0.
    \end{align*}
    In the same way,
    \begin{align*}
        \Theta_h^+(\tau, \mathbf{x}, \bar q_\gamma) \underset{h \to 0}{\to} \Theta^+(\tau, \mathbf{x}, \bar q_\gamma)
    \end{align*}
    because
    \begin{align*}
        \Theta_h^+(\tau, \mathbf{x}, \bar q_\gamma) & = \mathbb{E}_{\mathbf{X}=\mathbf{x}} \biggl[ \frac{K_h(T-\tau)}{f(T|\mathbf{X}=\mathbf{x})} \mathbb{E}_{\mathbf{X}=\mathbf{x}, T} \Bigl[ \bar q_\gamma^{\mathbf{x}, T} + (Y - \bar q_\gamma^{\mathbf{x}, T}) \Gamma^{\sign(Y - \bar q_\gamma^{\mathbf{x}, T})} \Bigr] \biggr] \quad \text{by tower property} \\
        & = \mathbb{E}_{\mathbf{X}=\mathbf{x}} \biggl[ \frac{K_h(T-\tau)}{f(T|\mathbf{X}=\mathbf{x})} \Theta^+(T, \mathbf{x}, \bar q_\gamma) \biggr] \\
        & = \int \frac{K_h(t-\tau)}{f(T=t|\mathbf{X}=\mathbf{x})} \Theta^+(t, \mathbf{x}, \bar q_\gamma) f(T=t|\mathbf{X}=\mathbf{x}) \, \mathrm{d}t \\
        & = \int K(s) \Theta^+(\tau+sh, \mathbf{x}, \bar q_\gamma) \, \mathrm{d}s \\
        & = \Theta^+(\tau, \mathbf{x}, \bar q_\gamma) + \frac{h^2}{2} \int s^2 K(s) \pdv[order=2]{\Theta^+}{T}(\tau+sha, \mathbf{x}, \bar q_\gamma) \, \mathrm{d}s \quad \text{by Assumption~\ref{ass:Theta_plus_c2_bounded}, with $a \in [0, 1]$}
    \end{align*}
    and
    \begin{align*}
        |\Theta_h^+(\tau, \mathbf{x}, \bar q_\gamma) - \Theta^+(\tau, \mathbf{x}, \bar q_\gamma)| & \leq \frac{h^2}{2} \int s^2 K(s) \biggl| \pdv[order=2]{\Theta^+}{T}(\tau+sha, \mathbf{x}, \bar q_\gamma) \biggr| \, \mathrm{d}s \\
        & \leq \frac{h^2}{2} M_{\partial^2 \Theta^+} M_{s^2 K} \quad \text{by Assumption~\ref{ass:Theta_plus_c2_bounded}} \\
        & \underset{h \to 0}{\to} 0.
    \end{align*}
    Therefore,
    \begin{align*}
        \theta^{+, \mathrm{DR}}_h(\tau, \mathbf{x}) \underset{h \to 0}{\to} \Theta^+(\tau, \mathbf{x}, \bar q_\gamma) \geq \theta^+(\tau, \mathbf{x}) \quad \text{by Equation~\eqref{eqn:minimizer_Theta_plus}}.
    \end{align*}

    \item if $\Theta^+$ is correctly specified but not $f(T|\mathbf{X})$ (denoted $\bar f(T|\mathbf{X})$):
    
    By tower property,
    \begin{align*}
        \theta^{+, \mathrm{DR}}_h(\tau, \mathbf{x}) & = \Theta^+(\tau, \mathbf{x}, \bar q_\gamma) + \mathbb{E}_{\mathbf{X}=\mathbf{x}} \biggl[ \frac{K_h(T-\tau)}{\bar f(T|\mathbf{X}=\mathbf{x})} \Bigl[ \bar q_\gamma^{\mathbf{x}, T} + (Y - \bar q_\gamma^{\mathbf{x}, T}) \Gamma^{\sign(Y - \bar q_\gamma^{\mathbf{x}, T})} - \Theta^+(T, \mathbf{x}, \bar q_\gamma) \Bigr] \biggr] \\
        & = \Theta^+(\tau, \mathbf{x}, \bar q_\gamma) + \mathbb{E}_{\mathbf{X}=\mathbf{x}} \biggl[ \frac{K_h(T-\tau)}{\bar f(T|\mathbf{X}=\mathbf{x})} \Bigl( \underbrace{\mathbb{E}_{\mathbf{X}=\mathbf{x}, T} \Bigl[ \bar q_\gamma^{\mathbf{x}, T} + (Y - \bar q_\gamma^{\mathbf{x}, T}) \Gamma^{\sign(Y - \bar q_\gamma^{\mathbf{x}, T})} \Bigr]}_{= \Theta^+(T, \mathbf{x}, \bar q_\gamma)} - \Theta^+(T, \mathbf{x}, \bar q_\gamma) \Bigr) \biggr] \\
        & = \Theta^+(\tau, \mathbf{x}, \bar q_\gamma) \geq \theta^+(\tau, \mathbf{x}) \quad \text{by Equation~\eqref{eqn:minimizer_Theta_plus}.}
    \end{align*}
\end{itemize}

Finally, if the conditional quantiles are correctly specified, then the inequality~\eqref{eqn:minimizer_Theta_plus} becomes an equality and the estimator $\theta^{+, \mathrm{DR}}_h(\tau, \mathbf{x})$ of $\theta^+(\tau, \mathbf{x})$ is therefore doubly robust.

\section{EXPERIMENTS}

The experiments were conducted using Amazon EC2 m6i.xlarge instances (only CPUs). Amazon EC2 g5.xlarge and g6.xlarge instances (CPUs and GPUs) were also tested without significant execution time improvement, as the neural network models are not big enough. The code was developed under the R Statistical Software v4.3.2 \parencite{rstatisticalsoftware}. Notable libraries that were used include \texttt{ggplot2} v3.4.4 \parencite{ggplot2} for data visualization, \texttt{torch} v0.12.0 \parencite{torch} for neural networks, and \texttt{foreach} v1.5.2 \parencite{foreach} for ``for" loops. See section~\ref{sec:libraries_licenses} for an exhaustive list of the libraries and corresponding licenses.

\subsection{Kernel Bandwidth \texorpdfstring{$h$}{h} Estimation}  \label{sec:bandwidth_estim}

The kernel bandwidth $h$ is estimated following Algorithm~\ref{alg:bootstrap_apo}, with an Epanechnikov kernel. We use a number of bootstrap samples $B = 100$ and a grid of bandwidths $\mathcal{H}$ that consists in 40 equally spaced values between 0.1 and 2.5 because, according to Theorem~\ref{theo:lim_tilde_theta_h_plus_moins}, the order of magnitude of the optimal $h$ is approximately 0.26 in the simulated dataset, as $n=900$, and 0.22 in the real dataset, as $n=1918$. It is also possible to perform parametric bootstrap because, as discussed in \textcite{silverman1987bootstrap} and \textcite{faraway1990bootstrap}, non-parametric bootstrap can lead to poor choices of the bandwidth due to bias underestimation.

\begin{algorithm}[h]
    \caption{Nonparametric bootstrap for the APO $\theta(\tau)$} \label{alg:bootstrap_apo}
    \begin{algorithmic}
        \Require Dataset $\mathcal{D} = \{ (\mathbf{X}_i, T_i, Y_i) \}_{i=1}^n$, treatment value $\tau$, number of bootstrap samples $B$, grid of bandwidths $\mathcal{H}$, CI level $\alpha$.
        \State Compute $\hat \theta^+_h(\tau)$ and $\hat \theta^-_h(\tau) $ from $\mathcal{D}$ and Equation~\eqref{eqn:tilde_kernel_apo_ulb}.
        \For{$b \in \{1, \dots, B\}$}
        \For{$k \in \{1, \dots, n\}$}
        \State Sample $(x^b_k, t^b_k, y^b_k)$ uniformly from $\mathcal{D}$ with replacement.
        \EndFor
        \For{$h \in \mathcal H$}
        \State Compute $\hat \theta^{\pm,b}_h(\tau)$  from $x^b_1, \ldots, x^b_n$, $t^b_1, \ldots, t^b_{n}$ and $y^b_1, \ldots, y^b_{n}$ and Equation~\eqref{eqn:tilde_kernel_apo_ulb}.
        \EndFor
        \EndFor
        \State Compute $\hat h^{\pm}(\tau) = \underset{h \in \mathcal H}{\argmin} \frac{1}{B} \sum_{b=1}^B \bigl(\hat \theta^{\pm,b}_h(\tau) - \hat \theta^{\pm}_h(\tau) \bigr)^2$.
        \State Compute the PEI $\Bigl[ \hat \theta^{-}_{\hat h^-(\tau)}(\tau), \theta^{+}_{\hat h^+(\tau)}(\tau) \Bigr]$ and the CI $\Bigl[ \hat \theta^{-,\lceil B \alpha /2 \rceil}_{\hat h^-(\tau)}(\tau), \hat \theta^{+, \lceil B ( 1 - \alpha /2) \rceil}_{\hat h^+(\tau)}(\tau) \Bigr]$.
    \end{algorithmic}
\end{algorithm}

\subsection{Density Estimation via Neural Networks}  \label{sec:dens_estim_nn}

\begin{figure}[h]
    \centering
    \vspace{.3in}
    \includegraphics[width=0.8\linewidth]{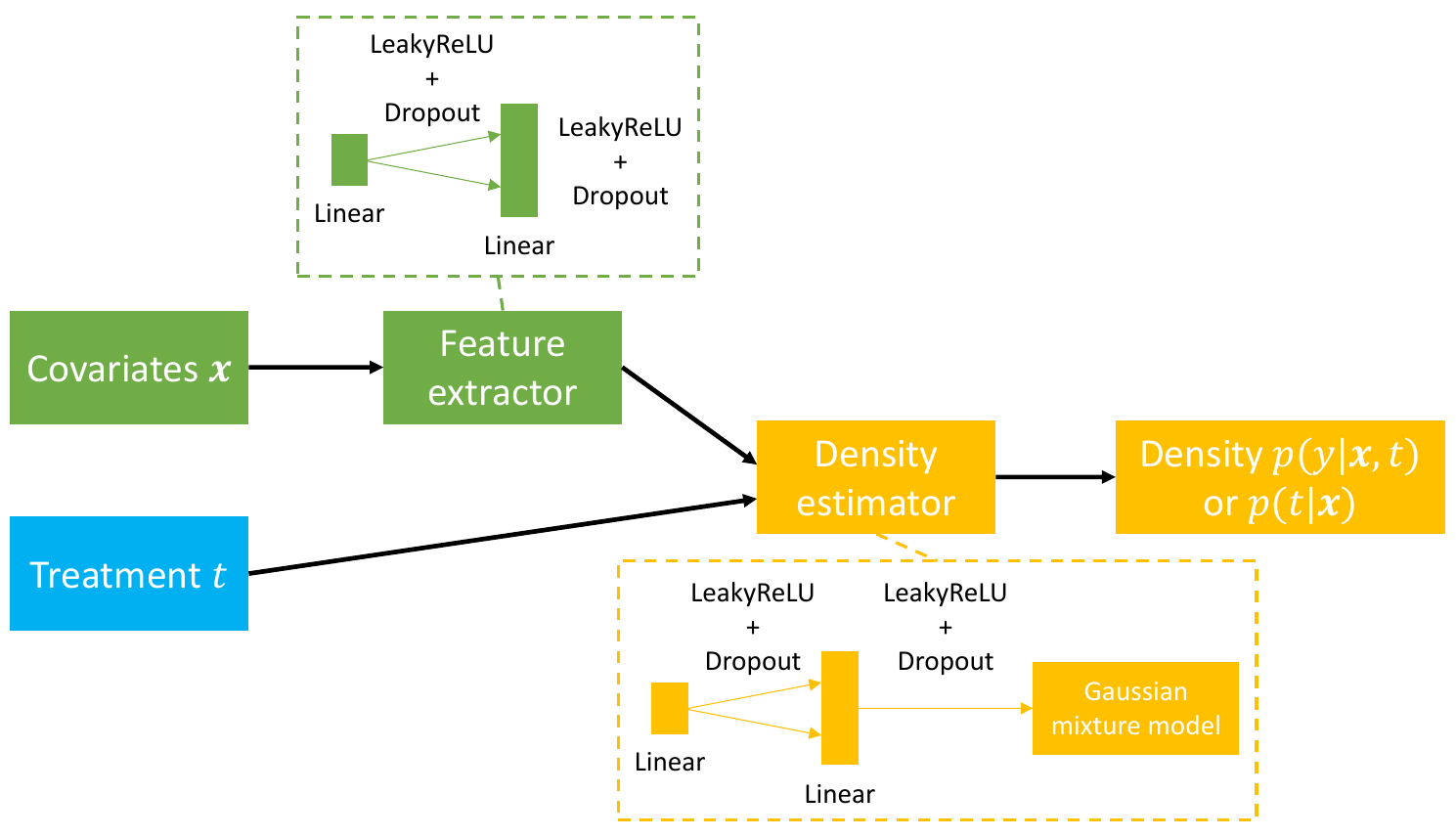}
    \vspace{.3in}
    \caption{Gaussian Mixture Model neural network architecture (inspired by \cite{jesson2022scalable}).}
    \label{fig:neural_network}
\end{figure}

The neural network architecture is detailed in Figure~\ref{fig:neural_network}. Notice that linear layers include a bias term. In the Gaussian Mixture Model module, we point out that we use 3 linear layers with $K$ hidden units (one for the weight, one for the mean, and one for the variance of each component) that are fed to the \texttt{distr\_mixture\_same\_family} function from the \texttt{torch} library via a categorical distribution for the \texttt{mixture\_distribution} parameter, and a normal distribution for the \texttt{component\_distribution} parameter. We fix some hyperparameters and fine-tune the number of hidden units and number of Gaussian components, as detailed in Table~\ref{tab:nn_hyperparam}, and we optimize the resulting network with Adam optimizer with a fine-tuned learning rate (chosen between $10^{-4}$ and $10^{-3}$, with a step of $10^{-4}$). Fine-tuning is made following Algorithm~\ref{alg:fine_tuning}. We recall that we perform 2-fold cross-fitting i.e., the data are randomly divided equally in two, with fine-tuning and model fitting on one half and predictions on the other half, and vice versa. In Algorithm~\ref{alg:fine_tuning}, we randomly choose $M=100$ triplets (learning rate, number of components $K$, number of hidden units), we set the number of random splits to $N=2$ and use a negative log-likelihood as loss function $l$. In practice, we also add a \textit{patience} (number of epochs training must continue after the loss stopped decreasing) of 5 epochs: we check if the mean of 5 consecutive losses is greater than the mean of the same number of consecutive losses 5 epochs before. When we re-estimate the conditional densities $f(Y|\mathbf{X}, T)$ and $f(T|\mathbf{X})$ on each bootstrap resample, we do not train the neural networks from scratch but start the training with weights estimated during the computation of the PEI on the original dataset (method known as \textit{transfer learning}). This allows gaining some computation time.

For the doubly robust estimators, we estimate $\theta^+(\tau, \mathbf{x})$ and $\theta^-(\tau, \mathbf{x})$ thanks to the same neural network architecture that was used to estimate $\eta(t, \mathbf{x})$ but, instead of regressing $Y$ on $\mathbf{X}$ and $T$, we regress $Y \Gamma^{\sign(Y - q_\gamma^{\mathbf{x}, \tau})}$ and $Y \Gamma^{-\sign(Y - q_{1-\gamma}^{\mathbf{x}, \tau})}$ on $\mathbf{X}$ and $T$.

\begin{table}[h]
    \caption{Neural network hyperparameters.} \label{tab:nn_hyperparam}
    \begin{center}
    \begin{tabular}{|l|ll|}
        \hline
         & \textbf{Hyperparameter} & \textbf{Value} \\
        \hline
        \multirow{4}{*}{\textbf{Feature extractor}} & Linear 1 (hidden units) & Fine-tuned (8, 16, 32 or 64) \\
         & Linear 2 (hidden units) & Fine-tuned (8, 16, 32 or 64) \\
         & Leaky ReLU (negative slope) & 0.04 \\
         & Dropout (probability) & 0.04 \\
        \hline
        \multirow{5}{*}{\textbf{Density estimator}} & Linear 1 (hidden units) & Fine-tuned (16, 32, 64 or 128) \\
         & Linear 2 (hidden units) & Fine-tuned (16, 32, 64 or 128) \\
         & Leaky ReLU (negative slope) & 0.04 \\
         & Dropout (probability) & 0.04 \\
         & Number of Gaussian components ($K$) & Fine-tuned (between 3 and 30) \\
        \hline
    \end{tabular}
    \end{center}
\end{table}

\begin{algorithm}[h]
    \caption{Fine-tuning algorithm for a Gaussian Mixture Model neural network}
    \label{alg:fine_tuning}
    \begin{algorithmic}
        \Require Dataset $\mathcal{D}$, list of hyperparameters $(L_1, \dots, L_M)$ (search space), number of random splits $N$, neural network model $\mathcal{N}$, $\ell$ loss function
        \For{$i \in [\![1,N]\!]$}
        \State Divide $\mathcal{D}$ randomly into $\mathcal{D}^{\mathrm{train}}$ (80\%), $\mathcal{D}^{\mathrm{valid}}$ (10\%), $\mathcal{D}^{\mathrm{test}}$ (10\%).
            \For{$c \in [\![1,M]\!]$}
            \State Train $\mathcal{N}$ with hyperparameters $L_c$ on $\mathcal{D}_{i}^{\mathrm{train}}$ and use $\mathcal{D}_{i}^{\mathrm{valid}}$ for early stopping.
            \State Compute  $l_{i, c }=\ell(\mathcal{D}_{i}^{\mathrm{test}})$.
            \EndFor
        \EndFor
        \State For all $c \in [\![1,M]\!]$, compute $l_c = \frac{1}{N} \sum_{i=1}^N l_{i, c}$.
        \State Choose $L_{\hat c}$ with $\hat c = \argmin_{c \in [\![1,M]\!]} l_c$.
    \end{algorithmic}
\end{algorithm}

\subsubsection{Modeling of the GPS \texorpdfstring{$\hat f(T=t|\mathbf{X}=\mathbf{x})$}{}}

In the same way as $\hat f(Y=y|\mathbf{X}=\mathbf{x}, T=t)$, the GPS can be written as a mixture of $K^\prime$ Gaussian components
\begin{equation*}
    \hat f(T=t|\mathbf{X}=\mathbf{x}) = \sum_{k=1}^{K^\prime} \tilde \pi_k(\mathbf{x}) \mathcal{N}(t|\tilde \mu_k(\mathbf{x}), \tilde \sigma_k^2(\mathbf{x})),
\end{equation*}
where $\tilde \pi_k(\mathbf{x})$, $\tilde \mu_k(\mathbf{x})$ and $\tilde \sigma_k^2(\mathbf{x})$ are, respectively, the weight, mean and variance of the $k$\textsuperscript{th} component.

\subsection{Conditional Quantile Estimation}  \label{sec:cond_quant_est}

To estimate the quantile function $Q(\upsilon; Y|\mathbf{X}=\mathbf{X}_i, T=T_i)$, we compute the conditional density $f(Y=y|\mathbf{X}=\mathbf{X}_i, T=T_i)$ and then search the root of the function $y \mapsto F(Y=y|\mathbf{X}=\mathbf{X}_i, T=T_i) - \upsilon$ thanks to the \texttt{uniroot} function from the \texttt{stats} library \parencite{rstatisticalsoftware}. The cumulative distribution $F$ is recovered thanks to the \texttt{cdf} attribute of the estimated Gaussian mixture model. As the outcome $Y$ is centered and scaled, we search the root in the range $[-10, 10]$.

\subsection{Implementation Choices for the Method from \texorpdfstring{\textcite{jesson2022scalable}}{}}  \label{sec:jesson_implementation}

We implement the algorithm from \textcite{jesson2022scalable} in the R language, as it is only available for Python (see \href{https://github.com/oatml/overcast}{https://github.com/oatml/overcast}). In particular, to estimate $f(Y=y|\mathbf{X}=\mathbf{x}, T=t)$ and $\eta(t, \mathbf{x})$, we use the same model as for our method and the architecture from Figure~\ref{fig:neural_network}. Moreover, as their method involves a Monte-Carlo integration (function $I(\cdot)$ from their paper), we sample 500 outcomes $Y_i$ from the estimated conditional density $f(Y=y|\mathbf{X}=\mathbf{x}, T=t)$. To get the estimated lower and upper bounds for the APO (denoted $\underline{\hat \mu}(t;\Lambda, \theta)$ and $\hat{\bar{\mu}}(t;\Lambda, \theta)$, respectively, in \cite{jesson2022scalable}), we average the estimated bounds for the CAPO on all observed covariates, not only a subset as suggested in their article.

\subsection{Details about the Simulated Dataset and Additional Results}  \label{sec:details_sim_data}

\subsubsection{Simulation Setup}

The joint distribution of $(\mathbf{X}, \mathbf{U})$ is a normal distribution $\mathcal{N}(\mathbf{0}, \mathbf{\Sigma})$, where
\begin{equation*}
    \mathbf{\Sigma} =
    \begin{pmatrix}
        \mathbf{\Sigma_X} & \mathbf{\Sigma_{XU}} \\
        \mathbf{\Sigma_{XU}}^\top & \mathbf{\Sigma_U}
    \end{pmatrix}.
\end{equation*}
$\mathbf{\Sigma_X}$ (resp., $\mathbf{\Sigma_U}$) is a tridiagonal matrix of size $p_\mathbf{X} \times p_\mathbf{X}$ (resp., $p_\mathbf{U} \times p_\mathbf{U}$), where the elements on the main diagonal are all equal to 1 and the elements on the subdiagonal and lower diagonal are all equal to $\rho_\mathbf{X} > 0$ (resp., $\rho_\mathbf{U} > 0$). $\mathbf{\Sigma_{XU}}$ is a $p_\mathbf{X} \times p_\mathbf{U}$ matrix with all coefficients equal to $\rho_\mathbf{XU} \geq 0$, where $\rho_\mathbf{XU} = \lambda \rho_\mathbf{XU}^\mathrm{max}$, with $0 \leq \lambda < 1$ and $\rho_\mathbf{XU}^\mathrm{max} = (1 - \rho_\mathbf{X})/p_\mathbf{U}$, to ensure that $\mathbf{\Sigma}$ is a diagonal dominant matrix and is, thus, invertible.

The properties of the multivariate normal distribution allow to say that
\begin{equation*}
    \mathbf{U}|\mathbf{X}=\mathbf{x} \sim \mathcal N\left( \mathbf{\Sigma_{XU}}^\top \mathbf{\Sigma_{X}}^{-1} \mathbf{x}, \,\mathbf{\Sigma_{U}} - \mathbf{\Sigma_{XU}}^\top \mathbf{\Sigma_{X}}^{-1}  \mathbf{\Sigma_{XU}} \right).
\end{equation*}

We define $T$ conditionally on  $\mathbf{X}=\mathbf{x}, \mathbf{U}=\mathbf{u}$ as $T = \langle \beta_\mathbf{X}, \mathbf{x} \rangle + \langle \beta_\mathbf{U}, \mathbf{u} \rangle + \varepsilon_T$ where $\varepsilon_T \sim \mathcal N(0, \sigma_{\varepsilon_T}^2)$, with $\sigma_{\varepsilon_T} > 0$, $\beta_\mathbf{X} \in \mathbb R^{p_\mathbf{X}}$ and $\beta_\mathbf{U} \in \mathbb R^{p_\mathbf{U}}$.
Moreover, the distribution of $T$ conditionally on $\mathbf{X}=\mathbf{x}$ is given by $\mathcal{N}(\mu_T, \sigma_T^2)$, where
\begin{align*}
    \mu_T & = \langle \beta_\mathbf{X}, \mathbf{x} \rangle + \left\langle \beta_\mathbf{U}, \, \mathbf{\Sigma_{XU}}^\top \mathbf{\Sigma_{X}}^{-1} \mathbf{x} \right\rangle \quad \text{and} \\
    \sigma_T^2 & = \sigma_{\varepsilon_T}^2 + \beta_\mathbf{U}^\top \bigl(\mathbf{\Sigma_{U}} - \mathbf{\Sigma_{XU}}^\top \mathbf{\Sigma_{X}}^{-1}  \mathbf{\Sigma_{XU}}\bigr) \beta_\mathbf{U}.
\end{align*}

Finally, for all $t \in \mathcal{T}$, we set the potential outcome to
\begin{equation*}
    Y(t) = t + \zeta  \langle \mathbf{X} , \gamma_\mathbf{X} \rangle \cdot e^{-t \langle \mathbf{X}, \gamma_\mathbf{X} \rangle} -  \langle \mathbf{U}, \gamma_\mathbf{U} \rangle \langle \mathbf{X}, \gamma_\mathbf{X} \rangle + \varepsilon_Y
\end{equation*}
where $\varepsilon_Y \sim \mathcal{N}(0, \sigma_{\varepsilon_Y}^2)$, $\sigma_{\varepsilon_Y} > 0$, $\zeta \in \mathbb{R}$, $\gamma_\mathbf{X} \in \mathbb{R}^{p_\mathbf{X}}$ and $\gamma_\mathbf{U} \in \mathbb{R}^{p_\mathbf{U}}$.

For all $\mathbf{x} \in \mathcal{X}$, the true CAPO is then
\begin{align*}
    \theta(\tau, \mathbf{x}) = \tau + \zeta \langle \mathbf{x}, \gamma_\mathbf{X} \rangle e^{-\tau \langle \mathbf{x}, \gamma_\mathbf{X} \rangle} - \left\langle \mathbf{\Sigma_{XU}}^\top  \mathbf{\Sigma_{X}}^{-1} \mathbf{x}, \gamma_\mathbf{U} \right\rangle \left\langle \mathbf{x} , \gamma_\mathbf{X} \right\rangle,
\end{align*}
and the true APO is given by
\begin{align*}
    \theta(\tau) = \tau \Bigl(1 - \zeta \cdot \gamma_\mathbf{X}^\top \mathbf{\Sigma_X} \gamma_\mathbf{X} \cdot e^{\frac{\tau^2}{2} \gamma_\mathbf{X}^\top \mathbf{\Sigma_X} \gamma_\mathbf{X}} \Bigr) - \gamma_\mathbf{U}^\top \mathbf{\Sigma_{XU}}^\top \gamma_\mathbf{X}.
\end{align*}

During the simulation process, in order to avoid isolated data points and violations of the positivity assumption, the observations that correspond to the 10\% biggest hat values of the $(\mathbf{X}, T, Y)$ design matrix are removed.

\begin{figure}[h]
    \centering
    \vspace{.3in}
    \includegraphics[width=0.6\linewidth]{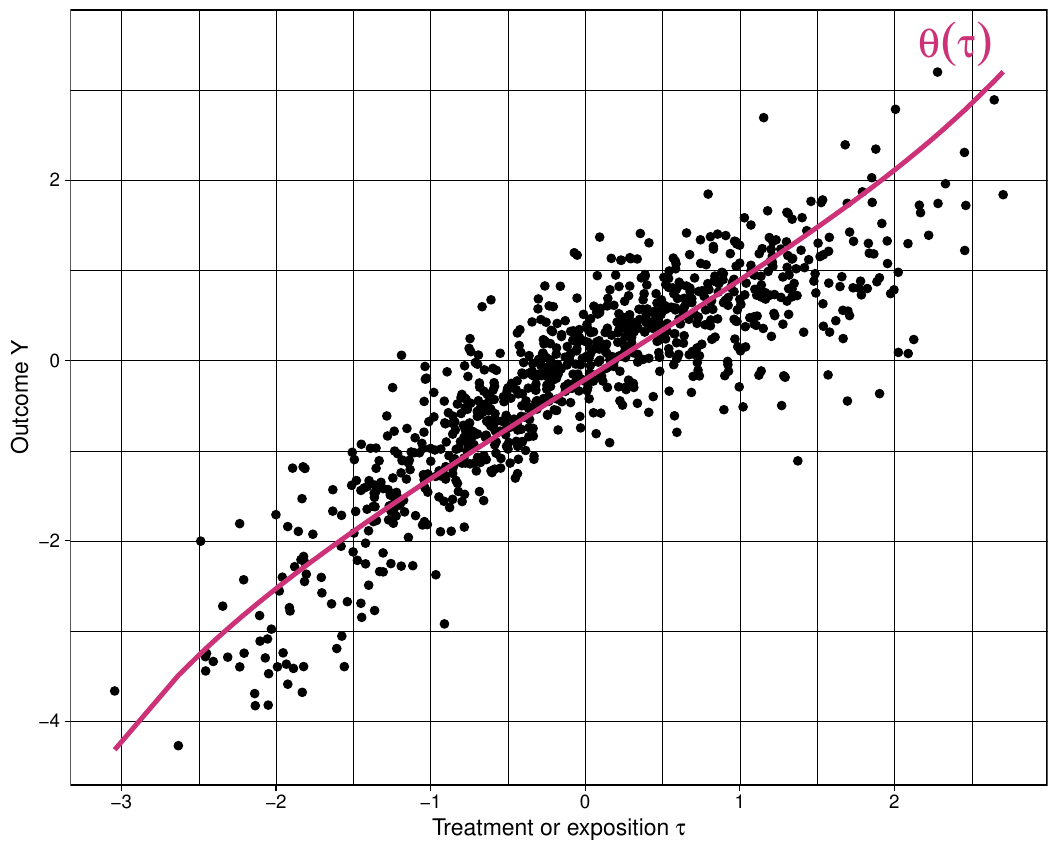}
    \vspace{.3in}
    \caption{Example of a simulated sample (treatment and outcome) with parameters from Table~\ref{tab:simu_param_values} and initial $n=1000$. The pink curve corresponds to the true unknown APO function $\theta(\tau)$.}
    \label{fig:simul_data_example}
\end{figure}

In our implementation, for reproducibility purpose, we set the random seeds to 1 (base R \texttt{set.seed} function and \texttt{torch\_manual\_seed} function from the \texttt{torch} library). Figure~\ref{fig:simul_data_example} is an example of a simulated sample with parameters from Table~\ref{tab:simu_param_values} and initial $n=1000$.

\subsubsection{Additional Sensitivity Analysis Results}

\begin{figure}[h]
    \centering
    \vspace{.3in}
    \includegraphics[width=0.7\linewidth]{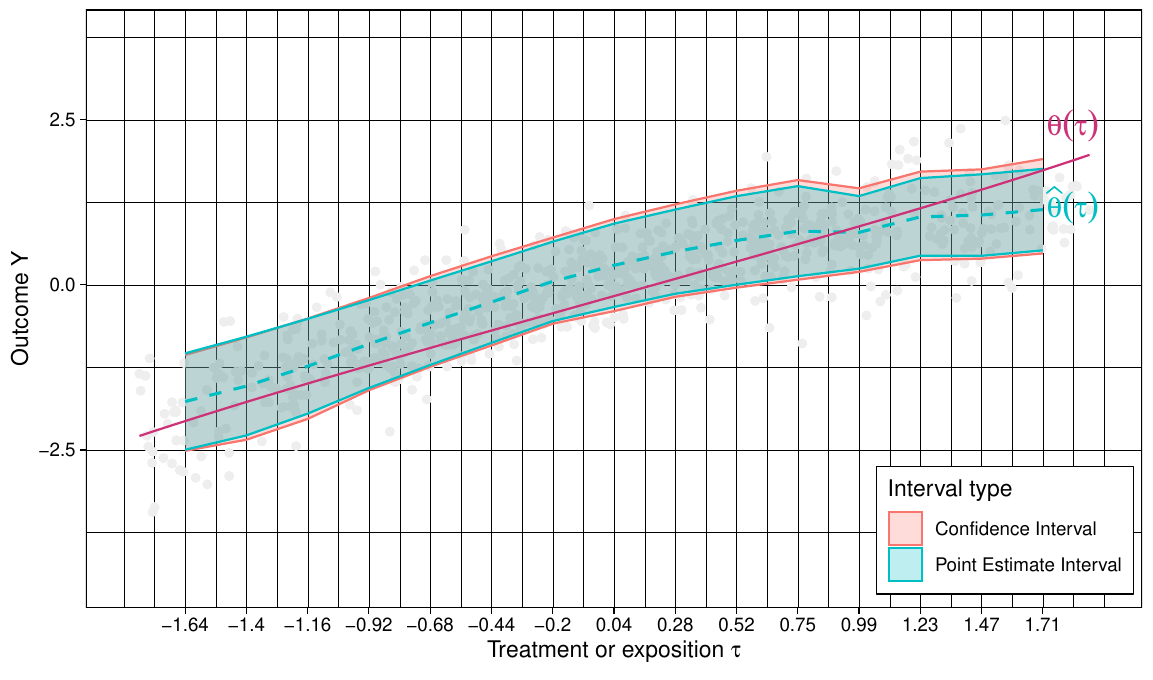}
    \vspace{.3in}
    \caption{Sensitivity analysis on the simulated dataset from Figure~\ref{fig:simul_data_example} for 15 values of $\tau$. The pink line corresponds to the true unknown APO function $\theta(\tau)$. The blue dotted curve corresponds to the estimated APO function $\hat \theta(\tau)$ under ignorability. The gray points correspond to the whole generated dataset. The red curves correspond to the estimated 95\%-level confidence intervals and the blue curves, to the Point Estimate Intervals obtained with our partially robust estimator.}
    \label{fig:simul_data_sens_analys_example}
\end{figure}

Figure~\ref{fig:simul_data_sens_analys_example} is a sensitivity analysis performed on the data from Figure~\ref{fig:simul_data_example}. Figures~\ref{fig:simu_exec_times_comparison}, \ref{fig:simu_ci_comparison_with_dr} and and \ref{fig:perc_boot_validity} were generated using the parameter values from Table~\ref{tab:simu_param_values}.

\begin{table}[h]
    \caption{Parameter values used to generate Figures~\ref{fig:simu_exec_times_comparison}, \ref{fig:simu_ci_comparison_with_dr} and \ref{fig:perc_boot_validity}.} \label{tab:simu_param_values}
    \begin{center}
    \begin{tabular}{|ll|ll|ll|}
        \hline
        \textbf{Parameter} & \textbf{Value} & \textbf{Parameter} & \textbf{Value} & \textbf{Parameter} & \textbf{Value} \\
        \hline
        $p_\mathbf{X}$ & 5 & $\lambda$ & 0.5 & $\gamma_\mathbf{U}$ & (0.4, 0.7, 0.7) \\
        $p_\mathbf{U}$ & 3 & $\beta_\mathbf{X}$ & (0.3, 0.3, 0.3, 0.3, 0.3) & $\zeta$ & -0.3 \\
        $\rho_\mathbf{X}$ & 0.3 & $\beta_\mathbf{U}$ & (0.2, 0.2, 0.2) & $\sigma_{\varepsilon_T}$ & 0.5 \\
        $\rho_\mathbf{U}$ & 0.3 & $\gamma_\mathbf{X}$ & (0.2, 0.2, 0.2, 0.2, 0.2) & $\sigma_{\varepsilon_Y}$ & 0.7 \\
        \hline
    \end{tabular}
    \end{center}
\end{table}

\subsubsection{Computation Time Issue}

Bootstrap procedures are known to become computationally intensive when the sample size becomes large. In order to reduce execution time, it is possible to perform parallel computing. However, for practical reasons, it was not possible to use this technique because the tensor objects from the \texttt{torch} library in \texttt{R} do not allow parallelization. Figure~\ref{fig:simu_exec_times_comparison} was therefore obtained with no parallel computing. Nevertheless, some parts in the code that did not involve \texttt{torch} tensors could be parallelized (essentially for the method from \textcite{jesson2022scalable}). Thus, when possible, except for Figure~\ref{fig:simu_exec_times_comparison}, we used parallel computation on Amazon EC2 c6i.16xlarge instances.

\subsubsection{Validity of the percentile bootstrap} \label{sec:validity_perc_boot}

\begin{figure}[h]
    \centering
    \vspace{.3in}
    \includegraphics[width=0.65\linewidth]{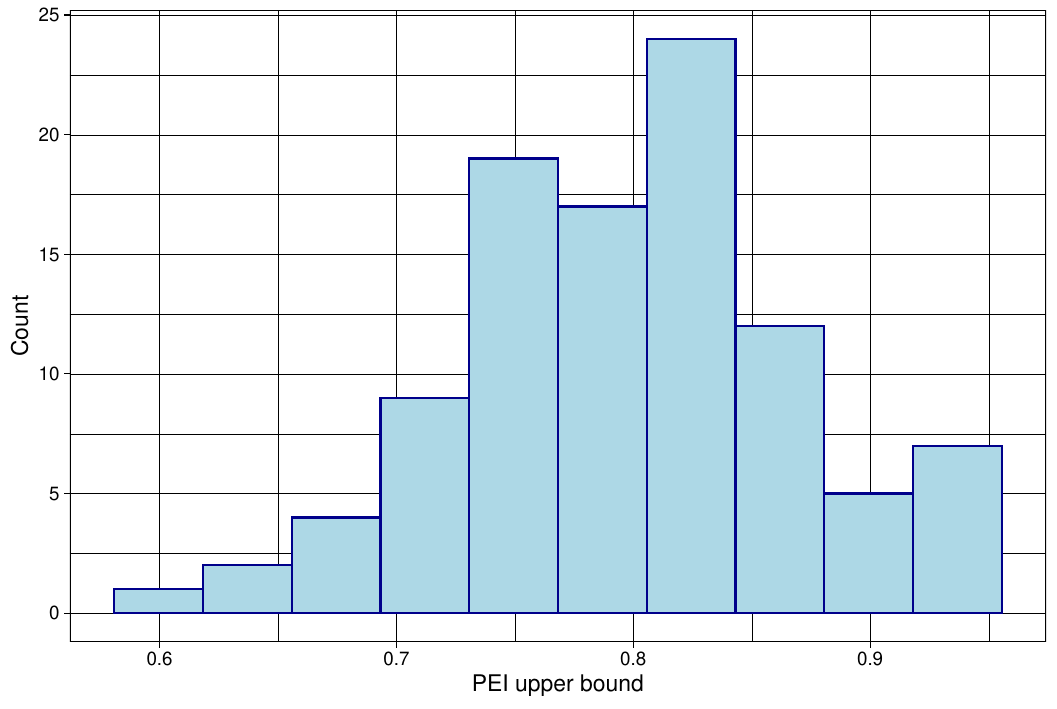}
    \vspace{.3in}
    \caption{Histogram of PEI upper bounds for treatment $\tau=0.038$. PEIs were computed with the partially robust estimator on 100 Monte-Carlo samples generated with the setup from Table~2.}
    \label{fig:perc_boot_validity}
\end{figure}

In the binary treatment case, \textcite{zhao2019sensitivity} justify theoretically their use of the percentile bootstrap method in Section~4.3. They also mention the fact that the estimators must be smooth enough for the method to be valid and they consider IPW estimators as such. In our case, the kernelization helps to smoothen our estimators. The distribution of the bounds must also be smooth enough for the percentile bootstrap to be valid. Therefore, we provide Figure~\ref{fig:perc_boot_validity} to support this claim, where we display a histogram of PEI upper bounds for treatment value $\tau = 0.038$ computed on 100 Monte-Carlo samples generated with the setup from Table~\ref{tab:simu_param_values}. The distribution appears to be smooth enough for the method to perform correctly. Moreover, the percentile bootstrap is particularly interesting when the distribution is skewed \parencite{helwig2017bootstrap}, which is the case here (left-skewed distribution).

\subsubsection{Effect of the Parameters from the Data Generation Process on \texorpdfstring{$\Gamma$}{}}

\begin{figure}[h]
    \centering
    \vspace{.3in}
    \includegraphics[width=0.6\linewidth]{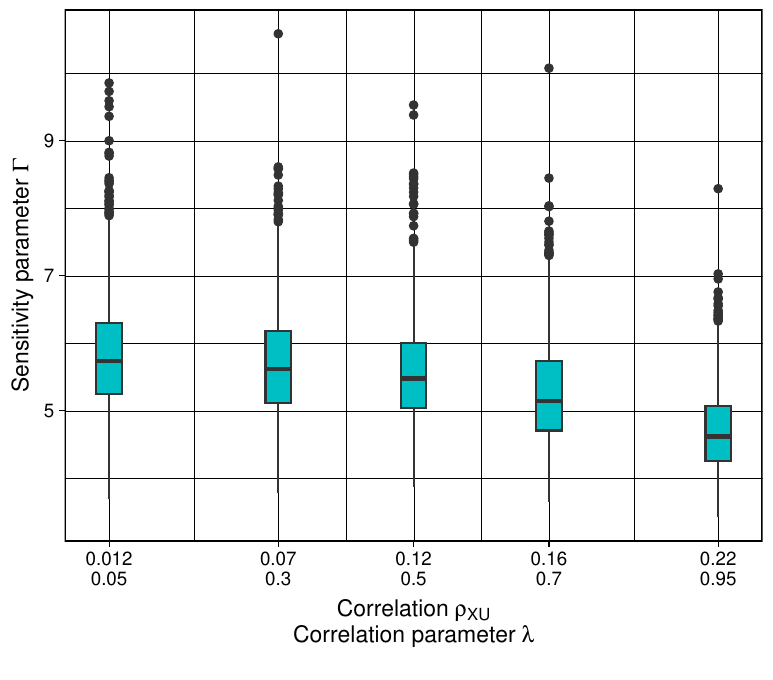}
    \vspace{.3in}
    \caption{Boxplots of estimated sensitivity parameter $\Gamma$ on 1000 Monte-Carlo samples for 5 values of correlation $\rho_\mathbf{XU}$ (setup from Table~\ref{tab:simu_param_values} except for $\rho_\mathbf{XU}$).}
    \label{fig:gamma_rho_XU}
\end{figure}

\begin{figure}[h]
    \centering
    \vspace{.3in}
    \includegraphics[width=0.6\linewidth]{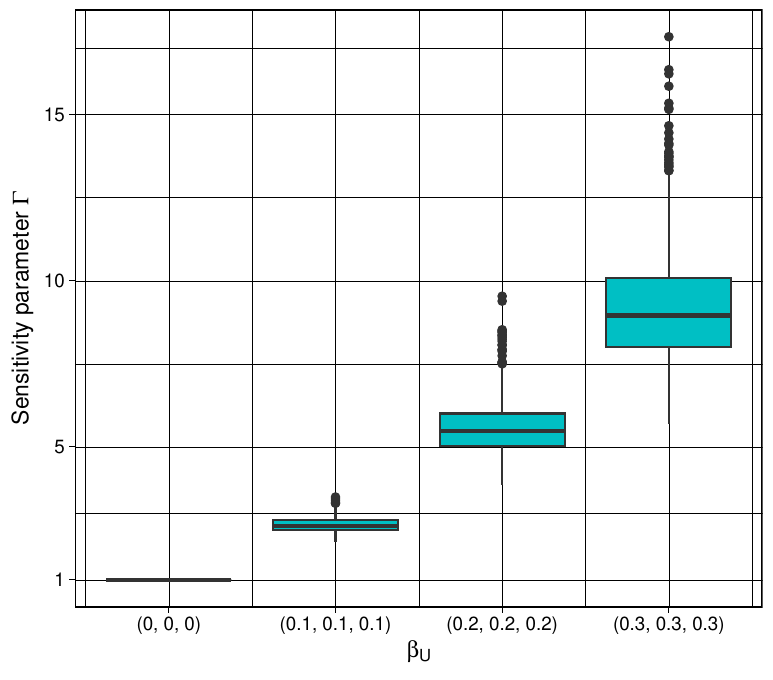}
    \vspace{.3in}
    \caption{Boxplots of estimated sensitivity parameter $\Gamma$ on 1000 Monte-Carlo samples for 4 values of $\beta_\mathbf{U}$ (setup from Table~\ref{tab:simu_param_values} except for $\beta_\mathbf{U}$).}
    \label{fig:gamma_beta_U}
\end{figure}

We perform an exploratory analysis of the influence of the parameters from the generation process of the dataset on the sensitivity parameter $\Gamma$. Figure~\ref{fig:gamma_rho_XU} displays the influence of the correlation between $\mathbf{X}$ and $\mathbf{U}$, i.e.\ $\rho_\mathbf{XU}$, on the chosen $\Gamma$. As expected this sensitivity parameter decreases as the correlation increases, because already observed covariates would explain unobserved confounders. Then, in Figure~\ref{fig:gamma_beta_U}, we vary the values of $\beta_\mathbf{U}$, which links the unobserved confounders to the treatment $T$. When $\mathbf{U}$ has no effect on the treatment, i.e.\ $\beta_\mathbf{U}$ is null, we expect $\Gamma$ to be equal to 1, which is equivalent to $\mathbf{X}$-ignorability. This is indeed what we observe in Figure~\ref{fig:gamma_beta_U}, with $\Gamma$ values increasing as $\beta_\mathbf{U}$ becomes larger.

\subsection{Details about the Real Dataset and Additional Results}  \label{sec:details_real_data}

The data are shared between three files: \texttt{County\_annual\_PM25\_CMR.csv}, \texttt{County\_RAW\_variables.csv} and \texttt{County\_SES\_index\_quintile.csv}.

The exposition $T$ is retrieved via the \texttt{PM2.5} variable, and the observed outcome $Y$, via the \texttt{CMR} variable.

We keep 10 continuous variables that correspond to the observed confounders $\mathbf{X}$: \texttt{healthfac\_2005\_1999}, \texttt{population\_2000}, \texttt{SES\_index\_2010}, \texttt{civil\_unemploy\_2010}, \texttt{median\_HH\_inc\_2010}, \texttt{femaleHH\_ns\_pct\_2010}, \texttt{vacant\_HHunit\_2010}, \texttt{owner\_occ\_pct\_2010}, \texttt{eduattain\_HS\_2010} and \texttt{pctfam\_pover\_2010}.

Only data from year 2010 are kept thanks to the \texttt{Year} variable. Then, \texttt{population\_2000} and \texttt{median\_HH\_inc\_2010} are log-normalized. Finally, \texttt{CMR}, \texttt{PM2.5} and all covariates are centered and scaled. As in the simulated data, we remove 10\% of the isolated data points that correspond to the 10\% biggest hat values of the design matrix of $(\mathbf{X}, T, Y)$.

\begin{figure}[h]
    \centering
    \vspace{.3in}
    \includegraphics[width=0.65\linewidth]{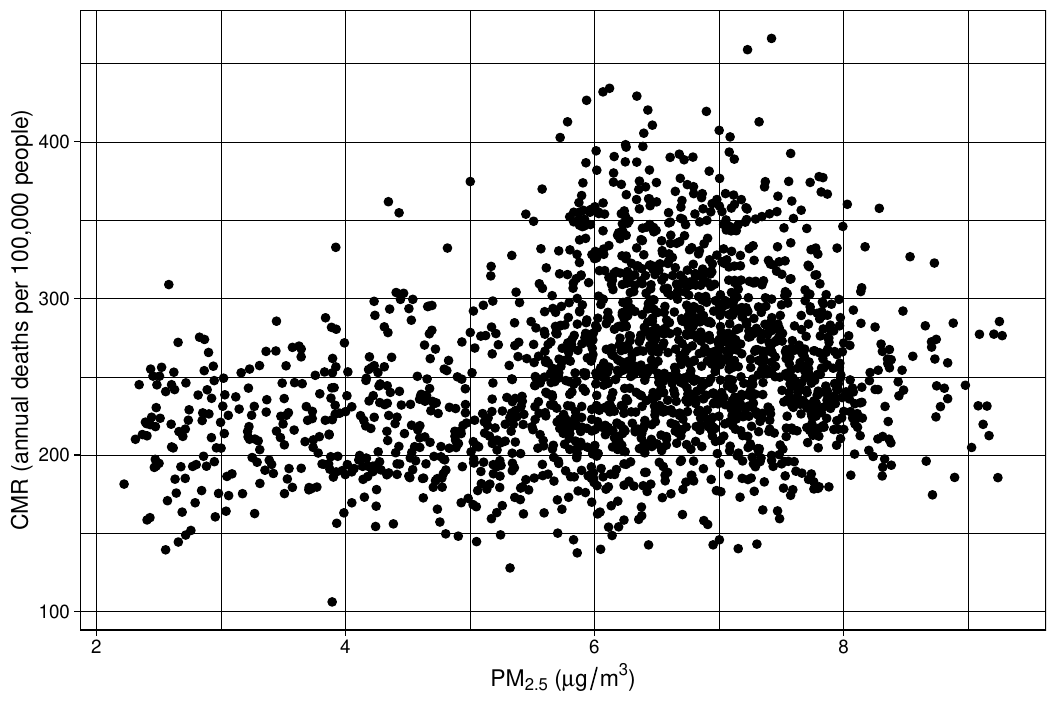}
    \vspace{.3in}
    \caption{CMR as a function of PM2.5.}
    \label{fig:real_data}
\end{figure}

Figure~\ref{fig:real_data} shows the distribution of the outcome (CMR) as a function of the exposition (PM2.5) without the outliers, and before centering and scaling.

\begin{figure*}[h]
    \centering
    \vspace{.3in}
    \includegraphics[width=\linewidth]{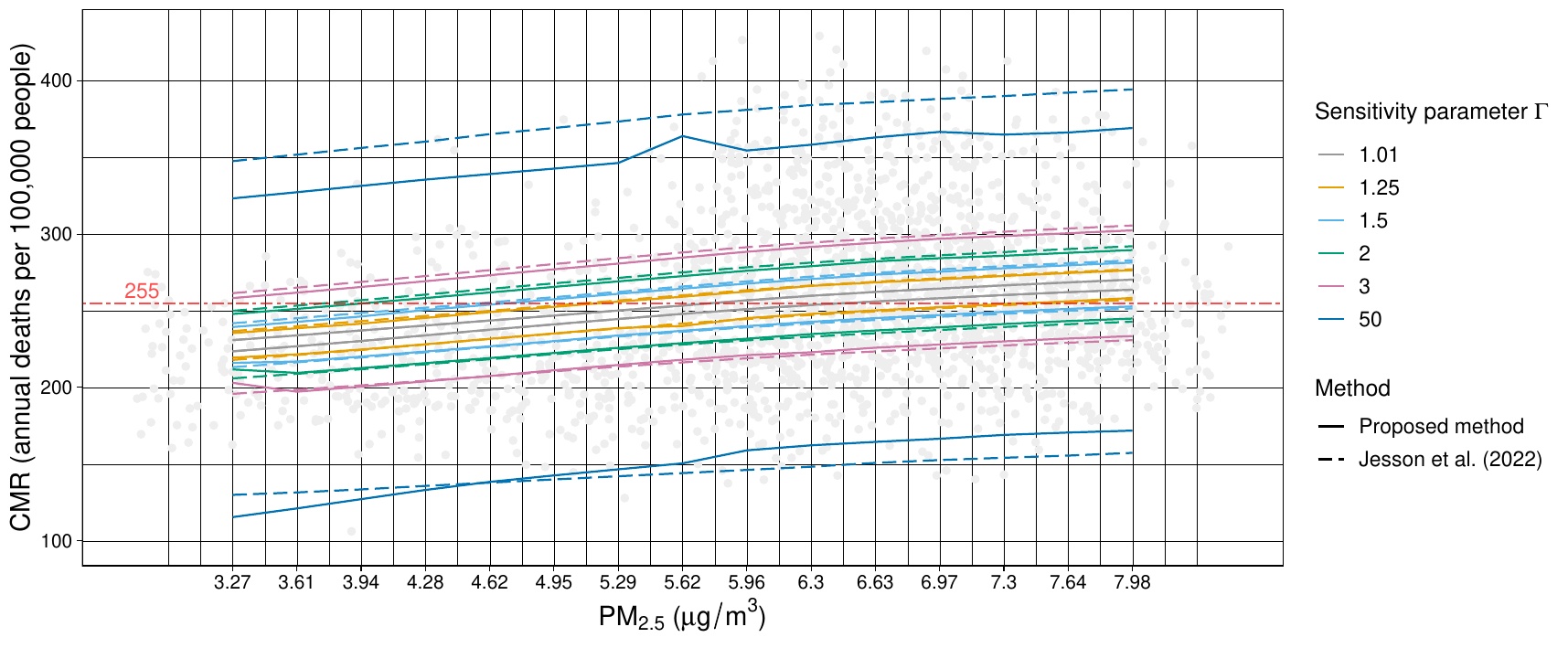}
    \vspace{.3in}
    \caption{Sensitivity analysis of the real dataset (95\%-level confidence intervals) with the proposed (partially robust) method and the one from \textcite{jesson2022scalable} for 6 values of $\Gamma$ and 15 values of exposition $\tau$ (PM2.5). The red dotted line corresponds to the average CMR (255 annual deaths per 100~000 people). The gray points are the real dataset with 82 observations removed to improve readability.}
    \label{fig:real_sensitivity_graph_with_gamma_50}
\end{figure*}

\begin{figure*}[h]
    \centering
    \vspace{.3in}
    \includegraphics[width=\linewidth]{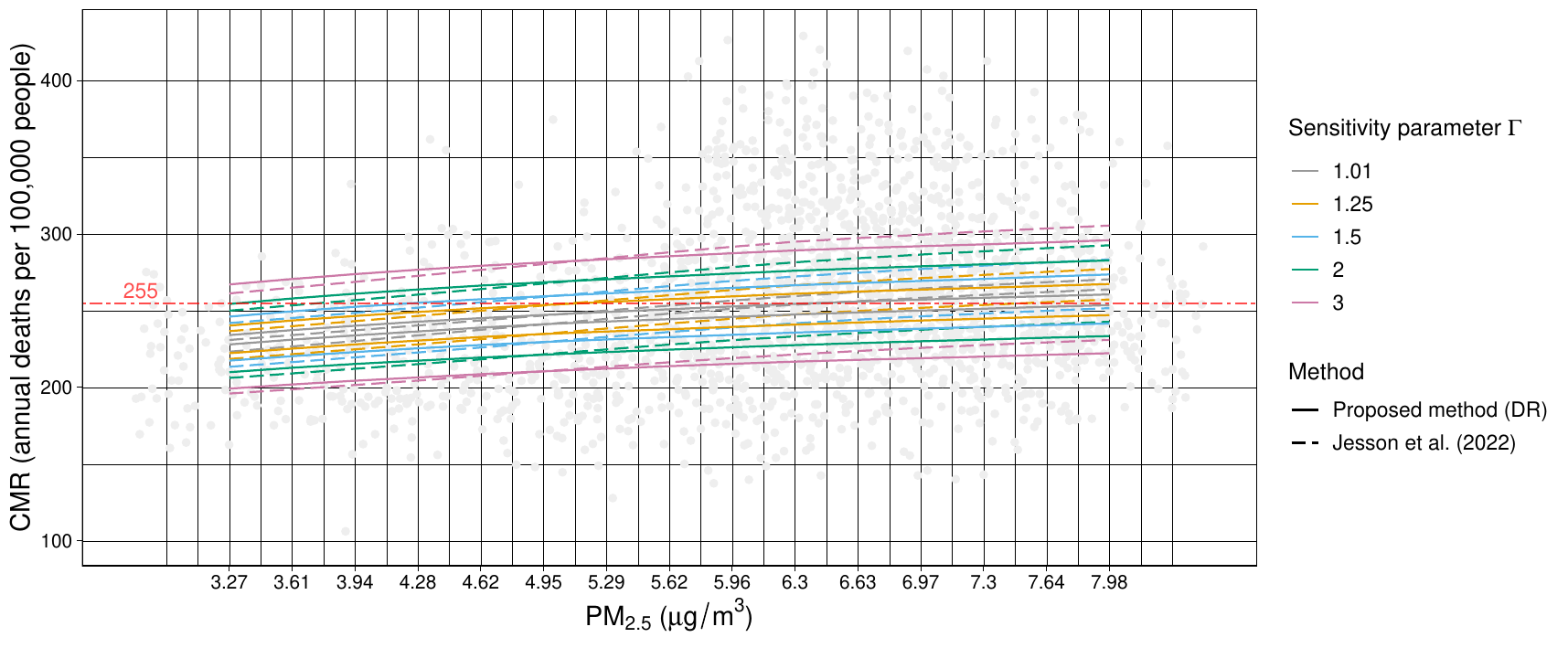}
    \vspace{.3in}
    \caption{Sensitivity analysis of the real dataset (95\%-level confidence intervals) with the proposed doubly robust method and the one from \textcite{jesson2022scalable} for 6 values of $\Gamma$ and 15 values of exposition $\tau$ (PM2.5). The red dotted line corresponds to the average CMR (255 annual deaths per 100~000 people). The gray points are the real dataset with 82 observations removed to improve readability.}
    \label{fig:real_sensitivity_graph_dr}
\end{figure*}

Figure~\ref{fig:real_sensitivity_graph_with_gamma_50} is the same sensitivity analysis as in Figure~\ref{fig:real_sensitivity_graph}, but with $\Gamma = 50$ as well.
However, high values of $\Gamma$ lead to extreme conditional quantiles, for which more suitable estimation methods than the one used in this paper should be considered. In Figure~\ref{fig:real_sensitivity_graph_dr}, we performed the same sensitivity analysis as in Figure~\ref{fig:real_sensitivity_graph}, but the proposed estimator was replaced with the doubly robust estimator. The performances of the doubly robust estimator are nearly the same as with the partially robust estimator but, as we apply propensity weight clipping, some bias appears and the general slopes of the sensitivity bounds are globally lower than the slopes of the bounds obtained with the method from \textcite{jesson2022scalable}.

\subsubsection{Estimation of \texorpdfstring{$\Gamma$}{} via Informal Benchmarking}  \label{sec:informal_benchmarking}

We estimate $\Gamma$ via informal benchmarking (see e.g., \cite{cinelli2020making}) using the sensitivity model defined in Equation~\eqref{eqn:cmsm_U}. We do not use the sensitivity model from \textcite{jesson2022scalable} as it involves $Y(\tau)$, and $f(T=\tau|\mathbf{X}=\mathbf{x}, Y(\tau)=y)$ would be harder to estimate than $f(T=\tau|\mathbf{X}=\mathbf{x}, \mathbf{U}=\mathbf{u})$.

For each of the 10 continuous observed confounders and each county, we compute the ratio from Equation~\eqref{eqn:cmsm_U}, where each observed confounder acts, one after another, as the unobserved confounder $\mathbf{U}$. In particular, we use the fitted neural network from Appendix~\ref{sec:dens_estim_nn} to estimate the GPS and we apply propensity weight clipping as explained in Subsection~\ref{sec:var_red_and_kernel_issue}. For each observed confounder $\mathbf{X}^{(j)}$, $1 \leq j \leq 10$, we then take the maximum and minimum estimated ratio across counties, denoted respectively $M_j$ and $m_j$, and estimate the confounding strength related to the $j$\textsuperscript{th} observed covariate $\Gamma_j$ by $\hat \Gamma_j = \max(M_j, 1/m_j)$. Table~\ref{tab:inf_bench} summarizes the estimated confounding strengths for each observed covariate.

\begin{table}[h]
    \centering
    \caption{Estimated confounding strengths for each observed covariate via informal benchmarking.}
    \begin{tabular}{|cc|cc|}
        \hline
        \textbf{Observed confounder} & $\hat \Gamma_j$ & \textbf{Observed confounder} & $\hat \Gamma_j$ \\
        \hline
        \texttt{pctfam\_pover\_2010} & 3.36 & \texttt{population\_2000} & 5.00 \\
        \texttt{healthfac\_2005\_1999} & 3.52 & \texttt{owner\_occ\_pct\_2010} & 5.08 \\
        \texttt{median\_HH\_inc\_2010} & 3.84 & \texttt{civil\_unemploy\_2010} & 5.23 \\
        \texttt{femaleHH\_ns\_pct\_2010} & 3.94 & \texttt{SES\_index\_2010} & 5.41 \\
        \texttt{eduattain\_HS\_2010} & 4.94 & \texttt{vacant\_HHunit\_2010} & 6.25 \\
        \hline
    \end{tabular}
    \label{tab:inf_bench}
\end{table}

Finally, even if this hypothesis can never be verified, we assume that the real confounding strength $\Gamma$ associated with the true unobserved confounder $\mathbf{U}$ lies in $[\min_j(\hat \Gamma_j), \max_j(\hat \Gamma_j)]$. Therefore, ``at best", $\mathbf{U}$ has a confounding strength equal to $\min_j(\hat \Gamma_j)$ (here, 3.36), and ``at worst", a confounding strength equal to $\max_j(\hat \Gamma_j)$ (here, 6.25). In practice, we estimate $\Gamma$ by taking the worst case, i.e.\ $\hat{\Gamma} = 6.25$, because we prefer overestimating $\Gamma$ than underestimating it.

\subsection{Libraries and Licenses}  \label{sec:libraries_licenses}

Libraries from Table~\ref{tab:libraries_licenses} are used in the proposed R implementation.

\begin{table}[H]
    \centering
    \caption{Libraries and corresponding licenses.}
    \begin{tabular}{|c|c|c|c|}
        \hline
        \textbf{Library} & \textbf{Authors} & \textbf{Version} & \textbf{License} \\
        \hline
        \texttt{foreach} & \textcite{foreach} & 1.5.2 & Apache License (== 2.0) \\
        \texttt{ggplot2} & \textcite{ggplot2} & 3.4.4 & MIT \\
        \texttt{latex2exp} & \textcite{latex2exp} & 0.9.6 & MIT \\
        \texttt{scales} & \textcite{scales} & 1.3.0 & MIT \\
        \texttt{tictoc} & \textcite{tictoc} & 1.2.1 & Apache License (== 2.0) \\
        \texttt{torch} & \textcite{torch} & 0.12.0 & MIT \\
        \hline
    \end{tabular}
    \label{tab:libraries_licenses}
\end{table}

\end{document}